\documentclass{amsart}
\usepackage{amssymb}

\usepackage{graphicx}
\usepackage{amscd}

\newtheorem{theorem}{Theorem}
\theoremstyle{plain}

\newtheorem{corollary}{Corollary}

\newtheorem{definition}{Definition}
\newtheorem{example}{Example}

\newtheorem{notation}{Notation}

\newtheorem{remark}{Remark}

\numberwithin{equation}{section}

\begin{document}

\begin{center}
\bigskip {\Huge Quantum physics with a hidden variable}

\bigskip

{\LARGE Antonio Cassa}

antonio.cassa@tin.it

http://xoomer.virgilio.it/cassa\_lazzereschi/antonio/welcome.htm

\bigskip

\bigskip

\bigskip {\LARGE Abstract}
\end{center}

Every quantum physical system can be considered the ''shadow'' of a special
kind of classical system.

The system proposed here is classical mainly because each observable
function has a well precise value on each state of the system: an
hypothetical observer able to prepare the system exactly in an assigned
state and able to build a measuring apparatus perfectly corresponding to a
required observable gets always the same real value.

The same system considered instead by an unexpert observer, affected by the
ignorance of a hidden variable, is described by a statistical theory giving
exactly and without exception the states, the observables, the dynamics and
the probabilities prescribed for the usual quantum system.

\bigskip

\bigskip

\section{\protect\LARGE Introduction}

\bigskip

In the following will be presented a system essentially classical: its
states are points of a manifold, its observables are functions on the
manifold, its dynamics are generated by functions and its differentiable
dynamics come from differentiable ''hamiltonian'' vector fields.

This system, on the other hand, is not perfectly classical because its
observable functions are, in general, far from being continuous or
differentiable and do not make a vector space; however when you consider
their ''mean value functions'' (that is integrated with respect to the
hidden variable) you get the ''expected value functions'' of the usual
quantum theory and these, when defined everywhere, are differentiable
functions making a vector space.

The system differs from a classical one also because you need two functions
to define a general dynamic: one is the hamiltonian ''expected value
function'' and the other is a differentiable function measuring an increment
in the speed of changement of the ''hidden variable''; when these two
functions are both smooth the dynamic comes from a smooth vector field.

\bigskip

The system to be examined will be introduced following a list of remarks:

\bigskip

\textbf{A}. The states of the system live inside an infinite dimensional
real Hilbert space $\mathcal{H}$ and the \textbf{state space} to consider is
the infinite dimensional spheric hypersurface $\mathbb{S=S(}\sqrt{2})$ of
radius $\sqrt{2}$ with its geometry.

This radius is chosen in such a way to have :
\begin{equation*}
\hbar =1
\end{equation*}
in the Schroedinger equation (choosing a different radius $r$ would imply a
Schroedinger equation with the constant $\frac{1}{2}r^{2}$ instead of $1$ in
the place of $\hbar $).

\bigskip

\textbf{B}. For every state $\varphi $ in $\mathbb{S}$ there is family of
states ''equivalent'' to $\varphi $ and differing by $\varphi $ only by a
''phase''. More precisely there is an \textbf{action of the group} $\mathbb{S%
}^{1}$ on $\mathbb{S}$ making the vector $\varphi $ to travel along the $%
\mathbb{S}^{1}$-orbit of states $\left[ \varphi \right] =\left\{ \rho
_{\theta }\varphi \right\} $ (or $\left\{ e^{i\theta }\cdot \varphi \right\}
$).

This action and in particular the operator $J=\rho _{\pi /2}$ allows to
consider, if we need it, the vector space $\mathcal{H}$ as a complex vector
space.

These $\mathbb{S}^{1}$-orbits have a twofold structure: ''from the outside''
they are riemannian circles embedded in $\mathbb{S}$, ''from the inside''
they are only probability spaces, that is sets $\left[ \varphi \right] $
furnished with a $\sigma $-algebra of measurable subsets and a measure
function $\mu _{\left[ \varphi \right] }$ taking value $1$ on the whole set $%
\left[ \varphi \right] $.

The states $\rho _{\theta }\varphi $ in the present theory will play the
role of \ the ''hidden states'' behind the ''apparent state'' $\left[
\varphi \right] $.

\bigskip

\textbf{C}. On the system can be carried out some propositions, that is
observable functions taking only the values $0$ or $1$; a proposition is
obviously characterized by the set $L$ where the function takes value $1$.

We will consider as \textbf{propositions }for our system the subsets $L$ of $%
\mathbb{S}$ satisfying the following conditions:

\begin{itemize}
\item  The measure $\mu _{\left[ \varphi \right] }(L\cap \left[ \varphi %
\right] )$ varies differentiably with the vector $\varphi $ (or the class $%
\left[ \varphi \right] $)

\item  Given two orthogonal vectors $\varphi $ and $\psi $ of $\mathbb{S}$
(not in the same $\mathbb{S}^{1}$-orbit) consider the vectors:
\begin{equation*}
\varphi (t)=\cos t\cdot \varphi +\sin t\cdot \psi
\end{equation*}

(the \textbf{sovrappositions} of $\varphi $ and $\psi $\ in $\mathbb{S}$) we
require that $\mu _{\left[ \varphi (t)\right] }(L\cap \left[ \varphi (t)%
\right] )$, as a function of the variable $t$, must take the form:
\begin{equation*}
\mu _{\left[ \varphi (t)\right] }(L\cap \left[ \varphi (t)\right] )=a\cdot
\cos ^{2}t+b\cdot \sin t\cos t+c\cdot \sin ^{2}t
\end{equation*}
.

\item  If $\mu _{\left[ \varphi (t)\right] }(L\cap \left[ \varphi (t)\right]
)$ is not always $0$ or always $1$, given $\varphi $ in $\mathbb{S}$, for
some orthogonal vector $\psi $ of $\mathbb{S}$ (not in $\left[ \varphi %
\right] $) the function $\mu _{\left[ \varphi (t)\right] }(L\cap \left[
\varphi (t)\right] )$ takes all the values in the interval $\left[ 0,1\right]
$.
\end{itemize}

All these subsets $L$ make a family $\mathcal{L}$ of subsets in $\mathbb{S}$
(called the \textbf{logic} of the system).

The family $\mathcal{L}$ is closed by complementation but is not a boolean
algebra (or a logic as in [V]), however it contains infinite\ boolean
algebras corresponding to the boolean algebras of commuting projectors in $%
\mathcal{H}$.

\bigskip

\textbf{D}. A function $f:\mathbb{S\rightarrow R}$ will be called an \textbf{%
observable} function if it allows, for every borel subset $B$ of $\mathbb{R}$%
, to check wether or not the function $f$ , on a given vector $\varphi $,
takes its value in $B$; that is if $f^{-1}(B)$ of $\mathbb{S}$ is a
proposition in $\mathcal{L}$ for every borel subset $B$.

All these functions make a family $\mathcal{O}$.

Each observable function has a well precise value on each state of the
system independently from the corresponding ''measuring process''; an
hypothetical expert observer able to prepare the system exactly in the state
$\varphi $ and able to build a measuring apparatus perfectly corresponding
to the observable function $f$ would get \textbf{always} the real value $%
f(\varphi )$.

The family $\mathcal{O}$ is not an algebra or a vector space, however it
contains infinite commutative functions algebras corresponding to the
commutative algebras of self-adjoint operators.

To an observable function $f$ it's associated its ''\textbf{mean value}''
function:

\begin{equation*}
\left\langle f\right\rangle (\varphi )=\int_{\left[ \varphi \right] }f(\psi
)\cdot d\mu _{\left[ \varphi \right] }
\end{equation*}

In general $\left\langle f\right\rangle $ is not defined on all $\mathbb{S}$%
, however when this happens the function $\left\langle f\right\rangle $ is
smooth (a \textbf{smooth kaehlerian}) function. All these functions (called
\textbf{kaehlerian}) make a space $\mathcal{K(}\mathbb{S)}$ definable
through the geometry of the space $\mathbb{S}$.

\bigskip

\textbf{E}.What is a symmetry for the system $\mathbb{S}$? The bijective
maps $\Phi :\mathbb{S\rightarrow S}$ respecting the scalar products, the
sovrappositions of orthogonal vectors and commuting with the actions $\rho
_{\theta }$ are good candidates: it is not difficult to verify that these
maps are exactly the unitary maps of $\mathcal{H}$ making the group $Unit(%
\mathcal{H}\mathbb{)}$.

But maybe is more suitable here to consider as \textbf{symmetries} the
diffeomorphisms $\Phi :\mathbb{S\rightarrow S}$ that are characterized by
the properties to bring $\mathbb{S}^{1}$-orbits in $\mathbb{S}^{1}$-orbits,
couples of orthogonal $\mathbb{S}^{1}$-orbits in couples of orthogonal $%
\mathbb{S}^{1}$-orbits and to commute with the actions $\rho _{\theta }$
(plus a technical condition); in this way a much wider group $Aut(\mathbb{S)}
$ of symmetries is obtained.

With respect to the group $Aut(\mathbb{S)}$ the \textbf{dynamics} of the
system are the one-parameter (continuous) groups $\Phi _{\cdot }:\mathbb{%
R\rightarrow }Aut(\mathbb{S)}$.It can be proved that with a natural topology
on $Aut(\mathbb{S)}$ all the one-parameter continuous groups in $Aut(\mathbb{%
S)}$ are given by two functions: a kaehlerian function $l$ giving origin to
a one-parameter unitary group that ''moves the $\mathbb{S}^{1}$-orbits'' and
a differentiable function $h$ on $\mathbb{S}$ (constant on the $\mathbb{S}%
^{1}$-orbits), incrementing the speed ''inside the $\mathbb{S}^{1} $%
-orbits''.

\bigskip

\textbf{F}. The system $\mathbb{S}$ is essentially a classical system to the
eyes of an hypotetical observer that we will call the \textbf{precise
observer}.

How appears the same system to an \textbf{imprecise observer} not having
under control the phase? Let'suppose that this observer is not able, for
pratical limits or intrinsic reasons, to produce the state $\varphi $ rather
than its rotated state $\rho _{\theta }\varphi $.

When he tries to prepare the system in the state $\varphi $ he can be
precise enough to prepare a state inside the $\mathbb{S}^{1}$-orbit $\left[
\varphi \right] =\left\{ \rho _{\theta }\varphi :0\leq \theta <2\pi \right\}
$ but he does not know which state in the $\mathbb{S}^{1}$-orbit is the
outcome; he cannot avoid to the state produced to be completely random in
its $\mathbb{S}^{1}$-orbit.

When he tries to measure the observable $f$ on the state $\varphi $ he can
get anyone of the values $\left\{ f(\psi );\psi \in \left[ \varphi \right]
\right\} $. After a large number of trials he gets his outcomes distributed
on the real line and, in the end, what he gets are only the numbers $\pi
(\varphi ,f,B)$ expressing the probabilities that the outcome falls in a
general borel subset $B$ (varying in the family $\mathcal{B(}\mathbb{R)}$ of
all borelian subsets of $\mathbb{R}$).

This probabilty, from the other side, measures the frequency for $\varphi $,
moving randomly in $\left[ \varphi \right] $, to fall in the set $%
f^{-1}(B)\cap \left[ \varphi \right] $. Therefore:

\begin{equation*}
\pi (\varphi ,f,B)=\mu _{\left[ \varphi \right] }(f^{-1}(B)\cap \left[
\varphi \right] )
\end{equation*}

\bigskip

\textbf{G}. The imprecise observer is compelled to consider the measuring
process intrinsically statistic; the space $\mathbb{S}$ keeps for him only
the meaning of the set of all possible preparations of the system and $%
\mathcal{O}$ keeps only the meaning of the set of all realizable measuring
apparatuses. All his experimental knowledge reduces to a map:

\begin{equation*}
\pi :\mathbb{S\times }\mathcal{O\times B(}\mathbb{R)\rightarrow }\left[ 0,1%
\right]
\end{equation*}

But now why he should consider different two preparations $\varphi _{1\text{
}}$and $\varphi _{2\text{ }}$if:
\begin{equation*}
\pi (\varphi _{1},f,B)=\pi (\varphi _{2},f,B)
\end{equation*}

for every apparatus $f$ and every borelian subset $B$ ?

Dually why he should consider different two apparatuses $f_{1\text{ }}$and $%
f_{2\text{ }}$if:
\begin{equation*}
\pi (\varphi ,f_{1},B)=\pi (\varphi ,f_{2},B)
\end{equation*}

for every preparation $\varphi $ and every borelian subset $B$ ?

His imprecision generates an equivalence relation $\mathcal{R}_{\mathbb{S}}$
among states in $\mathbb{S}$ and an equivalence relation $\mathcal{R}_{%
\mathcal{O}}$ among observables in $\mathcal{O}$: for the imprecise observer
the ''states'' he can define through his experiments are the equivalence
classes of $\mathcal{R}_{\mathbb{S}}$ in $\mathbb{S}$ and his ''state
space'' is the quotient space $\widehat{\mathbb{S}}=\mathbb{S}/\mathcal{R}_{%
\mathbb{S}}$, analogously his ''observable space'' is $\widehat{\mathcal{O}}=%
\mathcal{O}/\mathcal{R}_{\mathcal{O}}$. Over these objects he has a well
defined probability map:

\begin{equation*}
\widehat{\pi }:\widehat{\mathbb{S}}\mathbb{\times }\widehat{\mathcal{O}}%
\mathcal{\times B(}\mathbb{R)\rightarrow }\left[ 0,1\right]
\end{equation*}

\bigskip

\textbf{H}. Analogously he will consider his symmetries on $\widehat{\mathbb{%
S}}$ and not on $\mathbb{S}$; since the ''hidden'' symmetries respect the
equivalence relation $\mathcal{R}_{\mathbb{S}}$ the symmetry group $Aut(%
\mathbb{S)}$ acts naturally on $\widehat{\mathbb{S}}$ and so if we introduce
the subgroup $Aut_{I}(\mathbb{S)}$ of $Aut(\mathbb{S)}$ made by all the
symmetries acting identically on $\widehat{\mathbb{S}}$ the quotient group $%
\widehat{Aut(\mathbb{S)}}=Aut(\mathbb{S)}/Aut_{I}(\mathbb{S)}$ acts
effectively on $\widehat{\mathbb{S}}$ and it can be proved that gives the
right group of ''apparent'' symmetries for the imprecise observer.

Moreover every continuous dynamic in $Aut(\mathbb{S)}$ induces a dynamic in $%
\widehat{\mathbb{S}}$, made of transformations in $\widehat{Aut(\mathbb{S)}}$%
, continuous for the induced topology of $\widehat{Aut(\mathbb{S)}}$ and
conversely \ it is possible to prove that every continuous dynamic in $%
\widehat{Aut(\mathbb{S)}}$ comes from a continuous dynamic in $Aut(\mathbb{S)%
}$.

\bigskip

\textbf{I}. Briefly starting with the ingredients of ''classical'' system $%
\mathbb{S}$:

\begin{equation*}
(\mathbb{S},\mathcal{L},\mu _{\left[ \varphi \right] },\mathcal{O},Aut(%
\mathbb{S)})
\end{equation*}
in consideration by the precise observer, the ignorance of the phase induces
the imprecise observer to consider instead the statistical system:
\begin{equation*}
(\widehat{\mathbb{S}},\widehat{\mathcal{O}},\widehat{\pi },\widehat{Aut(%
\mathbb{S)}}).
\end{equation*}
of states, observables, probabilities and symmetries.

What it is proved in this paper, in short, it is exactly that this last
system is (it is isomorphic to)\textbf{\ the usual quantum system.}

Precisely the state space $\widehat{\mathbb{S}}$ is the complex projective
space of $\mathcal{H}$, the observable space $\widehat{\mathcal{O}}$ is
naturally isomorphic to the set of all self-adjoint operators of $\mathcal{H}
$ and the probability map $\widehat{\pi }$ becomes the probability map $%
\widehat{\pi }(\left[ \varphi \right] ,A,B)=\left\langle
E_{B}^{A}\right\rangle _{\varphi }$ (where $E_{B}^{A}$ is the projector
associated to the borel subset $B$ in the projector valued measure defined
by the self-adjoint operator $A$) of the canonical quantum theory; note also
that every observable function takes its essential values in the set of the
true outcomes of the corresponding quantum observable (the spectrum of its
associated self-adjoint operator).

Moreover $\widehat{Aut(\mathbb{S)}}$ becomes the symmetry group of $\mathbb{P%
}_{\mathbb{C}}(\mathcal{H})$, that is the group of unitary transformations
of $\mathcal{H}$ modulo the multiples of the identity.

This statement gives a rigorous content to to the assertion that a quantum
physical system can be considered the ''shadow'' of a classical system with
a ''hidden variable''.

\bigskip

Some of the characteristic features of the present hidden variable theory
are to be declared :

\bigskip

\begin{enumerate}
\item  \textbf{This theory is non local}: \ there is no room for (non-banal)
properties with the special independence required in the proof of the Bell
inequalities and wishful to represent apparatuses\ acting independently in
two spatially separated regions of the spacetime.

\item  \textbf{This theory is contextual: }behind a quantum proposition
there are in $\mathbb{S}$ infinite ''hidden'' classical propositions,
therefore the truth value $0$ or $1$ of the classical proposition on a
''hidden'' classical state depends not only on the ''hidden variable'' of
the state but also on the ''experimental context'' defined by the particular
classical proposition in consideration. The same holds for the observables.

\item  \textbf{This theory is ''relativistically invariant'': }it works
equally well\textbf{\ }either if you have assigned for the quantum system a
unitary representation of the Galilei group or a unitary representation of
the Poincar\`{e} group; it works also in the general relativistic case as
long as you have a quantum theory via a Hilbert space.
\end{enumerate}

\bigskip

I wish to thank, first of all, my son Andrea for his constant support, for
his sharp remarks and his challenging questions during ours periodical
talks. Moreover I am grateful to the scientists Maria Cristina Abbati, Renzo
Cirelli, Mauro Gatti and Alessandro Mani\`{a} of the Milan school for
theoretical physics for their help and for their enlightening geometrical
foundation of quantum mechanics.

I want to mention also Valter Moretti, Cesare Reina and Marco Toller with
whom I had, during these years, the opportunity to freely discuss the
problems of the quantum theory.

\bigskip

\bigskip

\section{{\protect\LARGE The system }$\mathbb{S}${\protect\LARGE , its
states and observables}}

\bigskip

\textit{In the following }$(\mathcal{H},\langle .,.\rangle )$\textit{\ will
denote an infinite dimensional \textbf{real Hilbert space} furnished with an
effective \textbf{action} }$\rho :\mathbb{S}^{1}\rightarrow Isom(\mathcal{H}%
,\langle .,.\rangle )$\textit{\ of the group }$\mathbb{S}^{1}$\textit{on }$%
\mathcal{H}$\textit{\ via isometries.}

\textit{For the action\ }$\rho $ \textit{we suppose valid the property: }
\begin{equation*}
\rho _{\theta }\varphi =\cos \theta \cdot \varphi +\sin \theta \cdot \rho
_{\pi /2}\varphi
\end{equation*}
\textit{where we denote, for simplicity, an element of }$\mathbb{S}^{1}$%
\textit{as a real number }$\theta $\textit{\ (implicitally modulo }$2\pi $%
\textit{) and we write }$\rho _{\theta }\varphi $ \textit{instead of }$\rho
(e^{i\theta })(\varphi )$\textit{.}

\textit{The special isometry }$\rho _{\pi /2}$\textit{\ will be denoted by }$%
J:\mathcal{H}\rightarrow \mathcal{H}$\textit{; obviously we get }$J^{2}=-id_{%
\mathcal{H}}$\textit{\ and }$\langle \varphi ,J\varphi \rangle =0$\textit{\
for every }$\varphi $\textit{\ in }$\mathcal{H}$\textit{.Trought }$J$\textit{%
\ the space }$\mathcal{H}$\textit{\ becomes a complex Hilbert space with }$%
\rho _{\theta }\varphi =e^{i\theta }\cdot \varphi $\textit{; we can also
consider on it the sesquilinear form defined by: }
\begin{equation*}
\langle \langle \varphi ,\psi \rangle \rangle =\langle \varphi ,\psi \rangle
+i\cdot \langle J\varphi ,\psi \rangle
\end{equation*}

\textit{However in general it appears preferable to consider }$\mathcal{H}$%
\textit{\ a real Hilbert space with the action }$\rho $\textit{\ instead of
a complex Hilbert space.}

\textit{Obviously two vectors }$\varphi ,\psi $\textit{\ in} $\mathcal{H}$
\textit{are orthogonal with respect to the sesquilinear form }$\langle
\langle \cdot ,\cdot \rangle \rangle $\textit{\ if and only if }$\psi $%
\textit{\ is orthogonal to }$\varphi $\textit{\ and }$J\varphi $\textit{\
with respect to the real scalar product }$\langle \cdot ,\cdot \rangle $%
\textit{.}

\textit{Afterwords we will refer always to to the ortogonality with respect
to the real scalar product }$\langle \cdot ,\cdot \rangle $\textit{, unless
dealing with complex objects like complex projectors or complex linear
subspaces.}

\bigskip

\begin{definition}
Our \textbf{space of states}\textit{\ }will be the (hypersurface) sphere of
radius $\sqrt{2\text{:}}$%
\begin{equation*}
\mathbb{S=S(}\sqrt{2}\mathbb{)=}\left\{ \varphi \in \mathcal{H}\text{ : }%
\left\| \varphi \right\| =\sqrt{2}\right\}
\end{equation*}
\end{definition}

\bigskip

\textit{The space} $\mathbb{S}$ \textit{is an infinite dimensional
riemannian manifold modelled on a Hilbert space; its radius }$\sqrt{2}$
\textit{is chosen in such a way to have :}
\begin{equation*}
\hbar =1
\end{equation*}
\textit{in the Schroedinger equation; choosing a different radius} $r$
\textit{would imply a Schroedinger equation with the constant} $\frac{1}{2}%
r^{2}$ \textit{in the place of the Plank constant, for example the more
elegant choice }$r=1$ \textit{would imply the uncommon choice for the Plank
constant} $\hbar =\frac{1}{2}$.

\textit{The space} $\mathcal{H}$ \textit{contains spherical hypersurfaces
where, formally, the Plank constant} $\hbar $ \textit{takes all possible
positive values.} \textit{In this perspective} on $\mathcal{H}$\textit{\ is
defined a non-negative ''quadratic'' function }$\widehat{\hbar }(\cdot ):%
\mathcal{H}$ $\rightarrow \mathbb{R}$ \textit{\ inducing a norm} $\left\|
\varphi \right\| =\sqrt{2\widehat{\hbar }(\varphi )}$\textit{and a real
scalar product} $\left\langle \varphi ,\psi \right\rangle =\widehat{\hbar }%
(\varphi +\psi )-\widehat{\hbar }(\varphi )-\widehat{\hbar }(\psi )$.

\textit{The real tangent space} $T_{\varphi }\mathbb{S}$ $=(\varphi )^{\perp
}$ \textit{contains the vector} $J\varphi $ \textit{and can be splitted in a
''\textbf{vertical part}''} $\mathcal{V}er_{\varphi }=\mathbb{R\cdot }%
J\varphi $ \textit{and in a ''\textbf{horizontal part}''} $\mathcal{H}%
or_{\varphi }=\left\{ \varphi ,J\varphi \right\} ^{\perp }$. \textit{The map}
$J$ \textit{sends} $\mathcal{H}or_{\varphi }$ \textit{in itself, we will
denote this restriction by }$J_{\varphi }$. \textit{Analogously a map:} $%
\rho _{\theta \varphi }:\mathcal{H}or_{\varphi }\rightarrow \mathcal{H}%
or_{\varphi }$ \textit{can be defined as} $\rho _{\theta \varphi }(X)=\cos
\theta \cdot X+\sin \theta \cdot J_{\varphi }X=e^{i\theta }\cdot X$.

\bigskip

\textit{The group} $\mathbb{S}^{1}$ \textit{acts on} $\mathbb{S}$ \textit{%
describing the} $\mathbb{S}^{1}$-\textbf{orbits} $\left[ \varphi \right]
=\left\{ \rho _{\theta }\varphi :\theta \in \mathbb{R}\right\} $; \textit{on
each} $\mathbb{S}^{1}$-\textit{orbit} $\left[ \varphi \right] $ \textit{%
there is only one measure} $\mu _{\left[ \varphi \right] }$ \textit{on the
natural borelian subsets having total measure equal to }$1$\textit{\ and
making measure preserving the natural correspondence }$\theta \longmapsto
\rho _{\theta }\varphi $\textit{\ between }$\mathbb{S}^{1}$, \textit{with
the normalized Haar measure, and }$\left[ \varphi \right] $ . \textit{Note
that} $T_{\varphi }\left[ \varphi \right] =\mathcal{V}er_{\varphi }=\mathbb{%
R\cdot }J\varphi $.

\textit{Given two vectors }$\varphi $\textit{\ and }$\psi $\textit{\ in the
same} $\mathbb{S}^{1}$\textit{-orbit we will denote by }$\psi /\varphi $%
\textit{\ the unique complex number }$u\in \mathbb{S}^{1}$ \textit{such that
}$\psi =\rho (u)(\varphi )$.

\textit{In each} $\mathbb{S}^{1}$\textit{-orbit is well defined a metric }$%
d_{\left[ \varphi \right] }:\left[ \varphi \right] \times \left[ \varphi %
\right] \rightarrow \left[ 0,\pi \right] $\textit{\ given by: }$d_{\left[
\varphi \right] }(\varphi ,\psi )=\left| Arg(\psi /\varphi )\right| $ (where
$Arg:\mathbb{S}^{1}\rightarrow ]-\pi ,\pi ]$ \textit{verifies} $%
Arg(e^{i\theta })=\theta $ \textit{for every } $\theta $ in $]-\pi ,\pi ]$).%
\textit{\ We will refer to this metric as the \textbf{phase distance in }}$%
\left[ \varphi \right] $\textit{.}

\bigskip

\begin{definition}
A subset $B$ of $\mathbb{S}$ will be called a \textbf{pseudo-borel subset}
if every intersection $B\cap \left[ \varphi \right] $ is Borel in $\left[
\varphi \right] $ ; a pseudo-borel subset of $\mathbb{S}$ will be called a
\textbf{pseudo-borel null subset} if every intersection $B\cap \left[
\varphi \right] $ is Borel null in $\left[ \varphi \right] .$
\end{definition}

\bigskip

\begin{definition}
Two pseudo-borel subsets $B$ and $C$ of $\mathbb{S}$ will be called \textbf{%
equivalent up to a pseudo-borel null subset} (or \textbf{null equivalent})
if their symmetric difference is a pseudo-borel null subset; moreover $B$
and $C$ will be called \textbf{equivalent in measure} if
\begin{equation*}
\mu _{\left[ \varphi \right] }(B\cap \left[ \varphi \right] )=\mu _{\left[
\varphi \right] }(C\cap \left[ \varphi \right] )
\end{equation*}
for every $\mathbb{S}^{1}$-orbit $\left[ \varphi \right] .$
\end{definition}

\bigskip

\textit{Given two orthogonal vectors }$\varphi $\textit{\ and }$\psi $%
\textit{\ in} $\mathbb{S}$ \ \textit{the map:} $\gamma _{\varphi \psi }:%
\mathbb{R\rightarrow S}$ \textit{defined by: }
\begin{equation*}
\gamma _{\varphi \psi }(t)=\cos t\cdot \varphi +\sin t\cdot \psi
\end{equation*}
\textit{parametrizes a \textbf{maximal circle} (a geodesics curve) in }$%
\mathbb{S}$.

\bigskip

\begin{definition}
A pseudo-borel subset $L$ of $\mathbb{S}$ will be called a \textbf{%
proposition} of $\mathbb{S}$ if :

\begin{itemize}
\item  the function $\varphi \longmapsto \mu _{\left[ \varphi \right]
}(L\cap \left[ \varphi \right] )$ is differentiable between $\mathbb{S}$ and
$\left[ 0,1\right] $

\item  given two orthogonal vectors $\varphi $ and $\psi $ in $\mathbb{S}$
the function $\mu _{\left[ \gamma _{\varphi \psi }(t)\right] }(L\cap \left[
\gamma _{\varphi \psi }(t)\right] )$ in the variable $t$ is a function of
the form
\begin{equation*}
a\cdot \cos ^{2}t+b\cdot \sin t\cos t+c\cdot \sin ^{2}t
\end{equation*}

\item  unless $L$ is equivalent in measure to $\emptyset $ or to $\mathbb{S}$%
, for every $\varphi $ in $\mathbb{S}$ there is a $\psi $ in $\mathbb{S}$
orthogonal to $\varphi $ such that the function $t\longmapsto \mu _{\left[
\gamma _{\varphi \psi }(t)\right] }(L\cap \left[ \gamma _{\varphi \psi }(t)%
\right] )$ takes all the values of $\left[ 0,1\right] $.
\end{itemize}
\end{definition}

\bigskip

\textit{This definition takes into account the behaviour of quantum
probabilities for a quantum property; infact if the property is represented
by the (complex) projector }$E$ \textit{you get}:

\begin{itemize}
\item  \textit{the map} $\varphi \mapsto \left\langle E\right\rangle
_{\varphi }$ $=\frac{1}{2}\left\langle \varphi ,E\varphi \right\rangle $%
\textit{\ is differentiable in }$\varphi $

\item  i\textit{f you take two orthogonal elements }$\varphi $\textit{\ and }%
$\psi $\textit{\ in} $\mathbb{S}$\textit{\ and consider the states
parametrized by the path }$\gamma (t)=\cos t\cdot \varphi +\sin t\cdot \psi $%
\textit{\ (the superposition states of }$\varphi $\textit{\ and }$\psi $%
\textit{\ in} $\mathbb{S}$\textit{) then the function of }$t$\textit{\ given
by }$\left\langle E\right\rangle _{\gamma (t)}$\textit{\ is: }
\begin{equation*}
\frac{1}{2}\left\langle \gamma (t),E(\gamma (t))\right\rangle =\cos
^{2}t\cdot \left\langle E\right\rangle _{\varphi }+\sin t\cos t\cdot
\left\langle \varphi ,E\psi \right\rangle +\sin ^{2}t\cdot \left\langle
E\right\rangle _{\psi }
\end{equation*}

\item  \textit{when }$\varphi $\textit{\ is in }$ImE$\textit{\ and }$\psi $%
\textit{\ is in }$\ker E$\textit{\ then the function above becomes equal to }%
$\cos ^{2}t$\textit{\ and takes all the values of }$\left[ 0,1\right] $%
\textit{. If }$E$\textit{\ is not }$0$\textit{\ or }$I$\textit{\ given }$%
\varphi $\textit{\ you can always find an orthogonal }$\psi $\textit{\ in
such a way that }$\gamma _{\varphi \psi }(t)$\textit{\ meets }$ImE$\textit{\
in }$\sigma $\textit{and }$\ker E$\textit{\ in }$\tau $\textit{, then }$%
\gamma _{\varphi \psi }$\textit{\ as }$\gamma _{\sigma \tau }$\textit{\
takes all the values of }$\left[ 0,1\right] $\textit{.}
\end{itemize}

\bigskip

\begin{definition}
The set $\mathcal{L}$ of all the propositions of $\mathbb{S}$ will be called
the \textbf{logic} of $\mathbb{S}$
\end{definition}

\bigskip

\textit{The functions of the form }$a\cdot \cos ^{2}t+b\cdot \sin t\cos
t+c\cdot \sin ^{2}t$\textit{\ make a three dimensional vector space (with
base }$\left\{ \cos ^{2}t,\sin t\cdot \cos t,\sin ^{2}t\right\} $\textit{\
or }$\left\{ 1,\sin t\cdot \cos t,\sin ^{2}t\right\} $\textit{). With
respect to this last base a function }$h(t)$\textit{\ of this space can be
written as:}
\begin{equation*}
h(t)=h(0)+\dot{h}(0)\cdot \sin t\cdot \cos t+\frac{1}{2}\ddot{h}(0)\cdot
\sin ^{2}t
\end{equation*}

\textit{A function in this space takes all the values of the interval }$%
\left[ 0,1\right] $\textit{\ if and only if can be expressed as }$\cos
^{2}(t+\beta )$\textit{.}

\bigskip

\begin{remark}
\begin{itemize}
\item  \textit{The empty set }$\emptyset $\textit{\ and all the sphere} $%
\mathbb{S}$ \textit{are propositions}

\item  if $L$ is a proposition its complement $\mathbb{S\setminus }L$ is
also a proposition

\item  every pseudo-borel subset of $\mathbb{S}$ equivalent in measure to a
proposition is a proposition

\item  every pseudo-borel subset of $\mathbb{S}$ null equivalent to a
proposition is a proposition

\item  if $L$ is a proposition and $U:\mathbb{S\rightarrow S}$ is a unitary
map then $U(L)$ is also a proposition.

\item  \textit{in general the intersection or the union of two propositions
\textbf{is not} a proposition.}

\item  \textit{later on it will be proved that if }$L$\textit{\ and }$M$%
\textit{\ are propositions with }$L\subset M$\textit{\ then also }$%
M\setminus L$\textit{\ is a proposition}.
\end{itemize}
\end{remark}

\bigskip

\begin{definition}
A function $f:$ $\mathbb{S}$ $\rightarrow \mathbb{R}$ will be called a
\textbf{pseudo-borelian} function on $\mathbb{S}$ if for every borel subset $%
B$ in $\mathbb{R}$ the inverse image $f^{-1}(B)$ is a pseudo-borel subset of
$\mathbb{S}$. Two pseudo-borelian functions on $\mathbb{S}$ will be called
\textbf{null equivalent} if they differ only on a null pseudo-borel subset
of $\mathbb{S}$.
\end{definition}

\bigskip

\begin{definition}
A function $f:$ $\mathbb{S}$ $\rightarrow \mathbb{R}$ will be called an
\textbf{observable} on $\mathbb{S}$ if for every borel subset $B$ in $%
\mathbb{R}$ its inverse image $f^{-1}(B)$ is a proposition of $\mathbb{S}$.
\end{definition}

\bigskip

\begin{notation}
The symbol $\mathcal{O}$ will denote the set of all the observable functions
on the space $\mathbb{S}$.
\end{notation}

\bigskip

\begin{remark}
\begin{itemize}
\item  the characteristic function $\chi _{A}$ of a pseudo-borel subset $A$
in $\mathbb{S}$ is an observable if and only if $A$ is a proposition

\item  every constant function on $\mathbb{S}$ is an observable

\item  if $f:$ $\mathbb{S}$ $\rightarrow \mathbb{R}$ is an observable on $%
\mathbb{S}$ and $b:\mathbb{R}\rightarrow \mathbb{R}$ is a borel function,
the function $b\circ f:\mathbb{S}$ $\rightarrow \mathbb{R}$ is also an
observable

\item  if $f:$ $\mathbb{S}$ $\rightarrow \mathbb{R}$ is an observable on $%
\mathbb{S}$ then $\left| f\right| $, $f^{+}$ and $f^{-}$ are observable
functions

\item  if $f:$ $\mathbb{S}$ $\rightarrow \mathbb{R}$ is a never zero
observable then the function $1/f$ is an observable (infact $1/f=b\circ f$
where $b:\mathbb{R}\rightarrow \mathbb{R}$ is the function defined by $%
b(0)=1 $ and $b(x)=1/x$ for $x\neq 0$)

\item  if $f:$ $\mathbb{S}$ $\rightarrow \mathbb{R}$ is an observable and $%
g: $ $\mathbb{S}$ $\rightarrow \mathbb{R}$ is a (pseudo-borelian) function
null equivalent to $f$ then $g$ is also an observable

\item  if $f:$ $\mathbb{S}$ $\rightarrow \mathbb{R}$ is an observable and $k$
is a constant the functions $k\cdot f$ and $k+f$ are observables.

\item  if $f:$ $\mathbb{S}$ $\rightarrow \mathbb{R}$ is an observable and $U:%
\mathbb{S\rightarrow S}$ is a unitary map then $f\circ U^{-1}$ is an
observable.

\item  in general the sum or the product of two observable functions \textbf{%
is not} an observable function.
\end{itemize}
\end{remark}

\bigskip

\begin{theorem}
For every observable function $f$ there exists a unique open subset $\Omega
_{f}$ of $\mathbb{R}$ such that $f^{-1}(\Omega _{f})$ is pseudo-borel null
and maximal among the open subsets of $\mathbb{R}$ with this property.
\end{theorem}

\begin{proof}
The family of all the open subsets $U$ of $\mathbb{R}$ such that $f^{-1}(U)$
is a pseudo-borel null subset of $\mathbb{S}$ is not empty and contains a
maximum element (the union of all its elements since it can be expressed as
a countable union).
\end{proof}

\bigskip

\begin{definition}
Let $f$ be an observable function the \textbf{essential image of }$f$ is the
(non empty) closed subset $Im_{e}(f)=\mathbb{R}\setminus \Omega _{f}$ of $%
\mathbb{R}$.
\end{definition}

\bigskip

A\textit{\ real value }$y$\textit{\ is not in }$Im_{e}(f)$\textit{\ if and
only if there is a neighborhood }$]y-\varepsilon ,y+\varepsilon \lbrack $%
\textit{\ of }$y$ \textit{such that }$f^{-1}(]y-\varepsilon ,y+\varepsilon
\lbrack )$\textit{\ is a pseudo-borel null subset of }$\mathbb{S}$.

$Im_{e}(f)\subset \overline{f(\mathbb{S)}}$; \textit{if }$y_{0}$\textit{\ is
an isolated value of }$Im_{e}(f)$\textit{\ then }$y_{0}$\textit{\ is a value
in }$f(\mathbb{S)}$.

\textit{Two null equivalent observable functions }$f,g$\textit{\ have the
same essential image.}

\textit{It is always possible to modify an observable function }$f$\textit{\
on a pseudo-borel null subset in such a way to have }$f^{-1}(\Omega
_{f})=\emptyset $\textit{\ and then }$Im_{e}(f)=\overline{f(\mathbb{S)}}$.

\textit{When }$Im_{e}(f)$\textit{\ is a discrete subset of} $\mathbb{R}$
\textit{and} $f^{-1}(\Omega _{f})=\emptyset $ \textit{then} $Im_{e}(f)=f(%
\mathbb{S)}$.

\bigskip

\begin{notation}
Let $f:$ $\mathbb{S}$ $\rightarrow \mathbb{R}$ be an observable, the family:
\begin{equation*}
\mathcal{A}_{f}=\left\{ f^{-1}(B);B\text{ is a borel subset of }\mathbb{R}%
\right\}
\end{equation*}
is a $\sigma $-algebra of subsets of $\mathbb{S}$ contained in $\mathcal{L}$
. Let's denote by $\widehat{\mathcal{A}_{f}}$ the bigger $\sigma $-algebra
in $\mathcal{L}$:
\begin{equation*}
\widehat{\mathcal{A}_{f}}=\left\{ A:A\text{ is null equivalent to a }%
f^{-1}(B)\text{ where }B\text{ is a borel subset of }\mathbb{R}\right\}
\end{equation*}
\end{notation}

\bigskip

\begin{notation}
Given two propositions $L$ and $M$ in $\mathcal{L}$ \ the boolean algebra of
subsets of $\mathbb{S}$ generated by $L$ and $M$ will be denoted by $%
\mathcal{B}_{L,M}$:
\begin{equation*}
\mathcal{B}_{L,M}=\left\{
\begin{array}{c}
\emptyset ,L,M,L\cap M,L\cup M,\complement L,\complement M,L\cap \complement
M,\complement L\cap M, \\
\complement L\cap \complement M,L\Delta M,\complement (L\Delta M),L\cup
\complement M,\complement L\cup M,\complement L\cup \complement M,\mathbb{S}
\end{array}
\right\}
\end{equation*}
\end{notation}

\bigskip

\begin{definition}
Two propositions $L$ and $M$ in $\mathcal{L}$ \ will be called \textbf{%
compatible} if their boolean algebra of subsets $\mathcal{B}_{L,M}$ is
contained in $\mathcal{L}$ .
\end{definition}

\bigskip

\begin{theorem}
Given two propositions $L$ and $M$ there is an observable function $f$ such
that $\mathcal{A}_{f}$ contains $L$ and $M$ \ if and only if the
propositions are compatible
\end{theorem}

\begin{proof}
($\Longrightarrow )$ Obvious.

($\Longleftarrow )$The sets $X_{1}=L\cap \complement M,$ $X_{2}=\complement
L\cap M,$ $X_{3}=L\cap M,$ $X_{4}=\complement L\cap \complement M$ are
pairwise disjoints (possibly empty) in $\mathcal{L}$ and every element of $%
\mathcal{B}_{L,M}$ is obtainable as a union of them.

The function $f:\mathbb{S\rightarrow R}$ defined by $f(\varphi )=n$ if $%
\varphi \in X_{n}$ is well defined, is an observable with $L=f^{-1}\left\{
1,3\right\} $ and $M=f^{-1}\left\{ 2,3\right\} $.
\end{proof}

\bigskip

\textit{Given two propositions }$L$\textit{\ and }$M$\textit{\ we will prove
later that are compatible if and only if }$L\cap M$\textit{\ is contained in}
$\mathcal{L}$,\textit{\ therefore when }$L\cap M$\textit{\ is not contained
in }$L$\textit{\ the characteristic functions }$\chi _{L},\chi _{M}$\textit{%
\ are observable functions but the functions }$\chi _{L}+\chi _{M}$\textit{,
}$\chi _{L}\cdot \chi _{M}$\textit{\ are not.} \textit{Anyway (exercise) it
is already possible to prove now that if }$\mathcal{B}_{L,M}$ \textit{is not
contained in }$\mathcal{L}$\textit{\ then there exist two propositions }$P$%
\textit{\ and }$Q$\textit{\ in} $\mathcal{B}_{L,M}\cap \mathcal{L}$ \
\textit{with }$P\cap Q\notin \mathcal{L}$; \textit{the corresponding
characteristic functions then }$\chi _{P},\chi _{Q}$\textit{\ are observable
functions but }$\chi _{P}+\chi _{Q}$\textit{, }$\chi _{P}\cdot \chi _{Q}$%
\textit{\ are not.}

\bigskip

\textit{Inside} $\mathcal{O}$ \textit{there are some natural algebras of
functions:}

\textit{\bigskip }

\begin{notation}
Let $\mathcal{A}$ be a $\sigma $-algebra of subsets of $\mathbb{S}$
contained in $\mathcal{L}$, the symbol:
\begin{equation*}
\mathcal{O}_{\mathcal{A}}=\left\{ f;\text{ }f\in \mathcal{O}\text{: }%
f^{-1}(B)\in \mathcal{A}\text{ for every borel subset }B\text{ of }\mathbb{R}%
\right\}
\end{equation*}

will denote the space of all the observables performable in the context of
the ''classical logic'' assigned by the family $\mathcal{A}$ .
\end{notation}

\bigskip

\begin{theorem}
Let $\mathcal{A}$ be a $\sigma $-algebra of subsets of $\mathbb{S}$
contained in $\mathcal{L}$, the space $\mathcal{O}_{\mathcal{A}}$ is an
algebra (over $\mathbb{R}$) of observable functions of $\mathbb{S}$
containing the constants. Moreover $\mathcal{O}_{\mathcal{A}}$ is closed
with respect to the left composition with the borel functions of $\mathbb{R}$
and if $f:$ $\mathbb{S}$ $\rightarrow \mathbb{R}$ is a never zero observable
in $\mathcal{O}_{\mathcal{A}}$ the function $1/f$ is also in $\mathcal{O}_{%
\mathcal{A}}$.
\end{theorem}

\begin{proof}
Taken two functions $f,g$ in $\mathcal{O}_{\mathcal{A}}$ let's consider the
map $\left( f,g\right) :\mathbb{S}$ $\rightarrow \mathbb{R}^{2}$ since $%
\mathcal{A}$ is closed by intersections the inverse image with respect to $%
\left( f,g\right) $ of every open rectangle of $\mathbb{R}^{2}$ is in $%
\mathcal{A}$ and then also the inverse image with respect to $\left(
f,g\right) $ of every borel subset of $\mathbb{R}^{2}$ is in $\mathcal{A}$.

Denoted by $S:\mathbb{R}^{2}\rightarrow \mathbb{R}$ and $P:\mathbb{R}%
^{2}\rightarrow \mathbb{R}$ the two borel functions given by the operations,
respectively, of addition and multiplication on the real numbers we can
deduce that the two functions: $f+g=S\circ \left( f,g\right) $ and $f\cdot
g=P\circ \left( f,g\right) $ are in $\mathcal{O}_{\mathcal{A}}$.
\end{proof}

\bigskip

\textit{If }$\mathcal{A}$ \textit{is a }$\sigma $\textit{-algebra of subsets
of }$\mathbb{S}$ \textit{contained in} $\mathcal{L}$ \textit{closed with
respect to the passage to a null equivalent element in} $\mathcal{L}$
\textit{then} $\mathcal{O}_{\mathcal{A}}$ \textit{is closed with respect to
the passage to a null equivalent function.}

\bigskip

\begin{definition}
A map $\nu :\mathbb{S\rightarrow S}$ will be called a \textbf{measure
equivalence }if:

\begin{itemize}
\item  is bijective

\item  sends every $\mathbb{S}^{1}$-orbit $\left[ \varphi \right] $ in itself

\item  every map $\nu |_{\left[ \varphi \right] }:\left[ \varphi \right]
\rightarrow \left[ \varphi \right] $ is a borel equivalence preserving the
measure ($\nu |_{\left[ \varphi \right] \ast }\mu _{\left[ \varphi \right]
}=\mu _{\left[ \varphi \right] }$)
\end{itemize}
\end{definition}

\bigskip

\textit{The family of all the measure equivalences of} $\mathbb{S}$ \textit{%
make a group of transformations of }$\mathbb{S}$.

\textit{If }$L$\textit{\ is a proposition and }$\nu $\textit{\ is a measure
equivalence of }$\mathbb{S}$\textit{\ the image }$\nu (L)$\textit{\ is also
a proposition.}

\textit{\ If }$f$\textit{\ is an observable on }$\mathbb{S}$ \textit{and }$%
\nu $\textit{\ is a measure equivalence of} $\mathbb{S}$ \textit{the
function }$f\circ \nu $\textit{\ is also an observable}

\bigskip

\begin{theorem}
Two propositions $L$ and $M$ are equivalent in measure if and only if there
exists a measure equivalence $\nu $ of $\mathbb{S}$ such that $\nu (L)$ and $%
M$ are null equivalent
\end{theorem}

\begin{proof}
($\Longleftarrow )$ Obvious.

($\Longrightarrow )$Since the sets $L\cap \left[ \varphi \right] $ and $%
M\cap \left[ \varphi \right] $ have the same measure in $\left[ \varphi %
\right] $ there exists a borel equivalence $\nu |_{\left[ \varphi \right] }:%
\left[ \varphi \right] \rightarrow \left[ \varphi \right] $ preserving the
measure such that $\nu |_{\left[ \varphi \right] }(L\cap \left[ \varphi %
\right] )$ and $M\cap \left[ \varphi \right] $ are equivalent up to a borel
null subset (it is a consequence, for example, of Thm. 9 p. 327 in [R] )$.$
The family $\left\{ \nu |_{\left[ \varphi \right] }\right\} $ makes a a
measure equivalence $\nu $ of $\mathbb{S}$.
\end{proof}

\bigskip

\bigskip

\section{\protect\LARGE Observables and kaehlerian functions}

\bigskip

\begin{definition}
A function $l:\mathbb{S}$ $\rightarrow \mathbb{R}$ is called \textbf{smooth
kaehlerian on }$\mathbb{S}$\textbf{\ }if:

\begin{itemize}
\item  is smooth on $\mathbb{S}$

\item  $l\circ \rho _{\theta }=l$ for every $\theta $ in $\mathbb{R}$

\item  for every couple of orthogonal vectors $\varphi $ and $\psi $ in $%
\mathbb{S}$, $l(\cos t\cdot \varphi +\sin t\cdot \psi )$ is a function of
the form $a\cdot \cos ^{2}t+b\cdot \sin t\cos t+c\cdot \sin ^{2}t$ in the
variable $t$.
\end{itemize}
\end{definition}

\bigskip

\textit{The smooth kaehlerian functions on} $\mathbb{S}$ \textit{are defined
in such a way to be exactly the liftings to} $\mathbb{S}$ \textit{of the
(smooth) kaehlerian functions defined on} $\mathbb{P}_{\mathbb{C}}(\mathcal{H%
}\mathbb{)}$ \textit{in [CMP], [G] and [CGM]}.

\bigskip

\begin{notation}
\textit{Let's denote by} $\mathcal{KS(}\mathbb{S)}$ t\textit{he vector space
of all smooth kaehlerian functions on}\textbf{\ }\textit{the space}\textbf{\
}$\mathbb{S}$.
\end{notation}

\bigskip

\textit{For every smooth function} $l:\mathbb{S}$ $\rightarrow \mathbb{R}$%
\textit{\ such that }$l\circ \rho _{\theta }=l$\textit{\ for every }$\theta $%
\textit{\ in }$\mathbb{R}$\textit{\ the vectors} $Grad_{\varphi }^{\mathbb{S}%
}l$ \ and $J_{\varphi }Grad_{\varphi }^{\mathbb{S}}l$ \textit{are horizontal.%
}

\textit{For every proposition }$L$\textit{\ the function }$\varphi \mapsto
\mu _{\left[ \varphi \right] }(L\cap \left[ \varphi \right] )$\textit{\ is
smooth kaehlerian.}

\textit{The function} $\left\langle A\right\rangle :\mathbb{S(}\sqrt{2}%
)\rightarrow \mathbb{R}$ \textit{defined by }$\left\langle A\right\rangle
(\varphi )=\frac{1}{2}\left\langle \varphi ,A\varphi \right\rangle $,\textit{%
\ for a bounded self-adjoint complex operator }$A$\textit{\ on }$\mathcal{H}$%
,$\ $\textit{\ is a\ smooth kaehlerian function such that for every }$X$%
\textit{\ in} $T_{\varphi }\mathbb{S}$ \textit{it holds}:\textit{\ }$%
X_{\varphi }\left\langle A\right\rangle =\left\langle A\varphi
,X\right\rangle $\textit{\ }.

\textit{We have}:$Grad_{\varphi }^{\mathbb{S}}(\left\langle A\right\rangle
)=pr_{\varphi ^{\perp }}(A\varphi )$ \textit{because:} $\left\langle
Grad_{\varphi }^{\mathbb{S}}(\left\langle A\right\rangle ),X\right\rangle
=\left\langle A\varphi ,X\right\rangle =$

$=\left\langle pr_{T_{\varphi }\mathbb{S}}(A\varphi ),X\right\rangle $
\textit{for every }$X$\textit{\ in} $T_{\varphi }\mathbb{S}$, \textit{then} $%
Grad_{\varphi }^{\mathbb{S}}(\left\langle A\right\rangle )=A\varphi
-\left\langle A\right\rangle _{\varphi }\cdot \varphi $.

\bigskip

\begin{definition}
A function $l:W\cap \mathbb{S}\rightarrow \mathbb{R},$ where $W$ is a
complex dense linear subspace of $\mathcal{H}$, is called \textbf{kaehlerian
}if:

\begin{itemize}
\item  for every $\varphi $ in $W\cap \mathbb{S}$ the map $d_{\varphi
}l:W\cap T_{\varphi }\mathbb{S\rightarrow R}$ given by $d_{\varphi
}l(X)=X_{\varphi }l$ is well defined, linear and continuous

\item  for every complex closed linear subspace $F$ of $W$ the restriction $%
l|_{F\cap \mathbb{S}}$ is a smooth kaehlerian function on $F\cap \mathbb{S}$

\item  the couple $(l,W)$ is maximal with respect to the two properties
given above.
\end{itemize}
\end{definition}

\bigskip

\begin{notation}
Let's denote by $\mathcal{K(}\mathbb{S)}$ the set of all kaehlerian
functions on\textbf{\ }$\mathbb{S}$.
\end{notation}

\bigskip

\textit{A smooth function} $l:\mathbb{S}$ $\rightarrow \mathbb{R}$ s\textit{%
uch that for some couple of orthogonal vectors }$\varphi $\textit{\ and }$%
\psi $\textit{\ in} $\mathbb{S}$ \textit{the function }$l(\cos t\cdot
\varphi +\sin t\cdot \psi )$\textit{\ has the form }$a\cdot \cos
^{2}t+b\cdot \sin t\cos t+c\cdot \sin ^{2}t$\textit{\ can be expressed as:}
\begin{equation*}
l(\cos t\cdot \varphi +\sin t\cdot \psi )=l(\varphi )+d_{\varphi }l(\psi
)\cdot \sin t\cdot \cos t+\frac{1}{2}H_{\varphi }^{l}(\psi ,\psi )\cdot \sin
^{2}t
\end{equation*}
\textit{Where }$H^{l}$\textit{\ is the hessian of the function }$l$\textit{\
(it is enough to remember that }
\begin{equation*}
d_{\varphi }l(\psi )=(l\circ \gamma _{\varphi \psi })^{\prime }(0)\mathit{\
and\ }H_{\varphi }^{l}(\psi ,\psi )=(l\circ \gamma _{\varphi \psi })^{\prime
\prime }(0)
\end{equation*}
\textit{\ since }$\gamma _{\varphi \psi }$\textit{\ is a geodesic curve).}

\textit{It is not difficult to prove that two smooth functions }$l_{1},l_{2}:%
\mathbb{S}$ $\rightarrow \mathbb{R}$ \textit{such that for every couple of
orthogonal vectors }$\varphi $\textit{\ and }$\psi $\textit{\ in} $\mathbb{S}
$ \textit{the functions }$l_{1}(\cos t\cdot \varphi +\sin t\cdot \psi )$%
\textit{, }$l_{2}(\cos t\cdot \varphi +\sin t\cdot \psi )$\textit{\ have the
form }$a\cdot \cos ^{2}t+b\cdot \sin t\cos t+c\cdot \sin ^{2}t$\textit{\ are
equal if (and only if) it is possible to find a vector }$\varphi _{0}$%
\textit{\ in }$\mathbb{S}$ \textit{where}:
\begin{equation*}
l_{1}(\varphi _{0})=l_{2}(\varphi _{0})\text{, }d_{\varphi
_{0}}l_{1}=d_{\varphi _{0}}l_{2}\text{, }H_{\varphi _{0}}^{l_{1}}=H_{\varphi
_{0}}^{l_{2}}
\end{equation*}

\bigskip

\begin{theorem}
Let $l:\mathbb{S}$ $\rightarrow \mathbb{R}$ be a function, $l$ is a smooth
kaehlerian\textbf{\ }function if and only if there exists one (and only one)
bounded self-adjoint complex linear operator $A:\mathcal{H}\mathbb{%
\rightarrow }\mathcal{H}$ \ such that:
\begin{equation*}
l(\varphi )=\left\langle A\right\rangle _{\varphi }=\frac{1}{2}\cdot
\left\langle \varphi ,A\varphi \right\rangle
\end{equation*}
for every $\varphi $ in $\mathbb{S}$.
\end{theorem}

\begin{proof}
($\Longleftarrow $) Obvious.

($\Longrightarrow $)(The proof \ mimics the proof given in [G]). Let's fix a
vector $\varphi _{0}$ in $\mathbb{S}$.

Since the map $dl_{\varphi _{0}}:T_{\varphi _{0}}\mathbb{S}\rightarrow
\mathbb{R}$ is a continuous linear map there exists a vector $Z$ $%
(=Grad_{_{\varphi _{0}}}l)$ in $T_{\varphi _{0}}\mathbb{S}$ such that $%
dl_{\varphi _{0}}(X)=\left\langle Z,X\right\rangle $ for every vector $X$ in
$T_{\varphi _{0}}\mathbb{S}$; moreover the Hessian $H_{\varphi
_{0}}^{l}:T_{\varphi _{0}}\mathbb{S}\times T_{\varphi _{0}}\mathbb{S}%
\rightarrow \mathbb{R}$ is a continuous bilinear symmetric map and therefore
there exists a bounded self-adjoint map $B:T_{\varphi _{0}}\mathbb{S}%
\rightarrow T_{\varphi _{0}}\mathbb{S}$ such that $H_{\varphi
_{0}}^{l}(X,Y)=\left\langle X,B(Y)\right\rangle $ for every $X,Y$ in $%
T_{\varphi _{0}}\mathbb{S}$.

There exists a unique continuous real linear map $A:\mathcal{H}\mathbb{%
\rightarrow }\mathcal{H}$ with the assigned value $A(\varphi _{0})=l(\varphi
_{0})\cdot \varphi _{0}+Z$ and such that: $A(X)=\frac{1}{2}\left\langle
Z,X\right\rangle \cdot \varphi _{0}+l(\varphi _{0})\cdot X+B(X)$ for every $%
X $ in $T_{\varphi _{0}}\mathbb{S}$.

It is not difficult to prove that $A$ is symmetric.

The function $\left\langle A\right\rangle :\mathbb{S}\rightarrow \mathbb{R}$
defined by $\left\langle A\right\rangle (\varphi )=\frac{1}{2}\left\langle
\varphi ,A\varphi \right\rangle $ is smooth and for every couple of
orthogonal vectors $\varphi ,\psi $ in $\mathbb{S}$ the function $%
\left\langle A\right\rangle (\gamma _{\varphi \psi }(t))$ has the form $%
a\cdot \cos ^{2}t+b\cdot \sin t\cos t+c\cdot \sin ^{2}t$. With some more
calculation it is possible to verify that:
\begin{equation*}
\left\langle A\right\rangle (\varphi _{0})=l(\varphi _{0})\text{, }%
d_{\varphi _{0}}\left\langle A\right\rangle =d_{\varphi _{0}}l\text{, }%
H_{\varphi _{0}}^{\left\langle A\right\rangle }=H_{\varphi _{0}}^{l}
\end{equation*}
therefore $\left\langle A\right\rangle =l$ and in particular $\left\langle
A\right\rangle \circ J=\left\langle A\right\rangle $.

Then written $A=B+C$ with $B=\frac{1}{2}(A-JAJ)$ and $C=\frac{1}{2}(A+JAJ)$
it holds $BJ=JB$, $CJ=-JC$, $\left\langle B\right\rangle \circ
J=\left\langle B\right\rangle $ and $\left\langle C\right\rangle \circ
J=-\left\langle C\right\rangle $. Then $\left\langle A\right\rangle \circ
J=\left\langle A\right\rangle $, implies $\left\langle C\right\rangle =0$ , $%
C=0$ and $AJ=JA$, that is $A$ is a (bounded) self-adjoint complex operator.

If another complex linear self-adjoint operator $B$ verifies $\left\langle
B\right\rangle (\varphi )=\left\langle A\right\rangle (\varphi )$ on $%
\mathbb{S}$ then $\left\langle X,AX\right\rangle =\left\langle
X,BX\right\rangle $ for every $X\neq 0$ and then $A=B$.
\end{proof}

\bigskip

\begin{theorem}
Let $W$ be a complex linear dense subspace of \ $\mathcal{H}$ $\ $and $%
l:W\cap \mathbb{S}$ $\rightarrow \mathbb{R}$ a function, the couple $(W,l)$
defines a kaehlerian\textbf{\ }function if and only if there exists one (and
only one) self-adjoint operator $A:W\mathbb{\rightarrow }\mathcal{H}$ \ such
that:
\begin{equation*}
\mathcal{D(}A)=W\text{ \ and }l(\varphi )=\left\langle A\right\rangle
_{\varphi }=\frac{1}{2}\cdot \left\langle \varphi ,A\varphi \right\rangle
\end{equation*}

for every $\varphi $ in $W\cap \mathbb{S}$.
\end{theorem}

\begin{proof}
($\Longrightarrow $) Taken $\varphi $ in $W\cap \mathbb{S}$ the map $\lambda
_{\varphi }:W\rightarrow \mathbb{R}$ defined by:
\begin{equation*}
\lambda _{\varphi }(X)=\left[ X-\frac{1}{2}\left\langle X,\varphi
\right\rangle \cdot \varphi \right] _{\varphi }\left( l\right) +\left\langle
X,\varphi \right\rangle \cdot l(\varphi )
\end{equation*}
is linear and continuous, therefore admits a continuous linear extension $%
\widetilde{\lambda _{\varphi }}:\mathcal{H}\mathbb{\rightarrow R}$ and is
possible to find a unique element $A_{0}(\varphi )$ in $\mathcal{H}$ such
that: $\left\langle X,A_{0}(\varphi )\right\rangle =\widetilde{\lambda
_{\varphi }}(X)$ for every $X$ in $\mathcal{H}$.

The map so obtained $A_{0}:W\cap \mathbb{S\rightarrow }\mathcal{H}$ can be
extended to a map $\widehat{A}_{0}:W\rightarrow \mathcal{H}$ with the
position $\widehat{A}_{0}(\varphi )=\frac{\left\| \varphi \right\| }{\sqrt{2}%
}A_{0}(\frac{\sqrt{2}}{\left\| \varphi \right\| }\cdot \varphi )$ for every $%
\varphi \neq 0$ and $\widehat{A}_{0}(0)=0$.

Fixed a closed complex linear subspace $F$ of $W$ we know, by the previous
theorem, that there exists a (complex) self-adjoint linear operator $%
A_{F}:F\rightarrow F$ such that
\begin{equation*}
l(\varphi )=\frac{1}{2}\cdot \left\langle \varphi ,A_{F}(\varphi
)\right\rangle
\end{equation*}
for every $\varphi $ in $F\cap \mathbb{S}$.

Then because $X=\left[ X-\frac{1}{2}\left\langle X,\varphi \right\rangle
\cdot \varphi \right] +\frac{1}{2}\left\langle X,\varphi \right\rangle \cdot
\varphi $ for $\varphi $ in $F\cap \mathbb{S}$ and $X$ in $F$, remembering
that $Y_{\varphi }l=\left\langle Y,A_{F}(\varphi )\right\rangle $, for every
$\varphi $ in $F\cap \mathbb{S}$ and $Y$ in $F\cap (\varphi )^{\perp }$, we
can prove that $\left\langle X,A_{F}(\varphi )\right\rangle =\left\langle X,%
\widehat{A}_{0}(\varphi )\right\rangle $. That is $A_{F}(\varphi )=pr_{F}(%
\widehat{A}_{0}(\varphi ))$. Hence:
\begin{equation*}
pr_{F}(\widehat{A}_{0}(\varphi ))=A_{F}(\varphi )\text{ for every }F\subset W%
\text{ and every }\varphi \text{ in }F\text{.}
\end{equation*}

Using this property systematically and the fact that $pr_{F}(Z)=0$ for every
$F\subset W$ implies $Z=0$ it is possible to prove that $\widehat{A}_{0}$ is
a complex, linear and hermitian operator defined on a dense linear subspace
of \ $\mathcal{H}$. Its closure $A=\overline{\widehat{A}}_{0}$ is the
desired self-adjoint operator since on every $\varphi $ in $W$ taken a
closed suspace $F\subset W$ with $\varphi $ in $F$ we have
\begin{equation*}
\frac{1}{2}\cdot \left\langle \varphi ,\overline{\widehat{A}}_{0}(\varphi
)\right\rangle =\frac{1}{2}\cdot \left\langle \varphi ,pr_{F}A(\varphi
)\right\rangle =\frac{1}{2}\cdot \left\langle \varphi ,A_{F}(\varphi
)\right\rangle =l(\varphi )\text{.}
\end{equation*}

Therefore $(\mathcal{D(}A)\cap \mathbb{S},\left\langle A\right\rangle )$
extends $(W\cap \mathbb{S},$ $l)$ but for the maximality it must hold $%
\mathcal{D(}A)\cap \mathbb{S=}W\cap \mathbb{S}$ and then $\mathcal{D(}A)%
\mathbb{=}W$.

As in the proof of the previous theorem it is possible to prove the unicity
on $W$ of such operator.

($\Longleftarrow $)For every self-adjoint complex operator $A:W\mathbb{%
\rightarrow }\mathcal{H}$ $\ $ the first two properties in the definition of
a (non smooth) kaehlerian function are easily verified for the function $%
\left\langle A\right\rangle :\mathcal{D(}A)\cap \mathbb{S(}\sqrt{2}%
)\rightarrow \mathbb{R}$ defined by $\left\langle A\right\rangle (\varphi )=%
\frac{1}{2}\left\langle \varphi ,A\varphi \right\rangle $. About the
maximality: if the couple $(W,l)$ extends $(\mathcal{D(}A),\left\langle
A\right\rangle )$ reasoning as above it is possible to find a self-adjoint
operator $A^{\prime }$ with $(\mathcal{D(}A^{\prime }),\left\langle
A^{\prime }\right\rangle )$ extending $(W,l)$; but $A$ and $A^{\prime }$ are
self-adjoint therefore $\mathcal{D(}A)=\mathcal{D(}A^{\prime })$ and $(W,l)=(%
\mathcal{D(}A),\left\langle A\right\rangle )$.
\end{proof}

\bigskip

\begin{notation}
\textit{The map }$A\mapsto \left\langle A\right\rangle $\textit{\ between
self-adjoint operators and kaehlerian\ functions on} $\mathbb{S}$\textit{\
is bijective; the restricted map between bounded self-adjoint operators and
smooth kaehlerian\ functions is a linear isomorphism.}Denoted by $SA(%
\mathcal{H)}$ the set of all (bounded or unbounded) self-adjoint operators
of $\mathcal{H}$ and by $SAB(\mathcal{H)}$ its subset (a vector space) of
all bounded operators, let's denote by $\alpha :$ $\mathcal{K(}\mathbb{%
S)\rightarrow }SA(\mathcal{H)}$ the bijective map defined by the previous
theorem ($\alpha (l)$ is the self-adjoint operator associated to the
kaehlerian function $l$).Its inverse is the map $\left\langle \cdot
\right\rangle :SA(\mathcal{H)\rightarrow K(}\mathbb{S)}$.This map induces an
isomorphism $\alpha |:$ $\mathcal{KS(}\mathbb{S)\rightarrow }SAB(\mathcal{H)}
$.
\end{notation}

\bigskip

\textit{Since we have proved that for every bounded self-adjoint operator }$%
A $\textit{\ we have: }$Grad_{\varphi }\left\langle A\right\rangle =A\varphi
-\left\langle A\right\rangle _{\varphi }\cdot \varphi $\textit{\ for every
smooth kaehlerian function }$l$\textit{\ it holds:}
\begin{equation*}
\alpha (l)(\varphi )=Grad_{\varphi }l+l(\varphi )\cdot \varphi
\end{equation*}

\bigskip

\begin{theorem}
For every proposition $L$ of $\mathbb{S}$ $\ $there exists one and only one
(complex) orthogonal projector $E$ of $\mathcal{H}$ such that for every $%
\varphi $ in $\mathbb{S}$ it holds:
\begin{equation*}
\mu _{\left[ \varphi \right] }(L\cap \left[ \varphi \right] )=\left\langle
E\right\rangle _{\varphi }
\end{equation*}
Conversely every orthogonal projector $E$ of $\mathcal{H}$ comes in this way
from a proposition $L$ and two propositions give origin to the same
projector if and only if they are measure equivalent.
\end{theorem}

\begin{proof}
When $L$ is null equivalent to $\emptyset $ or to $\mathbb{S}$ the thesis
follows immediately taking, respectively, $E=0$ and $E=I$. We will then\
suppose $L$ not null equivalent to $\emptyset $ or $\mathbb{S}$.

Since the function $\varphi \longmapsto \mu _{\left[ \varphi \right] }(L\cap %
\left[ \varphi \right] )$ is smooth kaehlerian there exists a bounded
self-adjoint operator $E$ such that $\mu _{\left[ \varphi \right] }(L\cap %
\left[ \varphi \right] )=\left\langle E\right\rangle _{\varphi }$; we have
only to prove that $E$ verifies $E^{2}=E$.

We have already proved in a remark above that for every $\varphi $ in $%
\mathbb{S}$ we have: $Grad_{\varphi }^{\mathbb{S}}\left\langle
E\right\rangle =E\varphi -\left\langle E\right\rangle \varphi $. Therefore$%
\left\| Grad_{\varphi }^{\mathbb{S}}\left\langle E\right\rangle \right\|
^{2}=2\cdot (\left\langle E^{2}\right\rangle _{\varphi }-\left\langle
E\right\rangle _{\varphi }^{2})$ and if we prove that $\left\| Grad_{\varphi
}^{\mathbb{S}}\left\langle E\right\rangle \right\| ^{2}=2\cdot (\left\langle
E\right\rangle _{\varphi }-\left\langle E\right\rangle _{\varphi }^{2})$ for
every $\varphi $ in $\mathbb{S}$ we get $E^{2}=E$.

Now $\left\| Grad_{\varphi }^{\mathbb{S}}\left\langle E\right\rangle
\right\| =$

$=\max_{\left\| \tau \right\| =1}\left[ \left| \left\langle Grad_{\varphi }^{%
\mathbb{S}}\left\langle E\right\rangle ,\tau \right\rangle \right| \right]
=\max_{\psi \perp \varphi ,\left\| \psi \right\| =\sqrt{2}}\left[ \left|
\left\langle Grad_{\varphi }^{\mathbb{S}}\left\langle E\right\rangle ,\frac{1%
}{\sqrt{2}}\psi \right\rangle \right| \right] =$

$=\frac{1}{\sqrt{2}}\max_{\psi \perp \varphi ,\left\| \psi \right\| =\sqrt{2}%
}\left[ \left| (\left\langle E\right\rangle \circ \gamma _{\varphi \psi
})^{\prime }(0)\right| \right] $.

Written $(\left\langle E\right\rangle \circ \gamma _{\varphi \psi })(t)=%
\frac{1}{2}\left[ (M(\psi )+m(\psi ))+(M(\psi )-m(\psi ))\cdot \cos
(2t+\alpha (\psi ))\right] $

where
\begin{equation*}
0\leq m(\psi )=\min (\left\langle E\right\rangle \circ \gamma _{\varphi \psi
})\leq \max (\left\langle E\right\rangle \circ \gamma _{\varphi \psi
})=M(\psi )\leq 1
\end{equation*}

It is not difficult to calculate:
\begin{equation*}
\left| (\left\langle E\right\rangle \circ \gamma _{\varphi \psi })^{\prime
}(0)\right| ^{2}=(M(\psi )-m(\psi ))^{2}\cdot \sin ^{2}(\alpha (\psi ))
\end{equation*}
and
\begin{equation*}
4(\left\langle E\right\rangle _{\varphi }-\left\langle E\right\rangle
_{\varphi }^{2})\geq (M(\psi )-m(\psi ))^{2}\cdot \sin ^{2}(\alpha (\psi ))%
\text{.}
\end{equation*}

By the definition of proposition we know there exists a vector $\psi _{0}$
in $\mathbb{S}$ orthogonal to $\varphi $ where $m(\psi _{0})=0$ and $M(\psi
_{0})=1$; we have: $\left\langle E\right\rangle _{\varphi }=\frac{1}{2}\left[
1+\cos (\alpha (\psi _{0}))\right] $ and $\left| (\left\langle
E\right\rangle \circ \gamma _{\varphi \psi _{0}})^{\prime }(0)\right|
^{2}=\sin ^{2}(\alpha (\psi _{0}))=4(\left\langle E\right\rangle _{\varphi
}-\left\langle E\right\rangle _{\varphi }^{2})$ with respect to this vector.

Therefore $\left\| Grad_{\varphi }^{\mathbb{S}}\left\langle E\right\rangle
\right\| ^{2}=2(\left\langle E\right\rangle _{\varphi }-\left\langle
E\right\rangle _{\varphi }^{2})$ and we have proved that $E$ is a projector.
Since $L$ prescribes $\left\langle E\right\rangle $ on $\mathbb{S}$ the
projector $E$ is the only one.

Conversely let $E$ be a projector on a complex closed linear subspace $F$ of
$\mathcal{H}$, if $E=0$ or $E=I$ the thesis follows immediately.We will
then\ suppose $E$ not $0$ and not $I$.

For every $\mathbb{S}^{1}$-orbit $\left[ \varphi \right] $ choose a borelian
subset $L_{\left[ \varphi \right] }$ such that $\mu _{\left[ \varphi \right]
}(L_{\left[ \varphi \right] })=\left\langle E\right\rangle _{\varphi }$. The
set $L=\bigcup_{_{\left[ \varphi \right] }}L_{\left[ \varphi \right] }$ is
pseudo-borelian in $\mathbb{S}$ and the map $\varphi \longmapsto \mu _{\left[
\varphi \right] }(L\cap \left[ \varphi \right] )$ is smooth as the map $%
\varphi \longmapsto \left\langle E\right\rangle _{\varphi }$.

For a vector $\psi $ in $\mathbb{S}$ orthogonal to $\varphi $ the map $%
t\longmapsto \gamma _{\left[ \gamma _{\varphi \psi }(t)\right] }(L\cap \left[
\gamma _{\varphi \psi }(t)\right] )$ has the form $a\cos ^{2}t+b\sin t\cos
t+c\sin ^{2}t$.

Moreover if $\varphi $ is in $F$ we can take $\psi $ in $F^{\perp }$ with $%
\gamma _{\left[ \gamma _{\varphi \psi }(t)\right] }(L\cap \left[ \gamma
_{\varphi \psi }(t)\right] )=\cos ^{2}t$ and analogously if $\varphi $ is in
$F^{\perp }$.

When $\varphi $ is not in $F$ or in $F^{\perp }$ we can find a vector $%
\alpha $ in $F\cap \mathbb{S}$ and a vector $\beta $ in $F^{\perp }\cap
\mathbb{S}$ in such a way that $\varphi =\cos t_{0}\cdot \alpha +\sin
t_{0}\cdot \beta $. If we consider the vector $\psi =-\sin t_{0}\cdot \alpha
+\cos t_{0}\cdot \beta $ we get $\gamma _{\left[ \gamma _{\varphi \psi }(t)%
\right] }(L\cap \left[ \gamma _{\varphi \psi }(t)\right] )=\cos
^{2}(t+\theta )$.

Two propositions $L$ and $L^{\prime }$ with $\mu _{\left[ \varphi \right]
}(L\cap \left[ \varphi \right] )=\left\langle E\right\rangle _{\varphi }=\mu
_{\left[ \varphi \right] }(L^{\prime }\cap \left[ \varphi \right] )$ are
obviously measure equivalent.
\end{proof}

\bigskip

\textit{To give an explicit proposition }$L$\textit{\ associated to a
projector }$E$\textit{\ we can proceed as follows: let's fix a map }$\sigma :%
\mathbb{P}_{\mathbb{C}}\mathbb{(}\mathcal{H}$ )$\rightarrow \mathbb{S}$
\textit{such that }$\sigma \left[ \varphi \right] \in \left[ \varphi \right]
$\textit{\ for every }$\mathbb{S}^{1}$-\textit{orbit }$\left[ \varphi \right]
$\textit{, the pseudo-borelian subset }$L_{\sigma }^{E}=\bigcup_{_{\left[
\varphi \right] }}e^{i]\pi -2\pi \left\langle E\right\rangle _{\varphi },\pi
]}\cdot \sigma \left[ \varphi \right] $\textit{\ in} $\mathbb{S}$ \textit{is
a proposition associated to }$E$. \textit{All the others propositions
associated to }$E$\textit{\ are the }$\nu (L_{\sigma }^{E})$\textit{\ (where
}$\nu $\textit{\ is a measure equivalence of} $\mathbb{S}$) \textit{and
their null equivalent}. \textit{Therefore in this theory it is possible to
claim that ''behind'' a quantum state }$\left[ \varphi \right] $\textit{\
there are infinite ''hidden'' states: the Hilbert vectors }$e^{i\theta
}\cdot \sigma \left[ \varphi \right] $\textit{\ (where }$e^{i\theta }$%
\textit{\ varies in the group} $\mathbb{S}^{1}$\textit{) and ''behind'' a
quantum proposition }$E$\textit{\ there are infinite ''hidden'' classical
propositions: essentially the propositions }$\nu (L_{\sigma }^{E})$\textit{\
(where }$\nu $\textit{\ varies in the group of all measure equivalences of} $%
\mathbb{S}$). \textit{The truth value }$0$\textit{\ or }$1$\textit{\ of one
of these hidden classical proposition on the hidden classical state is }$%
\chi _{\nu (L_{\sigma }^{E})}(e^{i\theta }\cdot \sigma \left[ \varphi \right]
)$\textit{\ and depends not only on the ''hidden variable'' }$e^{i\theta }$%
\textit{\ but also on the ''experimental context'' defined by the measure
equivalence }$\nu $\textit{.}

\bigskip

\textit{The suggestion that to each quantum measurement there could
correspond a collection of several deterministic ''hidden measurements'' has
appeared in several moments in the history of the hidden variable theories.
For a direction of development of this idea cfr. [A] and [CM]. }

\bigskip

\begin{notation}
Denoted by $PR\mathcal{(H}\mathbb{)}$ the set of all projector operators of $%
\mathcal{H}$ the previous theorem claims there is a surjective map $%
\varepsilon :\mathcal{L\rightarrow }PR\mathcal{(H}\mathbb{)}$ associating to
each proposition $L$ a projector $\varepsilon (L)$ such that $\left\langle
\varepsilon (L)\right\rangle _{\varphi }=\mu _{\left[ \varphi \right]
}(L\cap \left[ \varphi \right] )$ for every $\varphi $ in $\mathbb{S}$.
\end{notation}

\bigskip

\textit{A pseudo-borel subset }$A$\textit{\ of }$\mathbb{S}$ s\textit{uch
that }$\mu _{\left[ \varphi \right] }(A\cap \left[ \varphi \right]
)=\left\langle E\right\rangle _{\varphi }$\textit{\ (for every }$\varphi $%
\textit{\ in} $\mathbb{S}$)\textit{\ for a projector }$E$ \textit{is
necessarily a proposition (is measure equivalent to a proposition).}

\bigskip

\begin{theorem}
If $\ L$ and $M$ are propositions with $L$ $\subset $ $M$ then also $%
M\setminus L$ is a proposition.
\end{theorem}

\begin{proof}
$L$ $\subset $ $M$ implies $\varepsilon (L)\leq \varepsilon (M)$ therefore $%
\varepsilon (L)$ and $\varepsilon (M)$ commute and the difference $%
\varepsilon (M)-\varepsilon (L)$ is also a projector. Therefore there exists
a proposition $N$ such that $\varepsilon (N)=\varepsilon (M)-\varepsilon (L)$
and the pseudo-borel subset $M\setminus L$, because is measure equivalent to
$N$, is a proposition.
\end{proof}

\bigskip

\begin{theorem}
Given two propositions $L$ and $M$ if (and only if) $L\cap M$ is contained
in $\mathcal{L}$ then all the family $\mathcal{B}_{L,M}$ is contained in $%
\mathcal{L}$ ($L$ and $M$ are compatible).
\end{theorem}

\begin{proof}
We already know that $\emptyset $, $\mathbb{S}$, $L$, $M$, $L\cap M$, $%
\mathbb{\complement }L$, $\mathbb{\complement }M$ and $\mathbb{\complement }%
L\cup \mathbb{\complement }M$ are in $\mathcal{L}$.

Moreover $L\cap M\subset L$ implies $L\setminus L\cap M=$ $L\cap \mathbb{%
\complement }M$ and $\mathbb{\complement }L\cup M$ are in $\mathcal{L}$, $%
L\cap M\subset M$ implies $\mathbb{\complement }L\cap M$ and $L\cup \mathbb{%
\complement }M$ are in $\mathcal{L}$, $\mathbb{\complement }L\cap M\subset
\mathbb{\complement }L$ implies $\mathbb{\complement }L\cap \mathbb{%
\complement }M$ and $L\cup M$ are in $\mathcal{L}$ and finally $L\cap
M\subset L\cup M$ implies $L\Delta M$ and $\mathbb{\complement (}L\Delta M)$
are in $\mathcal{L}$.
\end{proof}

\bigskip

\textit{For a proposition} $L$ \textit{we have} $\varepsilon (L)=0$ \textit{%
if and only if} $L$\textit{\ is a pseudo-borel null subset.}

\textit{The following properties are easily proved:} $\varepsilon (\emptyset
)=0$, $\varepsilon (\mathbb{S})=I$, $\varepsilon (\mathbb{\complement }%
L)=I-\varepsilon (L)$, $\varepsilon (L)=\varepsilon (\nu L)$ \textit{for
every measure equivalence }$\nu $\textit{\ on}\ $\mathbb{S}$, $\varepsilon
(L)=\varepsilon (L^{\prime })\ $\textit{if }$L$\textit{\ and }$L^{\prime }$%
\textit{\ are null equivalent}, $\ L$ $\subset M$ \textit{implies }$%
\varepsilon (L)\leq \varepsilon (M)$ and $\varepsilon (M\setminus
L)=\varepsilon (M)-\varepsilon (L)$\textit{, }$L\cap M=\emptyset $\textit{\
implies} $\varepsilon (L)\perp \varepsilon (M)$.

\bigskip

\begin{theorem}
For every proposition $L$ and every unitary transformation $U$ it holds:
\begin{equation*}
\varepsilon (U(L))=U\circ \varepsilon (L)\circ U^{-1}
\end{equation*}
\end{theorem}

\begin{proof}
$\left\langle U\circ \varepsilon (L)\circ U^{-1}\right\rangle _{\varphi
}=\left\langle \varepsilon (L)\right\rangle _{U^{-1}\varphi }=\mu _{\left[
U^{-1}\varphi \right] }(L\cap \left[ U^{-1}\varphi \right] )=\mu _{\left[
\varphi \right] }(U(L)\cap \left[ \varphi \right] )$ therefore $\varepsilon
(U(L))=U\circ \varepsilon (L)\circ U^{-1}$.
\end{proof}

\bigskip

\begin{theorem}
For every observable function $f$ on $\mathbb{S}$ there exists one and only
one self-adjoint operator $T$ on $\mathcal{H}$ \ such that for every state $%
\varphi $ in $\mathbb{S}$ and every borel subset $B$ of $\mathbb{R}$ it
holds:
\begin{equation*}
\mu _{\left[ \varphi \right] }(f^{-1}(B)\cap \left[ \varphi \right]
)=\left\langle E_{B}^{T}\right\rangle _{\varphi }
\end{equation*}

Conversely \ every\ self-adjoint operator $T$ on $\mathcal{H}$ comes, in
this way, from an observable function.
\end{theorem}

\begin{proof}
For every $s$ in $\mathbb{R}$ let $L_{s}=f^{-1}((-\infty ,s])$ and $%
E_{s}=\varepsilon (L_{s})$. It is not difficult to prove that the family $%
\left\{ E_{s}\right\} _{s\in \mathbb{R}}$ is a spectral family of $\mathcal{H%
}$, therefore there is a self-adjoint complex operator $T$ such that $%
E_{(-\infty ,s]}^{T}=E_{s}$ for every $s$ in $\mathbb{R}$. That is $\mu _{%
\left[ \varphi \right] }(f^{-1}((-\infty ,s])\cap \left[ \varphi \right]
)=\left\langle E_{(-\infty ,s]}^{T}\right\rangle _{\varphi }$ for every $%
\varphi $ in $\mathbb{S}$ and every $s$ in $\mathbb{R}$.Then, for the usual
properties of borelian subsets of $\mathbb{R}$, it is possible to prove that
the analog properties hold for an interval $]r,s]$ and a general borel
subset $B$ of $\mathbb{R}$:
\begin{equation*}
\mu _{\left[ \varphi \right] }(f^{-1}(]r,s])\cap \left[ \varphi \right]
)=\left\langle E_{]r,s]}^{T}\right\rangle _{\varphi }\text{, \ \ }\mu _{%
\left[ \varphi \right] }(f^{-1}(B)\cap \left[ \varphi \right] )=\left\langle
E_{B}^{T}\right\rangle _{\varphi }
\end{equation*}

If $T^{\prime }$ is another operator with the same property, from $%
\left\langle E_{B}^{T}\right\rangle _{\varphi }=\left\langle
E_{B}^{T^{\prime }}\right\rangle _{\varphi }$for every $\varphi $ and $B$ it
follows first $E_{B}^{T}=E_{B}^{T^{\prime }}$ for every $B$ and then $%
T=T^{\prime }$.

Conversely let's suppose a self-adjoint operator $T$ is given.

Let's fix a map $\sigma :\mathbb{P}_{\mathbb{C}}\mathbb{(}\mathcal{H}%
)\rightarrow \mathbb{S}$ such that $\sigma \left[ \varphi \right] \in \left[
\varphi \right] $ for every $\mathbb{S}^{1}$-orbit $\left[ \varphi \right] $%
, it is possible to define a bijective map $\delta _{\left[ \varphi \right]
}:\left[ \varphi \right] \rightarrow \mathbb{S}^{1}$with $\delta _{\left[
\varphi \right] }(\psi )=\psi /\sigma \left[ \varphi \right] $; such a map
is a borel isomorphism preserving the measure ($\delta _{\left[ \varphi %
\right] \ast }\mu _{\left[ \varphi \right] }=\mu _{\mathbb{S}^{1}}$).

Also the map $\rho :$ $\mathbb{S}^{1}\rightarrow ]0,1]$ defined by $\rho (u)=%
\frac{1}{2\pi }(\pi +Arg(u))$ is a borel isomorphism preserving the measure (%
$\rho _{\ast }\mu _{\mathbb{S}^{1}}=\lambda _{]0,1]}$).

The cumulative function $F_{\left[ \varphi \right] }:\mathbb{R\rightarrow
\lbrack }0,1]$ defined by $F_{\left[ \varphi \right] }(s)=\left\langle
E_{(-\infty ,s]}^{T}\right\rangle _{\varphi }$is a monotone non-decreasing
function whose quasi-inverse $\widetilde{F_{\left[ \varphi \right] }}%
:]0,1[\rightarrow \mathbb{R}$ is a monotone non-decreasing function with the
property:
\begin{equation*}
(\widetilde{F_{\left[ \varphi \right] }}_{\ast }\lambda )(B)=\nu _{F_{\left[
\varphi \right] }}(B)
\end{equation*}

(cfr. [K-S] thm. 4 p. 94)(the symbol $\nu _{F}$ denotes the Borel measure
associated to the monotone function $F$) . The function $\widetilde{F_{\left[
\varphi \right] }}$ can be extended monotonically to a function (denoted in
the same way) $\widetilde{F_{\left[ \varphi \right] }}:]0,1]\rightarrow
\mathbb{R\cup }\left\{ +\infty \right\} $ with the position $\widetilde{F_{%
\left[ \varphi \right] }}(1)=+\infty $ and has the same property provided
that we decide that $\nu _{F_{\left[ \varphi \right] }}(\left\{ +\infty
\right\} )=0$.

The function $f:\mathbb{S\rightarrow R\cup }\left\{ +\infty \right\} $
defined on each $\left[ \varphi \right] $ by $f_{\left[ \varphi \right] }=%
\widetilde{F_{\left[ \varphi \right] }}\circ \rho \circ \delta _{\left[
\varphi \right] }$ is a pseudo-borelian function on $\mathbb{S}$ with the
property:
\begin{equation*}
f_{\left[ \varphi \right] \ast }\mu _{\left[ \varphi \right] }=\nu _{F_{%
\left[ \varphi \right] }}
\end{equation*}

that is: $\mu _{\left[ \varphi \right] }(f^{-1}(]r,s])\cap \left[ \varphi %
\right] )=F_{\left[ \varphi \right] }(s)-F_{\left[ \varphi \right]
}(r)=\left\langle E_{]r,s]}^{T}\right\rangle _{\varphi }$, this implies as
usual$:\mu _{\left[ \varphi \right] }(f^{-1}(B)\cap \left[ \varphi \right]
)=\left\langle E_{B}^{T}\right\rangle _{\varphi }$ for every borel subset $B$
of $\ \mathbb{R}$.

Since $f^{-1}(\left\{ +\infty \right\} )=\left\{ -\sigma \left[ \varphi %
\right] ;\text{ }\left[ \varphi \right] \in \mathbb{P}_{\mathbb{C}}\mathbb{(}%
\mathcal{H})\right\} $ is a pseudo-borel null subset of $\mathbb{S}$ we can
assign to $f$ on $f^{-1}(\left\{ +\infty \right\} )$ finite values in such a
way to obtain a pseudo-borel function $f:\mathbb{S\rightarrow R}$ keeping
the property:
\begin{equation*}
\mu _{\left[ \varphi \right] }(f^{-1}(B)\cap \left[ \varphi \right]
)=\left\langle E_{B}^{T}\right\rangle _{\varphi }\text{for every borel
subset }B\text{ of}\ \ \mathbb{R}\text{.}
\end{equation*}

As observed in a remark following the previous theorem this implies that
each $f^{-1}(B)$ is a proposition, that is $f$ is an observable function.
\end{proof}

\bigskip

\textit{Remembering the definition of a quasi-inverse function (cfr. [K-S]
thm. 4 p. 94) we can give the following explicit expression (out of the null
pseudo-borel subset }$-\sigma \left( \mathbb{P}_{\mathbb{C}}\mathbb{(}%
\mathcal{H})\right) $\textit{) of the function }$f$\textit{\ associated to
the operator }$T$\textit{\ (with the help of} $\sigma $) \textit{in the
preceding proof:}
\begin{equation*}
f(\varphi )=\min \left\{ s\in \mathbb{R};\text{ }\left\langle E_{(-\infty
,s]}^{T}\right\rangle _{\varphi }\geq \frac{1}{2\pi }(\pi +Arg(\varphi
/\sigma \left[ \varphi \right] ))\right\}
\end{equation*}

\textit{We will denote this function by} $f_{\sigma }^{T}$.

\bigskip

\begin{notation}
Let's denote by $\tau :\mathcal{O\rightarrow }SA(\mathcal{H}\mathbb{)}$ the
(surjective) map defined by the previous theorems.
\end{notation}

\bigskip

\textit{In the proof of the previous theorem we have proved in particular
that for an observable function }$f$\textit{\ it holds}: $f|_{\left[ \varphi %
\right] \ast }\mu _{\left[ \varphi \right] }=\nu _{F_{\left[ \varphi \right]
}^{\tau (f)}}$.

\textit{Note that:} $E_{B}^{\tau (f)}=\mathcal{\varepsilon }(f^{-1}(B))$
\textit{for every borel subset }$B$\textit{\ of} $\mathbb{R}$ \textit{and}
\textit{moreover} $\mathcal{\varepsilon }(L)=\tau (\chi _{L})$.

\textit{\ For every projector }$E$\textit{\ we have that} $L_{\sigma }^{E}$
\textit{is null equivalent to} $\left( f_{\sigma }^{E}\right) ^{-1}(\left\{
1\right\} )$ (\textit{with the notations chosen above)}.

\textit{Moreover:} $\tau (f)=0$ \textit{if only if }$f$\textit{\ is
(essentially) zero and }$\tau (k)=k\cdot I$\textit{\ for every constant
function }$k$\textit{\ on} $\mathbb{S}$ $.$

\bigskip

\begin{theorem}
For every observable function $f$ and every unitary transformation $U$ it
holds:
\begin{equation*}
\tau (f\circ U^{-1})=U\circ \tau (f)\circ U^{-1}
\end{equation*}
\end{theorem}

\begin{proof}
$E_{(-\infty ,s]}^{U\circ \tau (f)\circ U^{-1}}=U\circ E_{(-\infty
,s]}^{\tau (f)}\circ U^{-1}=U\circ \varepsilon (f^{-1}((-\infty ,s])\circ
U^{-1}=$%
\begin{equation*}
=\varepsilon ((f\circ U^{-1})^{-1}((-\infty ,s]))=E_{(-\infty ,s]}^{\tau
(f\circ U^{-1})}\text{ for every }s\text{ in }\mathbb{R}.
\end{equation*}
\end{proof}

\bigskip

\begin{theorem}
If $f$ is an observable function and $\nu $ is a measure equivalence then $%
\tau (f\circ \nu )=\tau (f)$
\end{theorem}

\begin{proof}
Infact $\mu _{\left[ \varphi \right] }((f\circ \nu )^{-1}(B)\cap \left[
\varphi \right] )=\mu _{\left[ \varphi \right] }(f^{-1}B\cap \left[ \varphi %
\right] )$ for every $\varphi $ and every B$.$
\end{proof}

\bigskip

\textit{Given a self-adjoint operator }$T$ \textit{there are infinite
observable functions} $f$ \textit{\ such that }$\tau (f)=T$,\textit{\ infact}
\textit{for every measure equivalence }$\nu $\textit{\ we have }$\tau
(f\circ \nu )=T$\textit{\ . These however are not the only ones since it is
enough to take }$\nu $\textit{\ in such a way to have }$\nu _{\ast }\mu _{%
\left[ \varphi \right] }=\mu _{\left[ \varphi \right] }$\textit{\ for every}
$\mathbb{S}^{1}$\textit{-orbit }$\left[ \varphi \right] $\textit{\ : it is
not necessary for }$\nu $\textit{\ to be a borel isomorphism for every} $%
\mathbb{S}^{1}$\textit{-orbit.}

\textit{Therefore ''behind'' a quantum observable }$T$\textit{\ there are
infinite ''hidden'' classical observable functions }$f$\textit{: the choice
of a particular function} \textit{depends on the ''experimental context''}
\textit{choosen} \textit{and gives} \textit{the value }$f(\varphi )$ \textit{%
of the hidden classical observable on the hidden classical state }$\varphi $.

\bigskip

\begin{theorem}
\textit{Given two compatible propositions }$L$\textit{\ and }$M$\textit{\
the corresponding projectors }$\varepsilon (L)$\textit{\ and }$\varepsilon
(M)$\textit{\ commute. }
\end{theorem}

\begin{proof}
\textit{Infact there is an observable function }$f$\textit{\ and two borel
subsets }$B,C$\textit{\ of }$\mathbb{R}$\textit{\ such that }$L=f^{-1}(B)$%
\textit{\ and }$M=f^{-1}(C)$\textit{, therefore }$\varepsilon
(L)=E_{B}^{\tau (f)}$\textit{\ and }$\varepsilon (M)=E_{C}^{\tau (f)}$%
\textit{\ are in the same spectral measure.}
\end{proof}

\bigskip

\textit{In other words given two non commuting projectors }$E$\textit{\ and }%
$F$\textit{\ \textbf{it is not possible} to find two propositions }$L$%
\textit{\ and }$M$\textit{\ with }$\varepsilon (L)=E$\textit{\ and }$%
\varepsilon (M)=F$\textit{\ and moreover with }$L\cap M$\textit{\ in }$%
\mathcal{L}$\textit{; this means that in this theory it is never possible to
check if a state }$\varphi $\textit{\ in }$S$\textit{\ }$\ $\textit{%
verifies, in the same time, two non compatible properties.}

\textit{If }$\varepsilon (L)$\textit{\ and }$\varepsilon (M)$\textit{\ don't
commute the ''precise observer'' can build an apparatus checking the
property }$L$\textit{\ and separately an apparatus checking the property }$M$%
\textit{\ but he will never be able to build an apparatus checking the
classical property }$L$\textit{\ }$AND$\textit{\ }$M$\textit{\ \ and
satisfying the properties stated for a proposition!}

\textit{The decision to add to }$\mathcal{L}$\textit{\ all the intersections
}$L\cap M$ \textit{or, better, to consider all the boolean algebra (}$\sigma
$\textit{-algebra) generated by} $\mathcal{L}$\textit{\ is equivalent to
consider possible some behaviour for the probabilities not predicted by the
usual Quantum Mechanics.}

\textit{\bigskip }

\textit{Assigned two projectors }$E$\textit{\ and }$F$\textit{\ we know that
it is possible to find two propositions }$L$\textit{\ and }$M$\textit{\ with
}$E=\varepsilon (L)$\textit{, }$F=\varepsilon (M)$\textit{; it is possible
to find }$L$ \textit{and} $M$ \textit{and} \textit{a pseudo-borel subset }$N$%
\textit{\ in} $\mathbb{S}$ \textit{such that for every }$\varphi $\textit{\
in} $\mathbb{S}$ \textit{it holds:}
\begin{equation*}
\mu _{\left[ \varphi \right] }(N\cap \left[ \varphi \right] )=\mu _{\left[
\varphi \right] }(L\cap \left[ \varphi \right] )\cdot \mu _{\left[ \varphi %
\right] }(M\cap \left[ \varphi \right] )\text{ ?}
\end{equation*}

\textit{The answer is \textbf{yes}, it is not difficult and you can also
take }$N=L\cap M$: l\textit{et's fix a map }$\sigma :\mathbb{P}_{\mathbb{C}}%
\mathbb{(}\mathcal{H})\rightarrow \mathbb{S}$ \textit{such that }$\sigma %
\left[ \varphi \right] \in \left[ \varphi \right] $\textit{\ for every }$%
\mathbb{S}^{1}$-\textit{orbit }$\left[ \varphi \right] $\textit{, the two
pseudo-borelian subsets}

\begin{center}
$
\begin{array}{c}
L=\bigcup_{_{\left[ \varphi \right] }}e^{i\pi ]1-2\left\langle
E\right\rangle _{\varphi },1]}\cdot \sigma \left[ \varphi \right] \\
M=\bigcup_{_{\left[ \varphi \right] }}e^{i\pi ]-2\left\langle E\right\rangle
_{\varphi }-2\left\langle F\right\rangle _{\varphi }+2\left\langle
E\right\rangle _{\varphi }\left\langle F\right\rangle _{\varphi
}+1,-2\left\langle E\right\rangle _{\varphi }+2\left\langle E\right\rangle
_{\varphi }\left\langle F\right\rangle _{\varphi }+1]}\cdot \sigma \left[
\varphi \right]
\end{array}
$\textit{\ }
\end{center}

\textit{\ are propositions with }$\varepsilon (L)=E$\textit{, }$\varepsilon
(M)=F$\textit{\ and }$\mu _{\left[ \varphi \right] }(L\cap M\cap \left[
\varphi \right] )=\left\langle E\right\rangle _{\varphi }\cdot \left\langle
F\right\rangle _{\varphi }$\textit{.}

\textit{The pseudo-borelian subset }$L\cap M$ \textit{is not, in general, a
proposition because if you take two orthogonal elements }$\varphi $\textit{\
and }$\psi $\textit{\ in} $\mathbb{S}$ \textit{(not in the same} $\mathbb{S}%
^{1}$\textit{-orbit) and consider the states parametrized by the path }$%
\gamma (t)=\cos t\cdot \varphi +\sin t\cdot \psi $\textit{\ (the
superposition states of }$\varphi $\textit{\ and }$\psi $\textit{\ in} $%
\mathbb{S}$\textit{) then the function: }$\mu _{\left[ \gamma (t)\right]
}(L\cap M\cap \left[ \gamma (t)\right] )=\left\langle E\right\rangle
_{\gamma (t)}\cdot \left\langle F\right\rangle _{\gamma (t)}$\textit{\ is
the product of two functions of the form }$a\cos ^{2}t+b\sin t\cos t+c\sin
^{2}t$\textit{\ and is not of the same form (unless one of the factor is
constant).}

\bigskip

\textit{If we stay in }$\mathcal{L}$ \textit{in particular there is not hope
to use the informations coming from two apparatuses corresponding to }$L$%
\textit{\ and }$M$\textit{\ \ ''acting in two spatially separated regions of
the spacetime'' to decide whether or not a state }$\varphi $\textit{\ is in }%
$L\cap M$\textit{\ !}

\textit{This is connected with the absence of meaningful propositions
furnished with the following special independence (the one considered by
Bell to prove his inequalities):}

\bigskip

\begin{definition}
Two propositions $L$ and $M$ will be called \textbf{totally independent} if
there exists a proposition $N$ such that for every $\varphi $ in $\mathbb{S}$
it holds:
\begin{equation*}
\mu _{\left[ \varphi \right] }(N\cap \left[ \varphi \right] )=\mu _{\left[
\varphi \right] }(L\cap \left[ \varphi \right] )\cdot \mu _{\left[ \varphi %
\right] }(M\cap \left[ \varphi \right] )
\end{equation*}
Two propositions $L$ and $M$ will be called \textbf{banally independent} if
one of them is null equivalent to $\emptyset $ or to $\mathbb{S}$.
\end{definition}

\bigskip

\begin{theorem}
Two propositions (\textit{in }$\mathcal{L}$) are totally independent if and
only if are banally independent.
\end{theorem}

\begin{proof}
$(\Longleftarrow )$ Obvious.

$(\Longrightarrow )$ Let $E=\varepsilon (L)$, $F=\varepsilon (M)$ and $%
G=\varepsilon (N)$; we have $\left\langle G\right\rangle _{\varphi
}=\left\langle E\right\rangle _{\varphi }\cdot \left\langle F\right\rangle
_{\varphi }$ for every non zero vector in $\mathcal{H}$ by hypothesis.

This implies $\ker G=\ker E\cup \ker F$ but this is possible only for $\ker
E\subset \ker F$ or $\ker F\subset \ker E$.

In the first case we have: $G=F$ and $\left\langle F\right\rangle _{\varphi
}\cdot (1-\left\langle E\right\rangle _{\varphi })=0$, therefore or $F=0$ or
$F\neq 0$ and $E=I$.

In the other case analogously or $E=0$ or $F=I$.
\end{proof}

\bigskip

\textit{The following theorem claims that an observable function }$f$\textit{%
\ takes its essential values in the spectrum of the associated self-adjoint
operator }$\tau (f)$\textit{. Let's remember that when the spectrum (the
essential image) is discrete we can choose the function }$f$\textit{\
(modifying it, if necessary, on a null pseudo-borel subset) in such a way
that its image is exactly the spectrum.}

\bigskip

\begin{theorem}
For every observable function $f$ we have: $Im_{e}(f)=spec(\tau (f))$.
\end{theorem}

\begin{proof}
A value $y$ is not in the spectrum of $\tau (f)$ if and only if there a
neighborhood $]y-\sigma ,y+\sigma ]$ of $y$ such that $E_{]y-\sigma
,y+\sigma ]}^{\tau (f)}=0$ (cfr. [W] thm. 7.22 pag. 200), therefore if and
only if $\varepsilon (f^{-1}(]y-\sigma ,y+\sigma ])=0$ or, equivalently,
when $f^{-1}(]y-\sigma ,y+\sigma ]$ is a pseudo-borel null subset of $%
\mathbb{S}$; but this is precisely the condition for the value $y$ not to
lie in $Im_{e}(f)$.
\end{proof}

\bigskip

\textit{The observable functions are rarely continuous,} \textit{for example}
\textit{when }$spec(\tau (f))$\textit{\ is not an interval of} $\mathbb{R}$
\textit{the observable function} $\ f:\mathbb{S\rightarrow R}$ \textit{is
not continuous.}

\bigskip

\begin{theorem}
Fixed an $\mathbb{S}^{1}$-orbit $\left[ \varphi _{0}\right] $, for every
borel function $b:\left[ \varphi _{0}\right] \rightarrow \mathbb{R}$ there
is an observable function $f$ such that $f|_{\left[ \varphi _{0}\right] }=b$.
\end{theorem}

\begin{proof}
Let's suppose first $\mathcal{H=}L_{\mathbb{C}}^{2}(\mathbb{R},\lambda )$
and $\varphi _{0}$ a positive real continuous function in $\mathbb{S}$.

As in the proof of a previous theorem let's fix a map $\sigma :\mathbb{P}_{%
\mathbb{C}}\mathbb{(}\mathcal{H})\rightarrow \mathbb{S}$ such that $\sigma %
\left[ \varphi \right] \in \left[ \varphi \right] $ for every $\mathbb{S}%
^{1} $-orbit $\left[ \varphi \right] $; associated to the self-adjoint
position operator $Q$ there is the cumulative function $F_{\left[ \varphi
_{0}\right] }:\mathbb{R\rightarrow \lbrack }0,1]$ defined by the equality: $%
F_{\left[ \varphi _{0}\right] }(s)=\left\langle E_{(-\infty
,s]}^{Q}\right\rangle _{\varphi _{0}}=\frac{1}{2}\int_{-\infty }^{s}\varphi
_{0}(r)^{2}\cdot d\lambda (r)$. This function is a homeomorphism between $%
\mathbb{R}$ and $\mathbb{]}0,1[$ therefore its quasi-inverse is equal to its
inverse and is again a homeomorphism and is a homeomorphism too the
restricted function $f_{\sigma }^{Q}|:\left[ \varphi _{0}\right] \backslash
\left\{ \varphi _{0}\right\} \rightarrow \mathbb{R}$. Let $f_{1}$ be an
observable function on $\mathbb{S}$ extending $f_{\sigma }^{Q}$ to the
pseudo-borel null subset $-\sigma \mathbb{P}_{\mathbb{C}}\mathbb{(}\mathcal{H%
})$.

The real function $c=b\circ \left( f_{\sigma }^{Q}|\right) ^{-1}$ is a borel
function and then the composed function $f_{2}=c\circ f_{1}$ is an
observable function with $f_{2}|_{\left[ \varphi _{0}\right] }=b$ out of $%
-\sigma \left[ \varphi _{0}\right] $, then the observable function $f$
defined as $f_{2}$ out of $-\sigma \mathbb{P}_{\mathbb{C}}\mathbb{(}\mathcal{%
H})$ and defined as $b(-\sigma \left[ \varphi _{0}\right] )$ on $-\sigma
\mathbb{P}_{\mathbb{C}}\mathbb{(}\mathcal{H})$ is the required function.

For any other vector $\varphi _{1}$ in $\mathbb{S}$ taken a unitary map $U$
such that $U(\varphi _{0})=\varphi _{1}$ the function $b\circ f\circ U^{-1}$
will have the desired property with respect to $\varphi _{1}$.

The proof above extends easily to any separable Hilbert space (with infinite
dimension).

For a general Hilbert space (with infinite dimension) it is possible to
proceed in an analogous way on a single addend remembering that any such
space is isomorphic to a direct Hilbert sum of a family $\left\{ L_{\mathbb{C%
}}^{2}(\mathbb{R}_{\beta },\lambda )\right\} _{\beta }$ of $L^{2}$-spaces.
\end{proof}

\bigskip

\begin{remark}
In particular taken $\varphi _{0}$ in $\mathbb{S}$ there exists an
observable function coincident on $\left[ \varphi _{0}\right] $ with the
function: $d(\varphi _{0},\varphi )=\left| Arg(\varphi /\varphi _{0})\right|
$ (the ''phase distance from the state $\varphi _{0}$'').
\end{remark}

\bigskip

\begin{definition}
Let $f$ be an observable function on $\mathbb{S}$, let's denote by:
\begin{equation*}
\mathcal{D(}\left\langle f\right\rangle )=\left\{ \varphi \in \mathbb{S}%
\text{; \ }f|_{\left[ \varphi \right] }\in L_{2}(\left[ \varphi \right] ,\mu
_{\left[ \varphi \right] })\right\}
\end{equation*}
the \textbf{domain} of the \textbf{mean value function} of $f:$%
\begin{equation*}
\left\langle f\right\rangle :\mathcal{D(}\left\langle f\right\rangle
)\rightarrow \mathbb{R}
\end{equation*}
defined by $\left\langle f\right\rangle (\varphi )=\int_{\left[ \varphi %
\right] }f|_{\left[ \varphi \right] }\cdot d\mu _{\left[ \varphi \right] }$.
\end{definition}

\bigskip

\textit{Note that since }$\mu _{\left[ \varphi \right] }\left( \left[
\varphi \right] \right) $\textit{\ is finite it holds: }$L_{2}(\left[
\varphi \right] ,\mu _{\left[ \varphi \right] })\subset L_{1}(\left[ \varphi %
\right] ,\mu _{\left[ \varphi \right] })$,\textit{\ then }$\left\langle
f\right\rangle $\textit{\ is well defined with: }$\left\langle
f\right\rangle (\varphi )=\int_{\left[ \varphi \right] }f|_{\left[ \varphi %
\right] }\cdot d\mu _{\left[ \varphi \right] }=$ =$\frac{1}{2\pi }%
\int_{0}^{2\pi }f(e^{i\theta }\cdot \varphi )\cdot d\lambda (\theta )$.

\textit{Let's remember that we have: }$D(T)=\left\{ \varphi \in \mathcal{H};%
\text{ }\int_{\mathbb{R}}t^{2}\cdot d\nu _{F_{\varphi }^{T}}<+\infty
\right\} $ \textit{for every self-adjoint operator }$T$\textit{\ \ (cfr. [W]
where }$F_{\varphi }^{T}:\mathbb{R\rightarrow R}$ \textit{is the function
defined by} $F_{\varphi }^{T}(r)=\left\langle E_{(-\infty
,r]}^{T}\right\rangle _{\varphi }$\textit{).}

\bigskip

\begin{theorem}
For every observable function $f$ we have:

\begin{itemize}
\item  $\mathcal{D(}\left\langle f\right\rangle )=\mathcal{D(\tau (}f))\cap
\mathbb{S}$

\item  $\left\langle f\right\rangle =\left\langle \mathcal{\tau (}%
f)\right\rangle |_{\mathcal{D(\tau (}f))\cap \mathbb{S}}$
\end{itemize}
\end{theorem}

\begin{proof}
$\int_{\mathbb{R}}t^{2}\cdot d\nu _{F_{\varphi }^{\mathcal{\tau (}f)}}=\int_{%
\mathbb{R}}t^{2}\cdot d(f|_{\left[ \varphi \right] \ast }\mu _{\left[
\varphi \right] })=\int_{\left[ \varphi \right] }\left( f|_{\left[ \varphi %
\right] }\right) ^{2}\cdot d\mu _{\left[ \varphi \right] }$ (cfr. [K-S] pag.
93) therefore the integrals are finite together.

For the equality in Thm. 7.14 (e) of [W] we have for $\varphi $ in $\mathcal{%
D(\tau (}f))\cap \mathbb{S}$ the equalities: $\left\langle \mathcal{\tau (}%
f)\right\rangle _{\varphi }=\int_{\mathbb{R}}t\cdot d\nu _{F_{\varphi }^{%
\mathcal{\tau (}f)}}=\int_{\mathbb{R}}t\cdot d(f|_{\left[ \varphi \right]
\ast }\mu _{\left[ \varphi \right] })=\int_{\left[ \varphi \right] }f|_{%
\left[ \varphi \right] }\cdot d\mu _{\left[ \varphi \right] }=\left\langle
f\right\rangle (\varphi )$.
\end{proof}

\bigskip

\begin{definition}
An observable function $f:\mathbb{S\rightarrow R}$ is \textbf{essentially
bounded }if \ there is a real number $M$ such that $f^{-1}((-\infty ,-M[\cup
]M,+\infty ))$ is a pseudo-borel null subset.
\end{definition}

\bigskip

\textit{An observable function }$f$\textit{\ is essentially bounded if and
only if its essential image is a bounded subset of }$R$\textit{.}

\textit{\bigskip }

\begin{notation}
Let's denote by $\mathcal{OB}$ the set of all essentially bounded observable
functions on $\mathbb{S}$.
\end{notation}

\bigskip

\textit{Let }$f$\textit{\ be an observable function, a real value }$y$%
\textit{\ is not in }$Im_{e}(f)$\textit{\ if and only if there exists an
essentially bounded observable }$g$\textit{\ such that }$g\cdot (f-y)$%
\textit{\ is null equivalent to }$1$, \textit{in other words if and only if
the function }$\frac{1}{f-y}$\textit{\ is defined almost everywhere} (%
\textit{exercise})\textit{.}

\bigskip

\begin{theorem}
For an observable function $f$ are equivalent:

\begin{enumerate}
\item  the function $f$ is essentially bounded

\item  the operator $\tau (f)$ is a bounded operator

\item  $\mathcal{D(}\left\langle f\right\rangle )=\mathbb{S}$
\end{enumerate}
\end{theorem}

\begin{proof}
$(2\Longleftrightarrow 3)$ For the Hellinger-Toeplitz thm. $\tau (f)$ is a
bounded operator if and only if $\mathcal{D(\tau (}f))=\mathcal{H}$; since $%
\mathcal{D(}\left\langle f\right\rangle )=\mathcal{D(\tau (}f))\cap \mathbb{S%
}$ this is equivalent to $\mathcal{D(}\left\langle f\right\rangle )=\mathbb{S%
}$.

$(1\Longrightarrow 3)$ If the function $f$ is essentially bounded then each
function $f|_{\left[ \varphi \right] }$ is in $L_{2}(\left[ \varphi \right]
,\mu _{\left[ \varphi \right] })$.

$(2\Longrightarrow 1)$ If $\tau (f)$ is a bounded operator then its spectrum
is bounded, therefore so is the essential image of $f$.
\end{proof}

\bigskip

\textit{The map} $\tau :\mathcal{O\rightarrow }SA(\mathcal{H}\mathbb{)}$
\textit{sends} $\mathcal{OB}$ \textit{in} $SAB(\mathcal{H}\mathbb{)}$.

\bigskip

\begin{theorem}
The mean value function $\left\langle f\right\rangle $ of an observable
function $f$ is a kaehlerian function, if the function $f$ is essentially
bounded then the function $\left\langle f\right\rangle $ is smooth
kaehlerian.

Every kaehlerian function is the mean value of an observable function and
every smooth kaehlerian function is the mean value of a bounded observable
function.
\end{theorem}

\begin{proof}
Since $\left\langle f\right\rangle =\left\langle \tau (f)\right\rangle $ the
function $\left\langle f\right\rangle $ is kaehlerian; when $f$ is
essentially bounded the function $\left\langle f\right\rangle $ is smooth
kaehlerian as $\left\langle \tau (f)\right\rangle $.

If $l$ is a kaehlerian function there exists a self-adjoint operator $T$
such that $l=\left\langle T\right\rangle $, therefore taken an observable
function $f$ such that $\tau (f)=T$ we have $\left\langle f\right\rangle
=\left\langle T\right\rangle =l$. When $l$ is smooth kaehlerian $T$ is
bounded and $f$ can be taken essentially bounded.
\end{proof}

\bigskip

\textit{The map} $\left\langle \cdot \right\rangle :\mathcal{O\rightarrow }%
\mathcal{K}(\mathcal{H}\mathbb{)}$ \textit{sends} $\mathcal{OB}$ \textit{in}
$\mathcal{KS}(\mathcal{H}\mathbb{)}$.

\bigskip

$\alpha (\left\langle f\right\rangle )=\tau (f)$.

\bigskip

\begin{theorem}
Let $f$ be an observable function for evey borelian function $b:\mathbb{%
R\rightarrow R}$ it holds:
\begin{equation*}
b(\tau (f))=\tau (b\circ f)
\end{equation*}
\end{theorem}

\begin{proof}
The operator $\tau (f)$ has spectral measure: $\left\{ E_{B}^{\tau
(f)}\right\} _{B\in \mathcal{B(}\mathbb{R)}}=\left\{ \varepsilon
(f^{-1}(B)\right\} _{B}$ and $\tau (b\circ f)$ has spectral measure given by
$\left\{ \varepsilon ((b\circ f)^{-1}(B)\right\} _{B}=\left\{ \varepsilon
(f^{-1}(b^{-1}B))\right\} _{B}$ but this is exactly the spectral measure of $%
b(\tau (f))$ (cfr. [W] prop. pag. 196), therefore $b(\tau (f))=\tau (b\circ
f)$.
\end{proof}

\bigskip

\textit{We have: }$\chi _{B}\circ \tau (f)=\varepsilon (f^{-1}B)$ \textit{and%
} $\left\langle b(\tau (f))\right\rangle =\left\langle b\circ f\right\rangle
$.

\bigskip

\bigskip

\section{\protect\LARGE Algebras of propositions and observables}

\bigskip

\begin{theorem}
Let $\mathcal{A}$ be a boolean algebra of subsets of $\mathbb{S}$ contained
in $\mathcal{L}$ and let $L$, $M$ be two elements of $\mathcal{A}$:

\begin{itemize}
\item  $\varepsilon (L\cap M)=\varepsilon (L)\wedge \varepsilon
(M)=\varepsilon (L)\cdot \varepsilon (M)$ and $\varepsilon (L\cup
M)=\varepsilon (L)\vee \varepsilon (M)=\varepsilon (L)+\varepsilon
(M)-\varepsilon (L)\cdot \varepsilon (M)$

\item  $\varepsilon (\mathcal{A}$ $)$ is a boolean algebra of commuting
projectors of $\mathcal{H}$ and $\varepsilon |:\mathcal{A\rightarrow }%
\varepsilon (\mathcal{A}$ $)$ is a boolean algebra morphism

\item  $\varepsilon (L)=\varepsilon (M)$ if and only if $L$ and $M$ are null
equivalent
\end{itemize}
\end{theorem}

\begin{proof}
Since $L$ and $M$ are compatible we know that there exists an observable
function $f$ and two borel subsets $B$ and $C$ of $\mathbb{R}$ such that $%
L=f^{-1}(B)$ and $M=f^{-1}(C)$; in particular this proves that $\varepsilon
(L)$ and $\varepsilon (M)$ commute, then:

\begin{itemize}
\item  $\varepsilon (L\cap M)=E_{B\cap C}^{\tau (f)}=E_{B}^{\tau (f)}\cdot
E_{C}^{\tau (f)}=\varepsilon (L)\varepsilon (M)=\varepsilon (L)\wedge
\varepsilon (M)$ and also $\varepsilon (L\cup M)=E_{B\cup C}^{\tau
(f)}=\varepsilon (L)+\varepsilon (M)-\varepsilon (L)\varepsilon
(M)=\varepsilon (L)\vee \varepsilon (M)$

\item  we already know that $\varepsilon (\complement L)=I-\varepsilon (L)$

\item  if $\varepsilon (L)=\varepsilon (M)$ then $\varepsilon (L\Delta
M)=\varepsilon (L\cup M)-\varepsilon (L\cap M)=\varepsilon (L)-\varepsilon
(L)=0$, therefore $L\Delta M$ is a pseudo-borel null subset.
\end{itemize}
\end{proof}

\bigskip

\begin{theorem}
Let $\mathcal{B}$ a boolean algebra of (commuting) projectors in a separable
Hilbert space $\mathcal{H}$, there exists a boolean algebra $\widetilde{%
\mathcal{B}}$ of subsets of $\mathbb{S}$ in $\mathcal{L}$ such that:

\begin{itemize}
\item  $\varepsilon (\widetilde{\mathcal{B}}$ $)=\mathcal{B}$

\item  $\varepsilon |:\widetilde{\mathcal{B}}$ $\rightarrow \mathcal{B}$ $\ $%
is a morphism of boolean algebras.
\end{itemize}
\end{theorem}

\begin{proof}
Let $\mathcal{B=}\left\{ E_{i}\right\} _{i\in I}$, since the $E_{i}$ commute
pairwise there exist a self-adjoint operator $T$ and a family of borel
functions $\left\{ b_{i}\right\} _{i\in I}$ such that $E_{i}=b_{i}\circ T$
for every $i\in I$ (cfr. [V] Thm. 3.9 p. 56).

Taken an observable function $f$ such that $\tau (f)=T$ let's consider the
boolean $\sigma $-algebra $\mathcal{A}_{f}$ of subsets of $\mathbb{S}$ in $%
\mathcal{L}$,\ the functions $b_{i}\circ f$ are all in $\mathcal{O}_{%
\mathcal{A}_{f}}$ and the propositions $L_{i}=f^{-1}(b_{i}^{-1}(1))$ are all
in $\mathcal{A}_{f}$. Obviously $\tau (b_{i}\circ f)=E_{i}$ , moreover $\tau
((b_{i}\circ f)^{2}-b_{i}\circ f)=0$ therefore there exists a null
pseudo-borel subset $N_{i}$ of $\mathbb{S}$ such that $(b_{i}\circ
f)^{2}-b_{i}\circ f=0$ on of $\mathbb{S\setminus }N_{i}$. That is $%
b_{i}\circ f$ is null equivalent to $\chi _{L_{i}}$ , this implies $%
E_{i}=\tau (b_{i}\circ f)=\tau (\chi _{L_{i}})=\varepsilon (L_{i})$ for
every $i\in I$ and therefore $\varepsilon (\mathcal{A}_{f})\supset \mathcal{B%
}$. It is enough now to choose $\widetilde{\mathcal{B}}=\left( \varepsilon
|_{\mathcal{A}_{f}}\right) ^{-1}(\mathcal{B)}$.
\end{proof}

\bigskip

\begin{remark}
A boolean algebra of pairwise commuting projectors\textit{\ can always be
realized in }$\mathcal{L}$ \textit{trought a boolean algebra of compatible
propositions. This realization is not unique.}
\end{remark}

\bigskip

\begin{corollary}
When $\mathcal{H}$ is a (separable) Hilbert space two projectors $E$ and $F$
commute if and only if there exist two compatible propositions $L$ and $M$
such that $\varepsilon (L)=E$ and $\varepsilon (M)=F$.
\end{corollary}

\bigskip

\begin{theorem}
Let $\mathcal{H}$ be a (separable) Hilbert space, two propositions $L$ and $%
M $ have commuting associated projectors $\varepsilon (L)$ and $\varepsilon
(M) $ if and only if there exists a measure equivalence $\nu $ such that $L$
and $\nu (M)$ are compatible.
\end{theorem}

\begin{proof}
If $L$ and $\nu (M)$ are compatible then $\varepsilon (L)$ and $\varepsilon
(\nu M)=\varepsilon (M)$ commute.

Conversely if $\varepsilon (L)$ and $\varepsilon (M)$ commute there are two
propositions $L^{\prime }$ and $M^{\prime }$ with $L^{\prime }\cap M^{\prime
}$ in $\mathcal{L}$ and $\varepsilon (L^{\prime })=\varepsilon (L)$, $%
\varepsilon (M^{\prime })=\varepsilon (M)$. Therefore there exist two
measure equivalences $\rho $ and $\sigma $ such that $L^{\prime }$is null
equivalent to $\rho (L)$ and $M^{\prime }$is null equivalent to $\sigma (M)$%
, then $\rho L\cap \sigma M$ and $L\cap \rho ^{-1}\sigma M$ are in $\mathcal{%
L}$.
\end{proof}

\bigskip

\begin{theorem}
Let $\mathcal{A}$ be a $\sigma $-algebra of subsets in $\mathbb{S}$
contained in $\mathcal{L}$ and $\left\{ L_{n}\right\} _{n\geq 1}\subset
\mathcal{A}$, there exists an observable $f$ in $\mathcal{O}_{\mathcal{A}}$
and a sequence $\left\{ B_{n}\right\} _{n\geq 1}\subset \mathcal{B(}\mathbb{R%
}\mathcal{)}$ such that $f^{-1}(B_{n})=L_{n}$ for every $n\geq 1$.
\end{theorem}

\begin{proof}
Let $\chi :\mathbb{S\rightarrow }\left\{ 0,1\right\} ^{\mathbb{N}^{+}}$be
the map defined by $\chi (\varphi )=\left( \chi _{L_{n}}(\varphi )\right)
_{n\geq 1}$, taken a borel equivalence $\beta :\left\{ 0,1\right\} ^{\mathbb{%
N}^{+}}\rightarrow \mathbb{R}$ between $\left\{ 0,1\right\} ^{\mathbb{N}%
^{+}} $ and a borel subset of $\mathbb{R}$ it is not difficult to prove that
the function $f=\beta \circ \chi :\mathbb{S\rightarrow R}$ is an observable
in $\mathcal{O}_{\mathcal{A}}$.

Taken a borel subset $B_{n}$ of $\mathbb{R}$ such that $\beta
^{-1}(B_{n})=\prod_{k\neq n}\left\{ 0,1\right\} \times \left\{ 1\right\}
_{n} $ it is possible to check that $f^{-1}(B_{n})=L_{n}$.
\end{proof}

\bigskip

\begin{corollary}
Let $\mathcal{A}$ be a $\sigma $-algebra of subsets in $\mathbb{S}$
contained in $\mathcal{L}$ and $\left\{ L_{n}\right\} _{n\geq 1}\subset
\mathcal{A}$, then $\varepsilon (\bigcup_{n\geq 1}L_{n})=\bigvee_{n\geq
1}\varepsilon (L_{n})$.
\end{corollary}

\begin{proof}
$\varepsilon (\bigcup_{n\geq 1}L_{n})=\varepsilon (f^{-1}(\bigcup_{n\geq
1}B_{n}))=E_{\cup _{n}B_{n}}^{\tau (f)}=\bigvee_{n}E_{B_{n}}^{\tau
(f)}=\bigvee_{n}\varepsilon (L_{n})$.
\end{proof}

\bigskip

\begin{theorem}
\begin{theorem}
When $\mathcal{H}$ is a separable Hilbert space assigned a sequence $%
\mathcal{E=}\left\{ E_{n}\right\} _{n\in \mathbb{N}}$ of pairwise orthogonal
(complex) projectors with $\sum_{n\in \mathbb{N}}E_{n}=I$ \ it is possible
to find a partition $\left\{ L_{n}\right\} _{n\in \mathbb{N}}$ of $\mathbb{S}
$ by propositions with $\varepsilon (L_{n})=E_{n}$ for every $n\in \mathbb{N}
$.
\end{theorem}
\end{theorem}

\begin{proof}
Proceeding as in the proof of a previous theorem we find an observable $f$,
a sequence of borel functions $\left\{ b_{n}\right\} _{n\in \mathbb{N}}$ and
propositions $L_{n}^{\prime }=f^{-1}(b_{n}^{-1}(1))$ such that $\varepsilon
(L_{n}^{\prime })=E_{n}$ for every $n\in \mathbb{N}$.

Since the $E_{n}$ are pairwise orthogonal the propositions have a pairwise
pseudo-borel null intersection $N_{nm}=L_{n}^{\prime }\cap L_{m}^{\prime }$
(whenever $n\neq m$); since $\sum_{n\in \mathbb{N}}E_{n}=I$ $\ $the union $%
\bigcup_{n\in \mathbb{N}}L_{n}^{\prime }$ has a pseudo-borel null complement
$N_{0}$ in $\mathbb{S}$.

The set $N=N_{0}\cup \bigcup_{n\neq m}N_{nm}$ is a pseudo-borel null subset,
taken $L_{1}=L_{1}^{\prime }\cup N$ and $L_{n}=L_{n}^{\prime }\setminus N$
for every $n\geq 2$, we have $L_{n}\cap L_{m}=\emptyset $ (whenever $n\neq m$%
) and $\bigcup_{n\in \mathbb{N}}L_{n}=\mathbb{S}$.
\end{proof}

\bigskip

\begin{remark}
\textit{The theorem above defines a map }$\delta :\mathcal{E\rightarrow L}$%
\textit{\ \ (defined by }$\delta (E_{n})=L_{n}$) \textit{such that }$%
(\varepsilon \circ \delta )(E)=E$\textit{\ for every }$E$\textit{\ in} $%
\mathcal{E}$ and \textit{transforming the sequence }$\left\{ E_{n}\right\}
_{n\in \mathbb{N}}$ \textit{of pairwise orthogonal (complex) projectors with}
$\sum_{n\in \mathbb{N}}E_{n}=I$ \textit{\ in a partition }$\left\{ \delta
(E_{n})\right\} _{n\in \mathbb{N}}$\ \textit{of }$\mathbb{S}$\textit{.}
\textit{This construction can not be generalized too much:}
\end{remark}

\bigskip

\begin{theorem}
Let $\mathcal{H}$ be a separable Hilbert space (of complex dimension at
least three), $\mathcal{E}$ $\ $a family of (complex) projectors and $\delta
:\mathcal{E\rightarrow L}$ a map such that:

\begin{enumerate}
\item  $(\varepsilon \circ \delta )(E)=E$ for every $E$ in $\mathcal{E}$

\item  if $E$ \ and $F$ are orthogonal in $\mathcal{E}$ $\ $then $\delta (E)$%
\ $\ $and $\delta (F)$\ are disjoint

\item  $\delta \left( \sum_{n\in \mathbb{N}}E_{n}\right) =\bigcup_{n}\delta
(E_{n})$\ if the $E_{n}$ are pairwise orthogonal in $\mathcal{E}$

\item  $\delta (I)$\ $=\mathbb{S}$
\end{enumerate}

then $\mathcal{E}$ \ \textbf{cannot} contain the family of all projectors on
(complex) lines.
\end{theorem}

\begin{proof}
Let's suppose the thesis is false; fixed a vector $\varphi _{0}$ in $\mathbb{%
S(}\sqrt{2})$ let's consider the function $G:$ $\mathbb{S(}1)\rightarrow
\left\{ 0,1\right\} $ defined by $G(u)=\chi _{\delta (pr_{\mathbb{C}%
u})}(\varphi _{0})$. For every orthonormal basis $\left\{ u_{n}\right\}
_{n\geq 1}$ in $\mathcal{H}$ the family $\left\{ \delta (pr_{\mathbb{C}%
u_{n}})\right\} _{n}$ is a partition of $\mathbb{S(}\sqrt{2})$, therefore
the vector $\varphi _{0}$ belongs to one and only to one of the sets $\delta
(pr_{\mathbb{C}u_{n}})$. This implies that $\sum_{n}G(u_{n})=1$ for every
orthonormal basis $\left\{ u_{n}\right\} _{n\geq 1}$, that is the function $%
G $ is a Gleason frame function (of weight $1$) (cfr. [Gl]).

Since we are in a separable Hilbert space of dimension at least three there
must exist a bounded self-adjoint operator $T$ such that $G(u)=\left\langle
T\right\rangle _{u}$ for every $u$ in $\mathbb{S(}1)$ (cfr. Thm. 3.5 of
[Gl]), but this implies that the function $\left\langle T\right\rangle $ is
constantly $0$ or costantly $1$ bringing in both cases to an absurd since
the vector $\varphi _{0}$ must belong to some of the sets $\delta (pr_{%
\mathbb{C}u})$ but cannot belong to all of them.
\end{proof}

\bigskip

\begin{remark}
T\textit{he previous theorem is stricly connected with the necessity of
considering the observables in their context (cfr. Ghirardi in [B] 4.6.5,
[G-D] and }[KS]\textit{).}
\end{remark}

\bigskip

\begin{theorem}
Let $\mathcal{A}$ be a (non empty) $\sigma $-algebra of subsets in $\mathbb{S%
}$ contained in $\mathcal{L}$ and $\left\{ f_{n}\right\} _{n\geq 1}\subset
\mathcal{O}_{\mathcal{A}}$, there exists an observable $f$ in $\mathcal{O}_{%
\mathcal{A}}$ such that $\mathcal{A}_{f_{n}}\subset \mathcal{A}_{f}$ for
every $n\geq 1$.
\end{theorem}

\begin{proof}
For the countable family $\left\{ f_{n}^{-1}(r,s);\text{ with }n\geq 1\text{
and }r,s\text{ in }\mathbb{Q}\right\} $ it is possible, for a previous
theorem, to find an observable $f$ in $\mathcal{O}_{\mathcal{A}}$ and a
countable family $\left\{ B_{n,r,s}\right\} $ in $\mathcal{B(}\mathbb{R}%
\mathcal{)}$ such that $f^{-1}(B_{n,r.s})=f_{n}^{-1}(r,s)$ for every $n$, $r$%
, $s$. Therefore $f_{n}^{-1}(B)\in \mathcal{A}_{f}$ $\ $for every $n\geq 1$
and every borel subset $B$ of $\mathbb{R}$.
\end{proof}

\bigskip

\begin{theorem}
Let $f$, $g$ be two observable functions, if $\mathcal{A}_{g}\subset
\mathcal{A}_{f}$ $\ $then there exists a borelian function $b:\mathbb{%
R\rightarrow R}$ such that $g=b\circ f$.
\end{theorem}

\begin{proof}
Let $\mathbb{Q=}\left\{ r_{n}\right\} _{n\geq 1}$ the set of rational nubers
in a sequence, since $\mathcal{A}_{g}\subset \mathcal{A}_{f}$ it is possible
to find a borel subset $B_{1}$ in $\mathbb{R}$ such that $%
f^{-1}(B_{1})=g^{-1}(-\infty ,r_{1})$; then, taken $B$ such that $%
f^{-1}(B)=g^{-1}(-\infty ,r_{2})$ it is possible to ''correct'' it by
defining $B_{2}=B\cap B_{1}$ if $r_{2}<r_{1}$ and $B_{2}=B\cup B_{1}$ if
instead $r_{1}<r_{2}$.

In both cases we get $f^{-1}(B_{2})=g^{-1}(-\infty ,r_{2})$ and, after the
''correction'', we have moreover that $B_{1}$ and $B_{2}$ are ordered as $%
r_{1}$, $r_{2}$.

Let $\left\{ r_{1},r_{2}\right\} =\left\{ r_{k_{1}},r_{k_{2}}\right\} $with $%
r_{k_{1}}<r_{k_{2}}$, taken $B$ with $f^{-1}(B)=g^{-1}(-\infty ,r_{3})$ it
is possible to ''correct'' it by defining $B_{3}=B\cap B_{k_{1}}$ if $%
r_{3}<r_{k_{1}}<r_{k_{2}}$, $B_{3}=B_{k_{1}}\cup (B\cap B_{k_{2}})$ if $%
r_{k_{1}}<r_{3}<r_{k_{2}}$and $B_{3}=B\cup B_{k_{2}}$ if instead $%
r_{k_{1}}<r_{k_{2}}<r_{3}$.

Again we get $f^{-1}(B_{3})=g^{-1}(-\infty ,r_{3})$ and $B_{1}$, $B_{2}$, $%
B_{3}$ are ordered as $r_{1}$, $r_{2}$, $r_{3}$.

Proceeding in this way it is possible to find a sequence $\left\{
B_{n}\right\} _{n\geq 1}\subset \mathcal{B(}\mathbb{R}\mathcal{)}$ such that
$f^{-1}(B_{n})=g^{-1}(-\infty ,r_{n})$ or every $n\geq 1$ and $B_{n}\subset
B_{m}$ whenever $r_{n}<r_{m}$.

It is not difficult to check that $f(\mathbb{S)\subset }X=\bigcup_{n}B_{n}%
\mathbb{\setminus }\bigcap_{n}B_{n}$.

The function $b:\mathbb{R\rightarrow R}$ defined by $b(x)=\inf \left\{ r_{n};%
\text{ }x\in B_{n}\right\} $ when $x\in X$ and $0$ elsewhere is well defined
and it is a borel function since it holds the equality: $b^{-1}(-\infty
,s)\cap X=\left( \bigcup_{r_{n}<s}B_{n}\right) \cap X$ for every $s$ in $%
\mathbb{R}$.

In the end it is possible to check that $f^{-1}(b^{-1}(-\infty
,s))=g^{-1}(-\infty ,s)$ for every $s$ in $\mathbb{R}$; this proves that $%
g=b\circ f$.
\end{proof}

\bigskip

\begin{corollary}
Let $\mathcal{A}$ be a (non empty) $\sigma $-algebra of subsets in $\mathbb{S%
}$ contained in $\mathcal{L}$ and $\left\{ f_{n}\right\} _{n\geq 1}\subset
\mathcal{O}_{\mathcal{A}}$, there exists an observable $f$ in $\mathcal{O}_{%
\mathcal{A}}$ and a sequence $\left\{ b_{n}\right\} _{n\geq 1}$ of borelian
functions such that $f_{n}=b_{n}\circ f$ for every $n\geq 1$.
\end{corollary}

\begin{proof}
It follows from the last two theorems.
\end{proof}

\bigskip

\begin{theorem}
Let $\mathcal{A}$ be a (non empty) $\sigma $-algebra of subsets in $\mathbb{S%
}$ contained in $\mathcal{L}$

\begin{itemize}
\item  $\tau (\mathcal{O}_{\mathcal{A}})$ is a commutative algebra of
operators

\item  the map:$\tau |:\mathcal{O}_{\mathcal{A}}\rightarrow SA(\mathcal{H)}$
is an algebra homomorphism

\item  $\ker (\tau |)$ is the set of observable functions that are zero out
of a null pseudo-borel subset of $\mathcal{A}$.
\end{itemize}
\end{theorem}

\begin{proof}
Taken two functions $f,g$ in $\mathcal{O}_{\mathcal{A}}$ the two
self-adjoint operators $\tau (f)$ and $\tau (g)$ commute since all the
projectors $E_{B}^{\tau (f)}$ and $E_{C}^{\tau (g)}$ in their spectral
measures commute; infact: since $f^{-1}B$ and $g^{-1}C$ are in $\mathcal{A}$
the projectors $E_{B}^{\tau (f)}=\varepsilon (f^{-1}B)$ and $E_{C}^{\tau
(g)}=\varepsilon (g^{-1}C)$ commute.

The $\sigma $-algebras $\mathcal{A}_{f}$ and $\mathcal{A}_{g}$ are in $%
\mathcal{A}$ therefore there exists a function $h$ in $\mathcal{O}_{\mathcal{%
A}}$ and two borel functions $b,c:\mathbb{R\rightarrow R}$ such that $%
f=b\circ h$ and $g=c\circ h$. Then we have: $\tau (f+g)=\tau ((b+c)\circ h)=$
$=(b+c)(\tau (h))=\tau (f)+\tau (g)$ and analogously $\tau (f\cdot g)=\tau
(f)\cdot \tau (g)$.

If $\tau (f)=0$ then $f^{-1}(\mathbb{R\setminus }\left\{ 0\right\} )$ is a
pseudo-borel null subset in $\mathcal{A}$.
\end{proof}

\bigskip

$\tau (1/f)=\tau (f)^{-1}$\textit{for an never zero observable }$f$.

\bigskip

\begin{theorem}
Let $\mathcal{H}$ be a separable Hilbert space and let $\mathcal{R}$ be a
commutative algebra in the family of self-adjoint operators of $\mathcal{H}$%
. There exists in $\mathcal{O}$ a commutative algebra $\widetilde{\mathcal{R}%
}$ of observables such that:

\begin{itemize}
\item  $\tau (\widetilde{\mathcal{R}})=\mathcal{R}$

\item  $\tau |:\widetilde{\mathcal{R}}\rightarrow \mathcal{R}$ is an algebra
homomorphism

\item  $\ker (\tau |)=\left\{ f\in \widetilde{\mathcal{R}}:\text{ }f\text{
is null equivalent to }0\right\} $
\end{itemize}
\end{theorem}

\begin{proof}
Let $\mathcal{R=}\left\{ T_{i}\right\} _{i\in I}$, since the operators
commute there exist a self-adjoint operator $T$ and a family of borel
functions $\left\{ b_{i}\right\} _{i\in I}$ such that $T_{i}=b_{i}\circ T$
for every $i\in I$ (cfr. [V] Thm. 3.9 p. 56).

Taken an observable function $f$ such that $\tau (f)=T$ let's consider the $%
\sigma $-algebra $\mathcal{A}=\mathcal{A}_{f}$,\ the functions $b_{i}\circ f$
are all in $\mathcal{O}_{\mathcal{A}}$ . Obviously $\tau (b_{i}\circ
f)=T_{i} $ therefore $\tau (\mathcal{O}_{\mathcal{A}})\supset \mathcal{R}$.
The proof is concluded taking $\widetilde{\mathcal{R}}=(\tau |_{\mathcal{O}_{%
\mathcal{A}}})^{-1}(\mathcal{R)}$.
\end{proof}

\bigskip

\textit{The algebra homomorphism }$\tau |:\widetilde{\mathcal{R}}\rightarrow
\mathcal{R}$ \textit{is essentially injective since measurable functions are
usually identified when differ only on a null subset. Therefore the theorem
above asserts that you can always realize a commutative operators algebra
trought a commutative algebra of observable functions.}

\textit{This realization is not unique and the choice of the algebra} $%
\widetilde{\mathcal{R}}$ \textit{is a way to declare the ''context'' of your
observables.}

\bigskip

\textit{Sometime, however, you can ask the algebra homomorphism }$\tau |$%
\textit{\ to be properly injective:}

\bigskip

\begin{theorem}
Let $T$ be a self-adjoint operator and let $\mathcal{B}$ be an algebra of
borelian functions $b$ on $\mathbb{R}$ with the following property:
\begin{equation*}
\overline{b^{-1}(0)}\supset spec(T)\Longrightarrow b=0
\end{equation*}
It is possible to find a (non empty) $\sigma $-algebra $\mathcal{A}$ of
subsets in $\mathbb{S}$ and an injective algebra homomorphism:
\begin{equation*}
\omega :\mathcal{B\circ }T=\left\{ b\circ T\text{; \ }b\in \mathcal{B}%
\right\} \rightarrow \mathcal{O}_{\mathcal{A}}
\end{equation*}
such that $\tau (\omega (b\circ T))=b\circ T$ for every $b$ in $\mathcal{B}$.
\end{theorem}

\begin{proof}
Let $f$ in $\mathcal{O}$ such that $\tau (f)=T$, the family $\mathcal{B\circ
}f\mathcal{=}\left\{ b\circ f\text{; \ }b\in \mathcal{B}\right\} $ is an
algebra of observable functions contained in $\mathcal{O}_{\mathcal{A}}$
where $\mathcal{A=A}_{f}$ and $\tau |:\mathcal{B\circ }f\mathcal{\rightarrow
B\circ }T$ is a surjective algebra homomorphism.

If $\tau (b\circ f)=0$ then $f^{-1}(\complement b^{-1}0)$ is pseudo-borel
null and also $f^{-1}(\complement \overline{b^{-1}0})$ is pseudo-borel null,
therefore $\overline{b^{-1}0}\supset Im_{e}(f)=spec(T)$ and by hypothesis $%
b\circ f=0$. That is $\tau |$ is an algebra isomorphism, its inverse is the
desired injective homomorphism $\omega $.
\end{proof}

\bigskip

\begin{example}
\begin{enumerate}
\item  If $spec(T)=\mathbb{R}$ the theorem hypothesis is verified for any
algebra $\mathcal{R}$ of continuous functions on $\mathbb{R}$

\item  If $spec(T)$ has a non-empty interior part the theorem hypothesis is
verified for the algebra $\mathcal{R}$ of analytic functions on $\mathbb{R}$

\item  If $spec(T)$ is an infinite subset the theorem hypothesis is verified
for the algebra $\mathcal{R}$ of polynomial functions.
\end{enumerate}
\end{example}

\bigskip

\bigskip

\section{\protect\LARGE Uncertainty relations}

\bigskip

\textit{This section adapts to the space }$\mathbb{S}$, \textit{with some
modifications, several results contained primarily in [CMP], [G], [CGM] and
is inserted with the main goal to show that the uncertainty relations follow
by themself (essentially because the dispersion is given by the norm of a
suitable vector).}

\bigskip

\begin{definition}
We will call \textbf{pre-symplectic form} on $\mathbb{S}$ the smooth 2-form $%
\omega $ defined on each tangent space by $\omega _{\varphi }:T\varphi
\mathbb{S\times }T\varphi \mathbb{S\rightarrow R}$ given by: $\omega
_{\varphi }(X,Y)=\left\langle JX,Y\right\rangle $.
\end{definition}

\bigskip

\textit{The form }$\omega $\textit{\ is bilinear, antisymmetric and closed
but is degenerate on the 1-dimensional subspace of vertical vectors.}

\bigskip

\begin{definition}
Taken two smooth functions $h,l$ on $\mathbb{S}$ we can define the following
two smooth functions $h\circ l,\left\{ h,l\right\} :\mathbb{S\rightarrow R}$
on $\mathbb{S}$ by the following expressions:
\begin{equation*}
\begin{array}{c}
(h\circ l)(\varphi )=\frac{1}{2}\left\langle Grad_{\varphi }h,Grad_{\varphi
}l\right\rangle +h(\varphi )\cdot l(\varphi ) \\
\left\{ h,l\right\} (\varphi )=\omega _{\varphi }(Grad_{\varphi
}h,Grad_{\varphi }l)
\end{array}
\end{equation*}
\end{definition}

\bigskip

\textit{Occasionally we will write }$l^{\circ n}$\textit{\ instead of }$%
\overbrace{l\circ l\circ ...\circ l}^{(n\text{ }times)}$.

\bigskip

\begin{theorem}
For every couple of smooth kaehlerian functions $h,l$ on $\mathbb{S}$ it
holds:

\begin{enumerate}
\item  $h\circ l$ and $\left\{ h,l\right\} $ are smooth kaehlerian functions

\item  $\alpha (h\circ l)=\frac{1}{2}\left[ \alpha (h)\cdot \alpha
(l)+\alpha (l)\cdot \alpha (h)\right] $

\item  $\alpha (\left\{ h,l\right\} )=-i\cdot \left[ \alpha (h)\cdot \alpha
(l)-\alpha (l)\cdot \alpha (h)\right] $
\end{enumerate}
\end{theorem}

\begin{proof}
2. Written $A=\alpha (h)$, $B=\alpha (l)$, $a=\left\langle A\right\rangle
_{\varphi }$ and $b=\left\langle B\right\rangle _{\varphi }$, it is possible
to prove, with some calculations, that: $\left\langle \frac{1}{2}\left[ AB+BA%
\right] \right\rangle _{\varphi }=ab+\frac{1}{2}\left\langle (A-aI)\varphi
,(B-bI)\varphi \right\rangle $.

3. In an analogous way it is possible to prove that:
\begin{equation*}
\left\langle -i\left[ AB-BA\right] \right\rangle _{\varphi }=Im\left\langle
\left\langle (A-aI)\varphi ,(B-bI)\varphi \right\rangle \right\rangle .
\end{equation*}

1. follows from 2. and 3.
\end{proof}

\bigskip

\textit{Obviously} $\mathcal{KS(H)}$ \textit{with the operations }$\left(
\cdot \right) \circ $\textit{\ }$\left( \cdot \right) $\textit{\ and }$%
\left\{ \cdot ,\cdot \right\} $\textit{\ becomes a Jordan-Lie algebra (cfr.
[E]).}

\textit{Each states }$\varphi $\textit{\ in }$\mathbb{S}$\textit{\ is
trivially dispersion free for the ''precise observer'', using the logic} $%
\mathcal{L}$, s\textit{ince his evaluation map }$\widehat{\varphi }:\mathcal{%
L\rightarrow \lbrack }0,1]$, d\textit{efined by }$\widehat{\varphi }(L)=\chi
_{L}(\varphi )$,\textit{\ takes only the values: }$0,1$\textit{\ (the
\textbf{dispersion} is defined by: }$\Delta _{\varphi }(L)=\sqrt{\widehat{%
\varphi }(L)-\widehat{\varphi }(L)^{2}}$ cfr. [J] ch. 6.3\textit{).}

\textit{The situation is different for the ''imprecise observer''; his
''evaluation map'' is }$\overline{\varphi }:\mathcal{L\rightarrow \lbrack }%
0,1]$ \textit{defined by} $\overline{\varphi }(L)=\left\langle \chi
_{L}\right\rangle (\varphi )$ \textit{therefore he gets a dispersion:} $%
\delta _{\varphi }(L)=\sqrt{\overline{\varphi }(L)-\overline{\varphi }(L)^{2}%
}=\sqrt{\left\langle \varepsilon (L)\right\rangle _{\varphi }-\left\langle
\varepsilon (L)\right\rangle _{\varphi }^{2}}$ $=\sqrt{\left\langle \left[
\varepsilon (L)-\left\langle \varepsilon (L)\right\rangle _{\varphi }\cdot I%
\right] ^{2}\right\rangle _{\varphi }}$ \textit{generally positive.}

\bigskip

\begin{definition}
Let $f$ be an essentially bounded observable function and let $\varphi $ be
a state in $\mathbb{S}$, the \textbf{dispersion }of $f$ in $\varphi $ is
given by the following expression:
\begin{equation*}
\delta _{\varphi }(f)=\sqrt{\left\langle \left[ f-\left\langle
f\right\rangle (\varphi )\right] ^{2}\right\rangle (\varphi )}
\end{equation*}
Let $l$ be a smooth function on $\mathbb{S}$ and let $\varphi $ be a state
in $\mathbb{S}$, the \textbf{dispersion }of $l$ in $\varphi $ is given by
the following expression:
\begin{equation*}
\delta _{\varphi }(l)=\sqrt{\left\langle \left[ l-l(\varphi )\right] ^{\circ
2}\right\rangle (\varphi )}
\end{equation*}
Let $T$ be a bounded self-adjoint operator and let $\varphi $ be a state in $%
\mathbb{S}$, the \textbf{dispersion }of $T$ in $\varphi $ is given by the
following expression:
\begin{equation*}
\delta _{\varphi }(T)=\sqrt{\left\langle \left[ T-\left\langle
T\right\rangle _{\varphi }\cdot I\right] ^{2}\right\rangle _{\varphi }}
\end{equation*}
\end{definition}

\bigskip

\begin{theorem}
For every essentially bounded observable function $f$ it holds:
\begin{equation*}
\delta _{\varphi }(f)=\delta _{\varphi }(\left\langle f\right\rangle
)=\delta _{\varphi }(\alpha (\left\langle f\right\rangle ))
\end{equation*}
\end{theorem}

\begin{proof}
It is enough to use the definitions.
\end{proof}

\bigskip

\begin{theorem}
For every essentially bounded observable function $f$ it holds:
\begin{equation*}
\delta _{\varphi }(f)=\frac{1}{\sqrt{2}}\cdot \left\| Grad_{\varphi
}\left\langle f\right\rangle \right\|
\end{equation*}
\end{theorem}

\begin{proof}
Written $A=\alpha (\left\langle f\right\rangle )$ and $a=\left\langle
A\right\rangle _{\varphi }$, it is not difficult to prove that: $\left(
\delta _{\varphi }(f)\right) ^{2}=$ $=\left( \delta _{\varphi }(A)\right)
^{2}=\frac{1}{2}\left\langle (A-aI)\varphi ,(A-aI)\varphi \right\rangle $.
\end{proof}

\bigskip

\begin{theorem}
(\textbf{Heisenberg uncertainty relation}) For every couple of smooth
kaehlerian functions $h,l$ on $\mathbb{S}$ it holds:
\begin{equation*}
\delta _{\varphi }(h)\cdot \delta _{\varphi }(l)\geq \sqrt{\left[ (h\circ
l)(\varphi )-h(\varphi )\cdot l(\varphi )\right] ^{2}+\frac{1}{4}\left[
\left\{ h,l\right\} (\varphi )\right] ^{2}}\geq \frac{1}{2}\left\{
h,l\right\} (\varphi )
\end{equation*}
\end{theorem}

\begin{proof}
By the Cauchy-Schwartz inequality (applied to the sesquilinear scalar
product $\left\langle \left\langle \cdot ,\cdot \right\rangle \right\rangle $%
) we have:
\begin{equation*}
\left( \delta _{\varphi }(h)\cdot \delta _{\varphi }(l)\right) ^{2}=\frac{1}{%
4}\left\| Grad_{\varphi }h\right\| ^{2}\cdot \left\| Grad_{\varphi
}l\right\| ^{2}\geq \frac{1}{4}\left| \left\langle \left\langle
Grad_{\varphi }h,Grad_{\varphi }l\right\rangle \right\rangle \right| ^{2}
\end{equation*}

and the proof is concluded observing that:
\begin{eqnarray*}
(h\circ l)(\varphi )-h(\varphi )\cdot l(\varphi )=\frac{1}{2}Re\left\langle
\left\langle Grad_{\varphi }h,Grad_{\varphi }l\right\rangle \right\rangle \\
\left\{ h,l\right\} (\varphi )=Im\left\langle \left\langle Grad_{\varphi
}h,Grad_{\varphi }l\right\rangle \right\rangle
\end{eqnarray*}
\end{proof}

\bigskip

\bigskip

\section{\protect\LARGE Symmetries and dynamics}

\bigskip

\begin{definition}
A diffeomorphism $\nu :\mathbb{S\rightarrow S}$ will be called an \textbf{%
internal equivalence }if:

\begin{itemize}
\item  it sends every $\mathbb{S}^{1}$-orbit in itself

\item  $\nu \circ \rho _{\theta }=\rho _{\theta }\circ \nu $ for every $\rho
_{\theta }$
\end{itemize}
\end{definition}

\bigskip

\begin{remark}
Every internal equivalence is a measure equivalence.
\end{remark}

\bigskip

\begin{notation}
We will denote by $Aut_{I}(\mathbb{S)=}\left\{ \nu ;\text{ }\nu \text{ is an
internal equivalence}\right\} $ the set of all internal equivalences, it is
a group of transformations of $\mathbb{S}$ containing $\mathbb{S}^{1}\cdot I$%
.
\end{notation}

\bigskip

\begin{theorem}
Let $\nu :\mathbb{S\rightarrow S}$ be a diffeomorphism, the following
properties are equivalent:

\begin{enumerate}
\item  $\nu $\textit{\ is an internal equivalence}

\item  \textit{there exists a differentiable map }$\varsigma :\mathbb{%
S\rightarrow S}^{1}$ \textit{constant on the }$\mathbb{S}^{1}$-orbits
\textit{such that} $\nu (\varphi )=\varsigma (\varphi )\cdot \varphi $

\item  \textit{there exists a differentiable function }$h:\mathbb{%
S\rightarrow R}$ \textit{constant on the }$\mathbb{S}^{1}$-orbits \textit{%
such that} $\nu (\varphi )=e^{-ih(\varphi )}\cdot \varphi $
\end{enumerate}
\end{theorem}

\begin{proof}
3)$\Longrightarrow $2) and 2)$\Longrightarrow $1) are obvious.

1)$\Longrightarrow $3) Since $\varphi $ and $\nu (\varphi )$ are in the same
$\mathbb{S}^{1}$-orbit there exists $\varsigma (\varphi )$ in $\mathbb{S}%
^{1} $ such that $\nu (\varphi )=\varsigma (\varphi )\cdot \varphi $. The
map $\varsigma :\mathbb{S\rightarrow S}^{1}$ so defined verifies $\varsigma
(u\cdot \varphi )=\varsigma (\varphi )$ and is differentiable; therefore the
map $\widehat{\varsigma }:\mathbb{P}_{\mathbb{C}}\mathbb{(}\mathcal{H)}%
\mathbb{\rightarrow S}^{1}$ defined by $\widehat{\varsigma }\left[ \varphi %
\right] =\varsigma (\varphi )$ is well defined and differentiable and admits
a continuous lifting $\widehat{h}:\mathbb{P}_{\mathbb{C}}\mathbb{(}\mathcal{%
H)}\mathbb{\rightarrow R}$ such that $\widehat{\varsigma }\left[ \varphi %
\right] =e^{-i\widehat{h}\left[ \varphi \right] }$ for every $\left[ \varphi %
\right] $ in $\mathbb{P}_{\mathbb{C}}\mathbb{(}\mathcal{H)}$; the function $%
\widehat{h}$ is moreover differentiable since the map $t\mapsto e^{it}$ is a
local diffeomorphism.

The function $h=\widehat{h}\circ \pi $, where $\pi :$ $\mathbb{S\rightarrow P%
}_{\mathbb{C}}\mathbb{(}\mathcal{H)}$ is the natural map, is the function
required.
\end{proof}

\bigskip

\begin{definition}
A \textbf{Hilbert} \textbf{automorphism }of $\mathbb{S}$ is a diffeomorphism
$U:\mathbb{S\rightarrow S}$ with the following properties:

\begin{itemize}
\item  $U\circ \rho _{\theta }=\rho _{\theta }\circ U$ $\ $($U$ respects the
action of $\mathbb{S}^{1}$)

\item  $U$ sends orthogonal vectors in orthogonal vectors ($U$ respects the
orthogonality)

\item  $U(\cos t\cdot \varphi +\sin t\cdot \psi )=\cos t\cdot U(\varphi
)+\sin t\cdot U(\psi )$ for every couple $\varphi $, $\psi $ of orthogonal
vectors in $\mathbb{S}$ ($U$ respects the sovrappositions).
\end{itemize}
\end{definition}

\bigskip

\begin{remark}
A Hilbert automorphism sends $\mathbb{S}^{1}$-orbits in $\mathbb{S}^{1}$%
-orbits and couples of orthogonal $\mathbb{S}^{1}$-orbits in couples of
orthogonal $\mathbb{S}^{1}$-orbits.
\end{remark}

\bigskip

\begin{theorem}
A diffeomorphism $U:\mathbb{S\rightarrow S}$ is a Hilbert automorphism if
and only if there exists a unitary trasformation $\mathcal{U}:\mathcal{%
H\rightarrow H}$ such that $U=\mathcal{U}|_{\mathbb{S}}$.
\end{theorem}

\begin{proof}
($\Leftarrow $) is obvious.

($\Longrightarrow $) The map $\widehat{U}:\mathbb{P}_{\mathbb{C}}\mathbb{(}%
\mathcal{H)\rightarrow }\mathbb{P}_{\mathbb{C}}\mathbb{(}\mathcal{H)}$
defined by $\widehat{U}\left[ \varphi \right] =\left[ U\varphi \right] $ is
a well defined bijective map respecting the antipodality, therefore (cfr.
[U] Thm. 5.1) there exists a unitary or antiunitary ($\mathbb{R}$-linear)
map $\mathcal{U}:\mathcal{H\rightarrow H}$ and a map $\varsigma :\mathbb{%
S\rightarrow S}^{1}$ such that $U\varphi =\varsigma (\varphi )\cdot \mathcal{%
U(\varphi )}$. The map $\varsigma $ is differentiable with $\varsigma
(u\varphi )=\varsigma (\varphi )$ when $\mathcal{U}$ is unitary and $%
\varsigma (u\varphi )=u^{2}\cdot \varsigma (\varphi )$ when $\mathcal{U}$ is
antiunitary.

Since $U$ respects the sovrappositions for every couple $\varphi $, $\psi $
of orthogonal vectors and every ($c$,$s$)$=$($\cos \theta $, $\sin \theta $)
we have:
\begin{equation*}
c\varsigma (c\varphi +s\psi )\mathcal{U\varphi +}s\varsigma (c\varphi +s\psi
)\mathcal{U\psi =}U(c\varphi +s\psi )=c\varsigma (\varphi )\mathcal{U\varphi
+}s\varsigma (\psi )\mathcal{U\psi }
\end{equation*}

therefore: $c\cdot \lbrack \varsigma (\varphi )-\varsigma (c\varphi +s\psi
)]=0$ and $s\cdot \lbrack \varsigma (\psi )-\varsigma (c\varphi +s\psi )]=0$
that is $\varsigma (\varphi )=\varsigma (c\varphi +s\psi )=\varsigma (\psi )$
for every couple $\varphi $, $\psi $ of orthogonal vectors and every ($c$,$%
s)=$($\cos \theta $, $\sin \theta $).

This proves that $\varsigma $ is constant and $\varsigma (u\varphi )$ cannot
be equal to $u^{2}\cdot \varsigma (\varphi )$ for a general $u$; therefore $%
\mathcal{U}$ is unitary.
\end{proof}

\bigskip

\begin{notation}
Let $\mathcal{U}nit\mathcal{(H)=}\left\{ U;\text{ }U\text{ is a Hilbert
automorphism}\right\} $, this is a group of measure equivalences of $\mathbb{%
S}$ with $\mathcal{U}nit\mathcal{(H)\cap }Aut_{I}(\mathbb{S)=S}^{1}\cdot I$.
\end{notation}

\bigskip

\begin{definition}
A diffeomorphism $\Phi :\mathbb{S\rightarrow S}$ is called a \textbf{%
semi-automorphism of} $\mathbb{S}$ if :

\begin{enumerate}
\item  $\Phi \circ \rho _{\theta }=\rho _{\theta }\circ \Phi $ for every $%
\theta $

\item  $\Phi $ sends couples of orthogonal $\mathbb{S}^{1}$-orbits in
couples of orthogonal $\mathbb{S}^{1}$-orbits.
\end{enumerate}
\end{definition}

\bigskip

\begin{notation}
The set $\Theta Aut(\mathbb{S)=}\left\{ \Phi ;\text{ }\Phi \text{ is a
semi-automorphism of\textbf{\ }}\mathbb{S}\right\} $ is a group of
transformations of $\mathbb{S}$. Every Hilbert automorphism and every
internal equivalence is a semi-automorphism.
\end{notation}

\bigskip

\begin{theorem}
There exists a unique group homomorphism $\sigma :\Theta Aut(\mathbb{%
S)\rightarrow }\left\{ 1,-1\right\} $ such that $\sigma (\Phi )=1$ if and
only if $\Phi $ is expressable as $\Phi =\Psi ^{2}\circ \Gamma ^{2}$ with $%
\Psi $ and $\Gamma $ in $\Theta Aut(\mathbb{S)}$.
\end{theorem}

\begin{proof}
Let $\Phi :\mathbb{S\rightarrow S}$ be a differentiable map, its horizontal
differential is the linear map $\Phi _{\ast }^{\mathcal{H}or}:\mathcal{H}%
or_{\varphi }\rightarrow \mathcal{H}or_{\Phi (\varphi )}$ defined by: $\Phi
_{\ast \varphi }^{\mathcal{H}or}(X)=\mathcal{H}or_{\Phi (\varphi )}(\Phi
_{\ast \varphi }(X))$.

Note that if $\Phi _{\ast \varphi }$ is injective then $\Phi _{\ast \varphi
}^{\mathcal{H}or}$ is not zero and if $h:\mathbb{S\rightarrow R}$ is a
differentiable function then $(e^{-ih}\cdot \Phi )_{\ast \varphi }^{\mathcal{%
H}or}=e^{-ih(\varphi )}\cdot \Phi _{\ast \varphi }^{\mathcal{H}or}$.

If $\Phi $ is a semi-automorphism the map $\widehat{\Phi }:\mathbb{P}_{%
\mathbb{C}}(\mathcal{H)\rightarrow }\mathbb{P}_{\mathbb{C}}(\mathcal{H)}$
given by $\widehat{\Phi }(\left[ \varphi \right] =\left[ \Phi (\varphi )%
\right] $ is well defined, bijective and preserves the antipodality
relation. Therefore (cfr. [U] Thm. 5.1) there exists a unitary or
antiunitary transformation $U:\mathcal{H\rightarrow H}$ such that $\widehat{U%
}=\widehat{\Phi }$; that is there is a map $\varsigma :\mathbb{S\rightarrow S%
}^{1}$ verifying \ the equality: $\Phi (\varphi )=\varsigma (\varphi )\cdot
U\varphi $. The map $\varsigma :\mathbb{S\rightarrow S}^{1}$ is necessarily
differentiable and, since $\mathbb{S}$ is simply connected, there exists a
continuous (and then differentiable) function $h:\mathbb{S\rightarrow R}$,
constant on the $\mathbb{S}^{1}$-orbits when $U$ is unitary, such that $\Phi
=e^{-ih}\cdot U.$

It's easy to check that a unitary or antiunitary transformation $\Phi :%
\mathbb{S\rightarrow S}$ verifies the equality:
\begin{equation*}
\Phi _{\ast }^{\mathcal{H}or}\circ J_{\varphi }=\sigma (\Phi )\cdot J_{\Phi
(\varphi )}\circ \Phi _{\ast }^{\mathcal{H}or}
\end{equation*}

with $\sigma (\Phi )=1$ if $\Phi $ is unitary or with $\sigma (\Phi )=-1$ if
$\Phi $ is antiunitary.

For a general semi-automorphism $\Phi =e^{-ih}\cdot U$ if we take the sign $%
\sigma (\Phi )=\sigma (U)$ the equality above is still verified (the sign $%
\sigma (\Phi )$ is well defined since $\Phi _{\ast }^{\mathcal{H}or}$ cannot
be zero).

Using the equality it's easy to check that $\sigma $ is a group homomorphism
and therefore $\sigma \left( \Psi ^{2}\circ \Gamma ^{2}\right) =1$ for every
$\Psi $ and $\Gamma $ in $\Theta Aut(\mathbb{S)}$.

Conversely if $\sigma (\Phi )=\sigma (e^{-ih}\cdot U)=1$ then $U$ is a
unitary transformation expressable as $U=e^{-iA}$ for a self-adjoint
endomorphism $A$ on $\mathcal{H}$, therefore $\Phi =\Psi ^{2}\circ \Gamma
^{2}$ with $\Psi (\varphi )=e^{-\frac{i}{2}A}\varphi $ and $\Gamma (\varphi
)=e^{-\frac{i}{2}h(\varphi )}\cdot \varphi $.

The unicity of the homomorphism $\sigma $ follows from the caracterization
of the elements with $\sigma (\Phi )=1$.
\end{proof}

\bigskip

\begin{remark}
\bigskip If $\Phi $, $\Psi $ are in $\Theta Aut(\mathbb{S)}$ with $\Psi
=e^{-ih}\cdot \Phi $ then $\sigma (\Psi )=\sigma (\Phi )$.
\end{remark}

\bigskip

\begin{definition}
A semi-automorphism $\Phi $ of\textbf{\ }$\mathbb{S}$ will be called an
\textbf{automorphism of }$\mathbb{S}$ if $\sigma (\Phi )=1$.
\end{definition}

\bigskip

\begin{remark}
The set $Aut(\mathbb{S)=}\left\{ \Phi ;\text{ }\Phi \text{ is an
automorphism of\textbf{\ }}\mathbb{S}\right\} $ is a normal subgroup of $%
\Theta Aut(\mathbb{S)}$ containing $\mathcal{U}nit\mathcal{(H)}$ and $%
Aut_{I}(\mathbb{S)}$ . For every $\Psi $ in $\Theta Aut(\mathbb{S)}$ the
element $\Psi ^{2}$ is in $Aut(\mathbb{S)}$; if \ $\Phi _{\cdot }:\mathbb{%
R\rightarrow }\Theta Aut(\mathbb{S)}$ is a one-parameter group then every $%
\Phi _{t}$ is in $Aut(\mathbb{S)}$, infact $\Phi _{t}=\left( \Phi
_{t/2}\right) ^{2}$ for every $t$ in $\mathbb{R}$.
\end{remark}

\bigskip

\begin{theorem}
A diffeomorphism $\Phi :\mathbb{S\rightarrow S}$ is an automorphism of $%
\mathbb{S}$ if and only if it can be expressed as the composition $\Phi
=U\circ \nu $ of a Hilbert automorphism $U$ and an internal equivalence $\nu
$.
\end{theorem}

\begin{proof}
From the proof of the preceding theorem we know that, when $\sigma (\Phi )=1$%
, then $\Phi =e^{-ih}\cdot U$ with $U$ unitary and $h$ constant on the $%
\mathbb{S}^{1}$-orbits. Therefore taken $\nu :\mathbb{S\rightarrow S}$
defined by: $\nu (\varphi )=e^{-ih(\varphi )}\cdot \varphi $, the map $\nu $
is an internal equivalence and $\Phi =U\circ \nu $. Conversely every
composition $\Phi =U\circ \nu $ of a Hilbert automorphism $U$ and of an
internal equivalence $\nu $ is a semi-automorphism that can be written as $%
\Phi =e^{-ih}\cdot U$. Therefore $\sigma (\Phi )=\sigma (U)=1.$
\end{proof}

\bigskip

\begin{corollary}
A diffeomorphism $\Phi :\mathbb{S\rightarrow S}$ is an automorphism of $%
\mathbb{S}$ if and only if there exists a self-adjoint operator $A$ of $%
\mathcal{H}$ $\ $(or, equivalently, a kaehlerian function $l$ with $\alpha
(l)=A$) and a differentiable function $h:\mathbb{S\rightarrow R}$ \textit{%
constant on the }$\mathbb{S}^{1}$-orbits such that:
\begin{equation*}
\Phi (\varphi )=e^{-ih(\varphi )}\cdot e^{-iA}\varphi =e^{-ih(\varphi
)}\cdot e^{-i\alpha (l)}\varphi
\end{equation*}
\end{corollary}

\begin{proof}
It follows from the definitions.
\end{proof}

\bigskip

\begin{remark}
\begin{itemize}
\item  $Aut(\mathbb{S)}$ is the smallest group of diffeomorphism of $\mathbb{%
S}$ containing $\mathcal{U}nit\mathcal{(H)}$ and $Aut_{I}(\mathbb{S)}$

\item  $Aut_{I}(\mathbb{S)}$ is a normal subgroup of $Aut(\mathbb{S)}$

\item  if $\Phi =U\circ \nu $ and $\Psi =V\circ \varpi $ with $U$ and $V$
unitary and $\nu $ and $\varpi $ internal equivalence then $\Phi \circ \Psi
=U\circ V\circ \rho $ for a suitable internal equivalence $\rho $

\item  every automorphism is a measure equivalence

\item  \textit{an automorphism }$\Phi $\textit{\ sends every proposition }$L$%
\textit{\ in a proposition }$\Phi (L)$

\item  \textit{if }$\Phi $\textit{\ is an automorphism then for every
observable }$f$\textit{\ the function}$\ f\circ \Phi ^{-1}$ is also an
observable

\item  an automorphism respects the phase distance of the $\mathbb{S}^{1}$%
-orbits
\end{itemize}
\end{remark}

\bigskip

\begin{remark}
On $Aut(\mathbb{S)}$ we will consider the topology induced by $\mathbb{S}^{%
\mathbb{S}}$ (the topology of ''pointwise convergence'').
\end{remark}

\bigskip

\begin{definition}
A \textbf{dynamic} in $\mathbb{S}$ is a continuous 1-parameter group of
automorphisms of $\mathbb{S}$ .
\end{definition}

\bigskip

\begin{theorem}
A differentiable map $\Phi _{\cdot }:\mathbb{R}\times \mathbb{S\rightarrow S}
$ is a dynamic in $\mathbb{S}$ if and only if there exists a self-adjoint
operator $A$ on $\mathcal{H}$ and a differentiable function $h:\mathbb{%
S\rightarrow R}$ \textit{constant on the }$\mathbb{S}^{1}$-orbits such that:
\begin{equation*}
\Phi _{t}(\varphi )=e^{-i\int_{0}^{t}h(e^{-irA}\varphi )\cdot dr}\cdot
e^{-itA}\varphi
\end{equation*}
\end{theorem}

\begin{proof}
($\Longleftarrow $) Obvious.

($\Longrightarrow $) The family $\left\{ \widehat{\Phi _{t}}\right\} _{t\in
\mathbb{R}}$ is a differentiable 1-parameter group of symmetries of $\mathbb{%
P}_{\mathbb{C}}\mathbb{(}\mathcal{H)}$ therefore (cfr. [Ba]) there exists a
self-adjoint operator $A$ on $\mathcal{H}$ such that $\widehat{\Phi _{t}}=%
\widehat{U^{t}}$ for every $t$ (where $U^{t}=e^{-itA}$).

Therefore there exists a map $\varsigma :\mathbb{R}\times \mathbb{%
S\rightarrow S}^{1}$ such that $\Phi _{t}(\varphi )=\varsigma (t,\varphi
)\cdot U^{t}\varphi $ for every $\varphi $ in $\mathbb{S}$ and every $t$ in $%
\mathbb{R}$; the map $\varsigma $ is constant on the $\mathbb{S}^{1}$-orbits
and is necessarily differentiable.

We can find a continuous lifting $\widetilde{\varsigma }:\mathbb{R}\times
\mathbb{S\rightarrow R}$ with respect to the space $\mathbb{S}$ and the
covering map $\varepsilon :\mathbb{R\rightarrow S}^{1}$ (given by $%
\varepsilon (r)=e^{-ir}$) such that $\widetilde{\varsigma }(0,\varphi )=0$
for every $\varphi $; since $\varepsilon $ is a local diffeomorphism the
lifting $\widetilde{\varsigma }$ is differentiable.

Using $\Phi _{t+s}=\Phi _{t}\circ \Phi _{s}$ we have:
\begin{equation*}
\varsigma (t+s,\varphi )=\varsigma (t,U^{s}\varphi )\cdot \varsigma
(s,\varphi )=\varsigma (s,U^{t}\varphi )\cdot \varsigma (t,\varphi )
\end{equation*}

therefore:
\begin{equation*}
\widetilde{\varsigma }(t+s,\varphi )=\widetilde{\varsigma }(t,U^{s}\varphi )+%
\widetilde{\varsigma }(s,\varphi )+2\pi \cdot k(t,s,\varphi )
\end{equation*}

where $k(t,s,\varphi )$ is an integer. Since $k(\cdot ,\cdot ,\cdot ):%
\mathbb{R\times R}\times \mathbb{S\rightarrow Z}$ is continuous, the
function $k$ must be constant and then equal to $0$.

In the same way it is possible to prove that $\widetilde{\varsigma }$ is
constant on the $\mathbb{S}^{1}$-orbits.

Therefore for the differentiable function $\eta (\cdot ,\cdot ):\mathbb{R}%
\times \mathbb{S\rightarrow R}$ defined by $\eta (t,\varphi )=\frac{\partial
\widetilde{\varsigma }}{\partial t}(t,\varphi )$ we get: $\int_{s}^{t+s}\eta
(r,\varphi )\cdot dr=\widetilde{\varsigma }(t,U^{s}\varphi )$ and $\eta
(s,\varphi )=\eta (0,U^{s}\varphi )$; then the function $h(\cdot ):\mathbb{%
S\rightarrow R}$ defined by $h(\varphi )=\eta (0,\varphi )$ is
differentiable, constant on the $\mathbb{S}^{1}$-orbits and:
\begin{equation*}
\varsigma (t,\varphi )=e^{-i\int_{0}^{t}h(U^{r}\varphi )\cdot dr}
\end{equation*}
\end{proof}

\bigskip

\begin{notation}
Let $l$ be a kaehlerian function on $\mathbb{S}$ and let $h$ be a
differentiable function \textit{constant on the }$\mathbb{S}^{1}$-orbits we
will denote by $\Phi _{l,h;\cdot }$ the smooth flow (the \textbf{Hamiltonian
flow defined by }$l$ \textbf{and} $h$\textbf{) }given by:
\begin{equation*}
\Phi _{l,h;t}(\varphi )=e^{-i\int_{0}^{t}h(e^{-ir\alpha (l)}\varphi )\cdot
dr}\cdot e^{-it\alpha (l)}\varphi
\end{equation*}
When $l$ is a smooth kaehlerian function we will denote by $X_{l,h}$ the
smooth vector field (the \textbf{Hamiltonian field defined by }$l$ \textbf{%
and} $h$\textbf{) }given by:
\begin{equation*}
X_{l,h}|_{\varphi }=-J_{\varphi }Grad_{\varphi }l-[l(\varphi )+h(\varphi
)]\cdot J\varphi =-i\cdot \alpha (l)\varphi -i\cdot h(\varphi )\cdot \varphi
\end{equation*}
\end{notation}

\bigskip

\begin{remark}
It's easy to check that ($l_{1},h_{1}$) and ($l_{2},h_{2}$) define the same
hamiltonian field if and only if there exists a real constant $c$ such that:
$l_{2}=l_{1}+c$ and $h_{2}=h_{1}-c$. A little bit more complicated is to
prove $\Phi _{l_{1},h_{1};\cdot }$ and $\Phi _{l_{2},h_{2};\cdot }$ define
the same hamiltonian flow under the same condition.
\end{remark}

\bigskip

\begin{theorem}
Let $l$ be a smooth kaehlerian function on $\mathbb{S}$ and let $h$ be a
differentiable function \textit{constant on the }$\mathbb{S}^{1}$-orbits,
the field $X_{l,h}$ is a complete field defining the Hamiltonian flow $\Phi
_{l,h;\cdot }$.
\end{theorem}

\begin{proof}
In the Hilbert space $\mathcal{H}$, written $\Phi _{t}=\Phi _{l,h;t}$, we
have for every vector $\varphi $ in $\mathbb{S}$ : $\dot{\Phi}%
_{t_{0}}(\varphi )=\frac{d}{dt}|_{t=t_{0}}(\Phi _{t}(\varphi ))=-ih(\Phi
_{t_{0}}(\varphi ))\cdot \Phi _{t_{0}}(\varphi )-i\alpha (l)\Phi
_{t_{0}}(\varphi )=X_{l,h}|_{\Phi _{t_{0}}(\varphi )}$, therefore each curve
$t\mapsto \Phi _{l,h;t}(\varphi )$ is an integral curve for the field $%
X_{l,h}$.
\end{proof}

\bigskip

\begin{remark}
Let $l$ be a smooth kaehlerian function on $\mathbb{S}$ with $\alpha (l)=A$
and let $h$ be a differentiable function \textit{constant on the }$\mathbb{S}%
^{1}$-orbits, the evolution $\left\{ \psi _{t}\right\} $ of a state $\psi
_{0}$ follows the (non generally linear) differential equation:
\begin{equation*}
\dot{\psi}=-iA\psi -ih(\psi )\cdot \psi =-J_{\psi }Grad_{\psi }l-[l(\psi
)+h(\psi )]\cdot J\psi
\end{equation*}
\end{remark}

\bigskip

\bigskip

\section{\protect\LARGE The imprecise observer}

\bigskip

\textit{In this section new structures are defined using the heuristic
hypothesis that the system considered previously is described now by an
observer intrinsically unable to distinguish between states in the same} $%
\mathbb{S}^{1}$\textit{-orbit (states differing only by ''the phase'').}

\textit{To better understand the situation let's describe in more detail
what could be a \textbf{precise observer}.}

\textit{Imagine an observer with a very efficient and completely automatized
laboratory: all the precise observer has to do is to give to his main
computer a program prescribing what is to be done and to press the ENTER key!%
}

\textit{In particular, given\ a physical system, for each state }$\varphi $%
\textit{\ of the system the precise observer has a program }$SP(\varphi )$%
\textit{\ describing to the computer how to prepare the system exactly in
the state }$\varphi $\textit{. Among these programs there are some act to
prepare the state }$\rho _{\theta }\varphi $\textit{\ from the state }$%
\varphi $\textit{\ or to prepare a specific assigned state in the }$\mathbb{S%
}^{1}$\textit{-orbit} $\left[ \varphi \right] =\left\{ \rho _{\theta
}\varphi :0\leq \theta <2\pi \right\} $.

\textit{Moreover for each observable }$f$\textit{\ measurable on the system
the precise observer has a program }$OP(f)$\textit{\ describing to the
computer how to prepare or activate the corresponding measuring apparatus.}

\textit{Given a state }$\varphi $\textit{\ and an observable }$f$\textit{\
the precise observer has the great satisfaction to check that every time the
measuring procedure }$OP(f)$\textit{\ is executed on the system prepared by }%
$SP(\varphi )$\textit{\ the measure displayed is always the same real number
}$f(\varphi )$\textit{: for the precise observer the measuring process is
completely deterministic.}

\textit{Among his observables there are some phase distance observables
allowing the precise observer to distinguish between states in the same} $%
\mathbb{S}^{1}$\textit{-orbit.}

\textit{The precise observer, in particular, concludes that, almost all the
times, the phase is decisive in the measuring process: he makes the
experience that a small changement of the phase may change completely the
measurement outcome.}

\bigskip

\textit{Let's consider now another observer studying the same physical
system but with a poorer ability (let's call him the \textbf{imprecise
observer}). The imprecise observer has all the programs of the precise
observer but the procedures of his laboratory and his computer are under a
curse: they can never reach the precision necessary to distinguish between
two different states in the same} $\mathbb{S}^{1}$\textit{-orbit.}

\textit{When the imprecise observer runs the procedure }$SP(\varphi )$%
\textit{\ his laboratory can be precise enough to prepare a state in the} $%
\mathbb{S}^{1}$\textit{-orbit }$\left[ \varphi \right] =\left\{ \rho
_{\theta }\varphi :0\leq \theta <2\pi \right\} $\textit{\ but he does not
know which state in the} $\mathbb{S}^{1}$\textit{-orbit is the outcome; the
imprecise observer cannot avoid to the state produced to be completely
random in its} $\mathbb{S}^{1}$\textit{-orbit.}

\textit{When the imprecise observer runs the measure procedure }$OP(f)$%
\textit{\ after }$SP(\varphi )$\textit{\ he can get anyone of the values }$%
\left\{ f(\psi );\psi \in \left[ \varphi \right] \right\} $\textit{. After a
large number of trials the imprecise observer gets his outcomes distributed
on the real line and, in the end, all he gets is representable by the
numbers }$\pi (\varphi ,f,B)=\mu (\left[ \varphi \right] \cap f^{-1}(B))$%
\textit{\ expressing the probability that the outcome falls in a general
borel subset }$B$\textit{\ of }$\mathbb{R}$\textit{.}

\textit{The precise observer could let us know what's wrong with the
imprecise observer; he could say that all the science and technology of the
imprecise observer \textbf{ignore} how to deal with the phases: his computer
simply skips over the program's lines prescribing some action able to define
the phase of a state of the system.}

\textit{In the imprecise observer's laboratory the phase of the system
simply comes from the past evolution of the system, therefore when a phase
is involved in a measure (and this happens almost all the times) the
imprecise observer gets the consequent result but without any control: from
his viewpoint the results come by chance.}

\textit{The precise observer could tell us what happens, for example, in the
idealized experiment of a lamp producing isolated photons in a state
assigned up to the phase and directed toward a half-silvered mirror
reflecting exactly half of them:}

\qquad ''\textit{I am able to control the phase of the photon's wave
function and I checked that the passage of the photon through the mirror is
strictly deterministic depending on this photon's phase.}

\qquad \textit{The imprecise observer instead is completely anaware of all
these phases coming and going and in any case is not able to measure or to
use them; therefore his photons are produced with a phase decided randomly
or, better, by the history of lamp+photon system and when the photon
interacts with the mirror it is this phase (its evolution) that decides what
happens next: the photon passes or does not pass the mirror''.}

\bigskip

\textit{If the imprecise observer is left unaware of his ignorance then he
comes to the drastic decision to consider his measuring process
intrinsically statistic. }

\textit{This implies a series of consequences. The space} $\mathbb{S}$ $%
\leftrightarrow \left\{ SP(\varphi )\right\} $\textit{keeps for him only the
meaning of set of all theorically possible preparations of the system and }$%
\mathcal{O}\leftrightarrow \left\{ OP(f)\right\} $\textit{keeps only the
meaning of set of all theorically possible measuring apparatuses. All his
experimental knowledge reduces to a probability map:}
\begin{equation*}
\pi :\mathbb{S\times }\mathcal{O\times B(}\mathbb{R)\rightarrow }\left[ 0,1%
\right]
\end{equation*}
\textit{remembering the probability }$\pi (\varphi ,f,B)$\textit{\ that the
outcome of the measuring apparatus associated to }$OP(f)$\textit{\ on the
system prepared following }$SP(\varphi )$\textit{\ falls in the borel set }$%
B $\textit{.}

\textit{But now why he should consider different two preparations }$\varphi
_{1\text{ }}$\textit{and }$\varphi _{2\text{ }}$\textit{if: }
\begin{equation*}
\pi (\varphi _{1},f,B)=\pi (\varphi _{2},f,B)
\end{equation*}

\textit{for every apparatus }$f$\textit{\ and every borelian subset }$B$%
\textit{\ ? And, dually, why he should consider different two apparatuses }$%
f_{1\text{ }}$\textit{and }$f_{2\text{ }}$\textit{if: }
\begin{equation*}
\pi (\varphi ,f_{1},B)=\pi (\varphi ,f_{2},B)
\end{equation*}

\textit{for every preparation }$\varphi $\textit{\ and every borelian subset
}$B$\textit{\ ?}

\textit{The ignorance of the phase generates therefore for the imprecise
observer an }\textbf{equivalence relation} $\mathcal{R}_{\mathbb{S}}$
\textit{among the states in} $\mathbb{S}$ \textit{and an }\textbf{%
equivalence relation} $\mathcal{R}_{\mathcal{O}}$ \textit{among the
observables in} $\mathcal{O}$:\textit{\ for the imprecise observer the real
''states'' he can distinguish through his experiments are the equivalence
classes of} $\mathcal{R}_{\mathbb{S}}$ \textit{in} $\mathbb{S}$ \textit{and
his ''\textbf{state space}'' is the quotient space} $\widehat{\mathbb{S}}=%
\mathbb{S}/\mathcal{R}_{\mathbb{S}}$. \textit{Analogously his ''\textbf{%
observable space}'' is} $\widehat{\mathcal{O}}=\mathcal{O}/\mathcal{R}_{%
\mathcal{O}}$.

\textit{Over these objects is still well defined the probability map}:
\begin{equation*}
\widehat{\pi }:\widehat{\mathbb{S}}\mathbb{\times }\widehat{\mathcal{O}}%
\mathcal{\times B(}\mathbb{R)\rightarrow }\left[ 0,1\right]
\end{equation*}

\textit{given by: }$\widehat{\pi }(\left[ \varphi \right] ,\left[ f\right]
,B)=\pi (\varphi ,f,B)$\textit{\ and now distinguishes the ''states''
through the ''observables'' and conversely.}

\bigskip

\textit{In the following we will show that the physical system experimented
by the imprecise observer with its states, observables, probabilities,
symmetries and dynamics is naturally isomorphic to the usual quantum system}
\textit{and this gives a rational basis to our claim that \textbf{the
imprecise observer, \ because of his ignorance of the phases, experiments
the physical system }}$\mathbb{S}$ \textit{\textbf{as a quantum system}.}

\bigskip

\begin{notation}
Since the probability $\pi (\varphi ,f,B)=\mu _{\left[ \varphi \right]
}(f^{-1}(B)\cap \left[ \varphi \right] )=\left\langle E_{B}^{\tau
(f)}\right\rangle _{\varphi }$ depends on the class $\left[ \varphi \right] $
and not from its representative $\varphi $ we have:
\begin{equation*}
\varphi \mathcal{R}_{\mathbb{S}}(\rho _{\theta }\varphi )\text{ for every }%
\theta \text{ in }\mathbb{R}
\end{equation*}
therefore the map denoted by $\mathbb{\chi }:\mathbb{P}_{\mathbb{C}}(%
\mathcal{H}\mathbb{)\rightarrow }\widehat{\mathbb{S}}$ and defined by $%
\mathbb{\chi }(\left[ \varphi \right] _{\mathbb{P}})=\left[ \varphi \right]
_{\mathcal{R}_{\mathbb{S}}}$ is well defined. If $f_{1}\mathcal{R}_{\mathcal{%
O}}f_{2}$ then $\left\langle E_{B}^{\tau (f_{1})}\right\rangle _{\varphi
}=\left\langle E_{B}^{\tau (f_{2})}\right\rangle _{\varphi }$ for every $%
\varphi $ in $\mathbb{S}$ and every $B$ in $\mathcal{B(}\mathbb{R}\mathcal{)}
$ then $\tau (f_{1})=\tau (f_{2})$ and therefore the map denoted by $\eta :%
\widehat{\mathcal{O}}\rightarrow SA(\mathcal{H}\mathbb{)}$ and defined by $%
\eta (\left[ f\right] _{\mathcal{R}_{\mathcal{O}}})=\tau (f)$ is well
defined.
\end{notation}

\bigskip

\begin{theorem}
The map $\mathbb{\chi }:\mathbb{P}_{\mathbb{C}}(\mathcal{H}\mathbb{%
)\rightarrow }\widehat{\mathbb{S}}$ \ is bijective.
\end{theorem}

\begin{proof}
The map is clearly surjective; if $\varphi ,\psi $ are in $\mathbb{S}$ with $%
\left\langle E_{B}^{T}\right\rangle _{\varphi }=\left\langle
E_{B}^{T}\right\rangle _{\psi }$ for every self-adjiont operator $T$ and
every borel subset $B$ of $\mathbb{R}$ then $\varphi $ is an element of a
complex linear subspace $F$ of $\mathcal{H}$ if and only if $\psi $ is in $F$%
. Therefore $\psi =e^{i\theta }\cdot \varphi $.
\end{proof}

\bigskip

\begin{remark}
In short the equivalence classes of $\mathcal{R}_{\mathbb{S}}$ are \textit{%
the} $\mathbb{S}^{1}$\textit{-orbits. If we consider on }$\widehat{\mathbb{S}%
}$ \ the quotient topology induced by $\mathbb{S}$ (this is topology that
the precise observer assignes to $\widehat{\mathbb{S}}$ ) the map $\chi $ is
a homeomorphism.
\end{remark}

\bigskip

\textit{The map} $\beta :\mathbb{S\rightarrow P}_{\mathbb{C}}(\mathcal{H}%
\mathbb{)}$ d\textit{efined by }$\beta (\varphi )=\left[ \varphi \right] $%
\textit{\ is a submersion where each linear map }$\beta _{\ast \varphi }|:%
\mathcal{H}or_{\varphi }\rightarrow T_{\left[ \varphi \right] }\mathbb{P}_{%
\mathbb{C}}(\mathcal{H}\mathbb{)}$\textit{\ is a topological isomorphism. It
is not difficult to check that if we consider on each }$T_{\left[ \varphi %
\right] }\mathbb{P}_{\mathbb{C}}(\mathcal{H}\mathbb{)}$ \textit{the scalar
product moved from} $\mathcal{H}or_{\varphi }$ \textit{by} $\beta _{\ast
\varphi }|$ \textit{we get on }$\mathbb{P}_{\mathbb{C}}(\mathcal{H}\mathbb{)}
$ \textit{a well defined metric tensor coincident with the
Kaehler-Fubini-Study metric tensor }$g_{\nu }$\textit{\ with }$\nu =1$%
\textit{\ considered in [CMP].}

\textit{Each tangent space} $T_{\left[ \varphi \right] }\mathbb{P}_{\mathbb{C%
}}(\mathcal{H}\mathbb{)}$ \textit{inherits from} $\mathcal{H}or_{\varphi }$
\textit{an isometric endomorphism} $J_{\left[ \varphi \right] }:T_{\left[
\varphi \right] }\mathbb{P}_{\mathbb{C}}(\mathcal{H}\mathbb{)\rightarrow }T_{%
\left[ \varphi \right] }\mathbb{P}_{\mathbb{C}}(\mathcal{H}\mathbb{)}$
\textit{given by} $\beta _{\ast \varphi }|\circ J_{\varphi }\circ (\beta
_{\ast \varphi }|)^{-1}$ \textit{(not depending on the representative} $%
\varphi $ \textit{chosen) defining on} $\mathbb{P}_{\mathbb{C}}(\mathcal{H}%
\mathbb{)}$ \textit{an (integrable) almost-complex structure.}

\bigskip

\begin{theorem}
The map $\eta :\widehat{\mathcal{O}}\rightarrow SA(\mathcal{H}\mathbb{)}$ is
bijective.
\end{theorem}

\begin{proof}
Every self-adjiont operator comes from an observable function therefore the
map $\eta $ is surjective. Let $\left[ f_{1}\right] _{\mathcal{R}_{\mathcal{O%
}}}$ , $\left[ f_{2}\right] _{\mathcal{R}_{\mathcal{O}}}$ be in $\widehat{%
\mathcal{O}}$ with $\tau (f_{1})=\tau (f_{2})$, then we have the following
equalities: $\pi (\varphi ,f_{1},B)=\left\langle E_{B}^{\tau
(f_{1})}\right\rangle _{\varphi }=\left\langle E_{B}^{\tau
(f_{2})}\right\rangle _{\varphi }=\pi (\varphi ,f_{2},B)$ for every $\varphi
$ and every borel subset $B$ of $\mathbb{R}$; that is $\left[ f_{1}\right] _{%
\mathcal{R}_{\mathcal{O}}}=\left[ f_{2}\right] _{\mathcal{R}_{\mathcal{O}}}$.
\end{proof}

\bigskip

\textit{If we identify }$\mathbb{P}_{\mathbb{C}}(\mathcal{H}\mathbb{)}$
\textit{with} $\widehat{\mathbb{S}}$ (\textit{through }$\mathbb{\chi }$)%
\textit{\ and} $\widehat{\mathcal{O}}$ \textit{with }$SA(\mathcal{H}\mathbb{)%
}$ (\textit{through }$\eta $)\textit{\ the probability map} $\widehat{\pi }:%
\widehat{\mathbb{S}}\mathbb{\times }\widehat{\mathcal{O}}\mathcal{\times B(}%
\mathbb{R)\rightarrow }\left[ 0,1\right] $ \textit{becomes the usual quantum
probability map }$p:$ $\mathbb{P}_{\mathbb{C}}(\mathcal{H}\mathbb{)}$ $%
\times SA(\mathcal{H}\mathbb{)\times }\mathcal{B(}\mathbb{R)\rightarrow }%
\left[ 0,1\right] $ \textit{given by} $p(\left[ \varphi \right]
,A,B)=\left\langle E_{B}^{A}\right\rangle _{\varphi }$, \textit{infact}:
\begin{equation*}
\widehat{\pi }(\mathbb{\chi }\left[ \varphi \right] _{\mathbb{P}},\left[ f%
\right] ,B)=\mu _{\left[ \varphi \right] }(\left[ \varphi \right] \cap
f^{-1}(B))\mathit{\ }=\left\langle E_{B}^{\tau (f)}\right\rangle _{\varphi
}=p(\left[ \varphi \right] _{\mathbb{P}},\eta (\left[ f\right] ),B)
\end{equation*}
\textit{The couple} $(\chi ^{-1},\eta )$ \textit{is an isomorphism between} $%
(\widehat{\mathbb{S}},\widehat{\mathcal{O}},\widehat{\pi })$ \textit{and} $(%
\mathbb{P}_{\mathbb{C}}(\mathcal{H}\mathbb{)},SA(\mathcal{H}\mathbb{)},p)$.

\bigskip

\textit{For the precise observer the symmetries of the physical system are
the automorphisms of }$\mathbb{S}$, \textit{that is} \textit{the elements of
the group} $Aut(\mathbb{S)}$. \textit{Let's make clear what is a symmetry
for the imprecise observer:}

\bigskip

\begin{definition}
A \textbf{semi-symmetry }\ for $(\widehat{\mathbb{S}},\widehat{\mathcal{O}},%
\widehat{\pi })$ is a couple ($\Lambda $, $\Omega $) of a bijective map $%
\Lambda :\widehat{\mathbb{S}}$ $\rightarrow \widehat{\mathbb{S}}$ \ and a
bijective map $\Omega :\widehat{\mathcal{O}}$ $\rightarrow \widehat{\mathcal{%
O}}$ $\ $such that:
\begin{equation*}
\widehat{\pi }(\Lambda \left[ \varphi \right] ,\Omega \left[ f\right] ,B)=%
\widehat{\pi }(\left[ \varphi \right] ,\left[ f\right] ,B)
\end{equation*}
for every $\left[ \varphi \right] $ in $\widehat{\mathbb{S}}$, every $\left[
f\right] $ in $\widehat{\mathcal{O}}$ \ and every borel subset $B$ of $%
\mathbb{R}$.
\end{definition}

\bigskip

\begin{notation}
We will denote by $\Theta Sym(\widehat{\mathbb{S}}$ , $\widehat{\mathcal{O}}%
) $ the set of all the semi-symmetries\textbf{\ }of $(\widehat{\mathbb{S}},%
\widehat{\mathcal{O}},\widehat{\pi })$; it is a group with respect to
composition of couples of bijective maps.
\end{notation}

\bigskip

\begin{theorem}
For every semi-symmetry $(\Lambda ,\Omega )$ there exists a unitary or
antiunitary transformation $U:\mathbb{S\rightarrow S}$ such that:

\begin{itemize}
\item  $\Lambda \left[ \varphi \right] =\widehat{U}\left[ \varphi \right]
=[U\varphi ]$ for every $[\varphi ]$ in $\widehat{\mathbb{S}}$

\item  $\Omega \lbrack f]=\widehat{U^{\ast }}[f]=[f\circ U^{-1}]$ for every $%
[f]$ in $\widehat{\mathcal{O}}$
\end{itemize}
\end{theorem}

\begin{proof}
Since $(\widehat{\mathbb{S}},\widehat{\mathcal{O}},\widehat{\pi })$ is
isomorphic to $(\mathbb{P}_{\mathbb{C}}(\mathcal{H}\mathbb{)},SA(\mathcal{H}%
\mathbb{)},p)$ we can suppose to be in this case. Note that if $E$ is a
projector in $\mathcal{H}$ the self-adjoint operator $\Omega (E)$ is also a
projector because $E_{\mathbb{R\setminus }\left\{ 0,1\right\} }^{\Omega
(E)}=0$.

Taken two antipodal classes $\left[ \varphi \right] $ and $\left[ \psi %
\right] $ in $\mathbb{P}_{\mathbb{C}}(\mathcal{H}\mathbb{)}$ it is possible
to find a projector $E$ such that $\left\langle E_{\left\{ 1\right\}
}^{E}\right\rangle _{\left[ \varphi \right] }=0$ and $\left\langle
E_{\left\{ 1\right\} }^{E}\right\rangle _{\left[ \psi \right] }=1$,
therefore $\left\langle E_{\left\{ 1\right\} }^{\Omega (E)}\right\rangle
_{\Lambda \left[ \varphi \right] }=0$ and $\left\langle E_{\left\{ 1\right\}
}^{\Omega (E)}\right\rangle _{\Lambda \left[ \psi \right] }=1$ and then the
classes $\Lambda \left[ \varphi \right] $ and $\Lambda \left[ \psi \right] $
are antipodal too; that is $\Lambda $ is a bijective transformation of $%
\mathbb{P}_{\mathbb{C}}(\mathcal{H}\mathbb{)}$ preserving the antipodality.

This implies that (cfr. [U] Thm. 5.1) there exists a unitary or antiunitary
map $U:\mathcal{H\rightarrow H}$ such that $\Lambda =\widehat{U}$, that is $%
\Lambda \left[ \varphi \right] =[U\varphi ]$ for every $[\varphi ]$ in $%
\mathbb{P}_{\mathbb{C}}(\mathcal{H}\mathbb{)}$.

It holds: $\left\langle E_{B}^{U\circ A\circ U^{-1}}\right\rangle _{\left[
U\varphi \right] }=\left\langle E_{B}^{A}\right\rangle _{\left[ \varphi %
\right] }=\left\langle E_{B}^{\Omega (A)}\right\rangle _{\Lambda \left[
\varphi \right] }=\left\langle E_{B}^{\Omega (A)}\right\rangle _{\left[
U\varphi \right] }$ for every $\varphi $ and every borel subset $B$ of $%
\mathbb{R}$; then $\Omega (A)=U\circ A\circ U^{-1}$ for every self-adjoint
operator $A$. Remembering that $\tau (f\circ U^{-1})=U\circ \tau (f)\circ
U^{-1}$ we get the thesis for $(\widehat{\mathbb{S}},\widehat{\mathcal{O}},%
\widehat{\pi })$.
\end{proof}

\bigskip

\begin{remark}
If $(\chi ^{-1},\eta )$ and $(\chi _{1}^{-1},\eta _{1})$ \textit{are two
isomorphism between} $(\widehat{\mathbb{S}},\widehat{\mathcal{O}},\widehat{%
\pi })$ \textit{and} $(\mathbb{P}_{\mathbb{C}}(\mathcal{H}\mathbb{)},SA(%
\mathcal{H}\mathbb{)},p)$ then the couple $(\chi _{1}^{-1}\circ \chi ,\eta
_{1}\circ \eta ^{-1})$ is an automorphism of $(\mathbb{P}_{\mathbb{C}}(%
\mathcal{H}\mathbb{)},SA(\mathcal{H}\mathbb{)},p)$ and therefore, reasoning
as in the proof of the previous theorem, comes from a unitary or antiunitary
map $U:\mathcal{H\rightarrow H}$. Hence on $\widehat{\mathbb{S}}$ there is a
unique natural topological and differentiable structure; in particular this
means that the imprecise observer is able to reconstruct the topology of $%
\widehat{\mathbb{S}}$ assigned by the precise observer.
\end{remark}

\bigskip

\begin{theorem}
There exists a unique homomorphism $\sigma :\Theta Sym(\widehat{\mathbb{S}}$
, $\widehat{\mathcal{O}})\rightarrow \left\{ 1,-1\right\} $ such that $%
\sigma (\Lambda ,\Omega )=1$ if and only if there is another element $%
(\Lambda _{1},\Omega _{1})$ such that \ $(\Lambda _{1},\Omega
_{1})^{2}=(\Lambda ,\Omega )$.
\end{theorem}

\begin{proof}
It is clear that the unicity follows from the characterization of the
elements where $\sigma $ takes the value $+1$. Since $(\widehat{\mathbb{S}},%
\widehat{\mathcal{O}},\widehat{\pi })$ is isomorphic to $(\mathbb{P}_{%
\mathbb{C}}(\mathcal{H}\mathbb{)},SA(\mathcal{H}\mathbb{)},p)$ we can
suppose to be in this case. As in the proof of a previous theorem we can
prove that for each $\Lambda =\widehat{U}$ , where $U$ is unitary or
antiunitary, there is a well defined sign $\sigma (\Lambda )=\sigma (U)=+1$
for $U$ unitary and $\sigma (\Lambda )=\sigma (U)=-1$ for $U$ antiunitary
such that in $\mathbb{S}$ it holds the property:
\begin{equation*}
U_{\ast \varphi }^{\mathcal{H}or}\circ J_{_{\varphi }}=\sigma (U)\cdot
J_{_{U\varphi }}\circ U_{\ast \varphi }^{\mathcal{H}or}
\end{equation*}
and therefore in $\mathbb{P}_{\mathbb{C}}(\mathcal{H}\mathbb{)}$ it holds
the property :
\begin{equation*}
\Lambda _{\ast \lbrack \varphi ]}\circ J_{[\varphi ]}=\sigma (\Lambda )\cdot
J_{\Lambda \lbrack \varphi ]}\circ \Lambda _{\ast \lbrack \varphi ]}
\end{equation*}
It's easy to prove then that the map $\sigma :\Theta Sym(\mathbb{P}_{\mathbb{%
C}}(\mathcal{H}\mathbb{)},SA(\mathcal{H}\mathbb{)})\rightarrow \left\{
+1,-1\right\} $ defined by $\sigma (\Lambda ,\Omega )=\sigma (\Lambda )$ is
a group homomorphism.

If $(\Lambda ,\Omega )=(\Lambda _{1},\Omega _{1})^{2}$ for an element $%
(\Lambda _{1},\Omega _{1})$ then $\sigma (\Lambda ,\Omega )=\left[ \sigma
(\Lambda _{1},\Omega _{1})\right] ^{2}=1$.

Conversely if $\sigma (\Lambda ,\Omega )=\sigma (\Lambda )=1$ then $\Lambda =%
\widehat{U}$ where $U=e^{-iA}$ for a self-adjoint transformation $A$ and $%
\Omega (T)=\widehat{U^{\ast }}[T]=U\circ T\circ U^{-1}$; therefore taken $%
\Lambda _{1}=$ $\widehat{e^{-\frac{i}{2}A}}$ and $\Omega _{1}(T)=e^{-\frac{i%
}{2}A}\circ T\circ e^{\frac{i}{2}A}$ we have an element in $\Theta Sym(%
\mathbb{P}_{\mathbb{C}}(\mathcal{H}\mathbb{)},SA(\mathcal{H}\mathbb{)})$
with $(\Lambda _{1},\Omega _{1})^{2}=(\Lambda ,\Omega )$.
\end{proof}

\bigskip

\begin{definition}
A semi-symmetry $(\Lambda ,\Omega )$ will be called a \textbf{symmetry} of $(%
\widehat{\mathbb{S}},\widehat{\mathcal{O}},\widehat{\pi })$ if $\sigma
(\Lambda ,\Omega )=1$.
\end{definition}

\bigskip

\begin{notation}
We will denote by $Sym(\widehat{\mathbb{S}}$ , $\widehat{\mathcal{O}})$ the
set of all symmetries of $(\widehat{\mathbb{S}},\widehat{\mathcal{O}},%
\widehat{\pi })$; $Sym(\widehat{\mathbb{S}}$ , $\widehat{\mathcal{O}})$ is a
normal subgroup of $\Theta Sym(\widehat{\mathbb{S}}$ , $\widehat{\mathcal{O}}%
)$.
\end{notation}

\bigskip

\begin{notation}
\textit{Every automorphism }$\Phi $ \textit{of} $\mathbb{S}$ \textit{brings
equivalence classes of }$\mathcal{R}_{\mathbb{S}}$ (\textit{the} $\mathbb{S}%
^{1}$\textit{-orbits) in equivalence classes, therefore defines an \textbf{%
induced transformation} }$\widehat{\Phi }$ \textbf{of }$\widehat{\mathbb{S}%
\text{ }}$ by $\widehat{\Phi }\left[ \varphi \right] =$ $[\Phi (\varphi )]$.
\ S\textit{ince an automorphism }$\Phi $ \textit{of }$\ \mathbb{S}$ \textit{%
is a measure equivalence t}he map $\Phi ^{\ast }:\mathcal{O\rightarrow O}$ ,%
\textit{\ given by }$\Phi ^{\ast }(f)=f\circ \Phi ^{-1}$, \textit{brings
equivalence classes of }$\mathcal{R}_{\mathcal{O}}$ in\textit{\ equivalence
classes of }$\mathcal{R}_{\mathcal{O}}$ and therefore is well defined an
\textit{\textbf{induced transformation} }$\widehat{\Phi ^{\ast }}$ \textbf{%
of }$\widehat{\mathcal{O}}$ by $\widehat{\Phi ^{\ast }}[f]=[f\circ \Phi
^{-1}]$.
\end{notation}

\bigskip

\begin{remark}
Note that $\widehat{\pi }(\widehat{\Phi }\left[ \varphi \right] ,\widehat{%
\Phi ^{\ast }}[f],B)=\widehat{\pi }(\left[ \varphi \right] ,\left[ f\right]
,B)$\textit{and then }$\left( \widehat{\Phi },\widehat{\Phi ^{\ast }}\right)
$ \textit{is a} semi-symmetry\textbf{\ }\ for $(\widehat{\mathbb{S}},%
\widehat{\mathcal{O}},\widehat{\pi })$ and using the definitions it's easy
to check that $\sigma (\widehat{\Phi },\widehat{\Phi ^{\ast }})=1$. We
express this fact by saying that \textbf{''a symmetry for the precise
observer appears as a symmetry to the imprecise observer''.}
\end{remark}

\bigskip

\begin{remark}
\textit{An automorphism }$\Phi $\textit{\ of} $Aut(\mathbb{S)}$ \textit{%
induces on} $\widehat{\mathbb{S}}$ \textit{the identity map if and only if
is in }$Aut_{I}(\mathbb{S)}$\textit{, then the actions }$\widehat{\Phi }$
\textit{on} $\widehat{\mathbb{S}}$ \textit{belong to the transformation group%
} $\widehat{Aut(\mathbb{S)}}=Aut(\mathbb{S)}/Aut_{I}(\mathbb{S)}$ \textit{%
(acting effectively on} $\widehat{\mathbb{S}})$.
\end{remark}

\bigskip

\textit{Remember that in quantum mechanics a \textbf{symmetry} of} $\mathbb{P%
}_{\mathbb{C}}(\mathcal{H}\mathbb{)}$ \textit{can be described as a
diffeomorphism} $\Lambda :\mathbb{P}_{\mathbb{C}}(\mathcal{H}\mathbb{%
)\rightarrow P}_{\mathbb{C}}(\mathcal{H}\mathbb{)}$ \textit{induced by
unitary transformations of }$\mathcal{H}$; \textit{therefore the \textbf{%
symmetry group} of} $\mathbb{P}_{\mathbb{C}}(\mathcal{H}\mathbb{)}$ \textit{%
is:}
\begin{equation*}
\mathbf{PROJ}(\mathbb{P}_{\mathbb{C}}(\mathcal{H}\mathbb{))=}\left\{ \Lambda
;\text{ }\Lambda \text{ is a symmetry of }\mathbb{P}_{\mathbb{C}}(\mathcal{H}%
\mathbb{)}\right\}
\end{equation*}
(\textit{with the topology of pointwise convergence) and} \textit{is
naturally isomorphic to the group} $Unit(\mathcal{H}\mathbb{)}/(\mathbb{S}%
^{1}\cdot I)$.

\textit{The following theorem proves that the symmetries induced on} $%
\widehat{\mathbb{S}}$ \textit{by the automorphisms} \textit{of} $\mathbb{S}$
\textit{are precisely the natural symmetries for the imprecise observer and }
\textit{through the identification} $\mathbb{\chi }:\mathbb{P}_{\mathbb{C}}(%
\mathcal{H}\mathbb{)\rightarrow }\widehat{\mathbb{S}}$ \textit{the} \textit{%
symmetries for the imprecise observer} \textit{become the symmetries of \ }$%
\mathbb{P}_{\mathbb{C}}(\mathcal{H}\mathbb{)}$:

\bigskip

\begin{theorem}
\begin{enumerate}
\item  The map $\Sigma :\widehat{Aut(\mathbb{S)}}\rightarrow Sym(\widehat{%
\mathbb{S}},\widehat{\mathcal{O}})$ defined by the expression: $\Sigma (%
\left[ \Phi \right] )=\left( \widehat{\Phi },\widehat{\Phi ^{\ast }}\right) $
is a group isomorphism.

\item  The map $\Pi :Sym(\widehat{\mathbb{S}},\widehat{\mathcal{O}}%
)\rightarrow \mathbf{PROJ}(\mathbb{P}_{\mathbb{C}}(\mathcal{H}\mathbb{))}$
defined by: $\Pi (\Lambda ,\Omega )=\mathbb{\chi }^{-1}\circ \Lambda \circ
\mathbb{\chi }$ is a group isomorphism.
\end{enumerate}
\end{theorem}

\begin{proof}
1. If $\Sigma (\left[ \Phi \right] )=e_{Sym(\widehat{\mathbb{S}},\widehat{%
\mathcal{O}})}$ then $\Phi $ sends every $\mathbb{S}^{1}$-orbit in itself
and therefore $\left[ \Phi \right] =e_{\widehat{Aut(\mathbb{S)}}}$. Moreover
$\Sigma $ is surjective since every symmetry in $Sym(\widehat{\mathbb{S}},%
\widehat{\mathcal{O}})$ comes from a unitary transformation.

2. Since $(\widehat{\mathbb{S}},\widehat{\mathcal{O}},\widehat{\pi })$ is
isomorphic to $(\mathbb{P}_{\mathbb{C}}(\mathcal{H}\mathbb{)},SA(\mathcal{H}%
\mathbb{)},p)$ we can suppose to be in this case with $\Pi (\Lambda ,\Omega
)=\Lambda $. Since every element in $\mathbf{PROJ}(\mathbb{P}_{\mathbb{C}}(%
\mathcal{H}\mathbb{))}$ comes from a unitary transformation the map $\Pi $
is surjective. If $\Pi (\Lambda ,\Omega )=\Lambda =id_{\mathbb{P}_{\mathbb{C}%
}(\mathcal{H}\mathbb{)}}$ then for every self-adjoint operator $A$ we have $%
\left\langle E_{B}^{\Omega (A)}\right\rangle _{\varphi }=\left\langle
E_{B}^{A}\right\rangle _{\varphi }$ for every $\varphi $ in $\mathbb{S}$ and
every borel set $B$; therefore $\Omega (A)=A$ and $(\Lambda ,\Omega
)=(id,id) $.
\end{proof}

\bigskip

\begin{remark}
The isomorphisms $\Sigma $ and $\Pi $ respect also the actions of the
''symmetries'' on the ''observables''.
\end{remark}

\bigskip

\begin{remark}
We will consider on $\widehat{Aut(\mathbb{S)}}$ the \textbf{topology}
induced by $\widehat{\mathbb{S}}^{\widehat{\mathbb{S}}}$(the topology of
''pointwise convergence'') and on $Sym(\widehat{\mathbb{S}}$ , $\widehat{%
\mathcal{O}})$ the topology that makes the isomorphism $\Sigma $ a
homeomorphism. From now on we will simply identify $Sym(\widehat{\mathbb{S}}$
, $\widehat{\mathcal{O}})$ with $\widehat{Aut(\mathbb{S)}}$.
\end{remark}

\bigskip

\begin{definition}
A \textbf{dynamic} in $(\widehat{\mathbb{S}}$ , $\widehat{\mathcal{O}})$ is
a continuous 1-parameter group in $\widehat{Aut(\mathbb{S)}}$ (in $Sym(%
\widehat{\mathbb{S}},\widehat{\mathcal{O}})$).
\end{definition}

\bigskip

\textit{It is easy to check that a one-parameter continuous group} $\Phi
_{\cdot }:\mathbb{R\rightarrow }Aut(\mathbb{S)}$ \textit{induces a
one-parameter continuous group}: $\widehat{\Phi }_{\cdot }:\mathbb{%
R\rightarrow }\widehat{Aut(\mathbb{S)}}$ ,\textit{\ therefore every dynamic
in the physical system }$\mathbb{S}$ \textit{induces an ''apparent'' }%
\textbf{dynamic on} $\widehat{\mathbb{S}}$ \textbf{seen by the imprecise
observer}\textit{. Two dynamics }$\Phi _{\cdot }$\textit{\ and }$\Psi
_{\cdot }$\textit{\ on} $\mathbb{S}$ \textit{give origin to the same dynamic
on} $\widehat{\mathbb{S}}$ $\ $\textit{for the imprecise observer (}$%
\widehat{\Phi }_{\cdot }=\widehat{\Psi }_{\cdot }$\textit{) if and only if
there is a continuous map} $\nu _{\cdot }:\mathbb{R\rightarrow }Aut_{I}(%
\mathbb{S)}$ \textit{such that }$\Psi _{t}=\Phi _{t}\circ \nu _{t}$\textit{\
for every }$t$\textit{\ in} $\mathbb{R}$.

\textit{In this way on the set} $\left\{ \Phi _{\cdot };\Phi _{\cdot }:%
\mathbb{R\rightarrow }Aut(\mathbb{S)}\text{ \ \textit{is a dyn. on }}\mathbb{%
S}\right\} $ \textit{of all dynamics on} $\mathbb{S}$ \textit{is defined}
\textit{an equivalence relation} $\mathcal{R}_{Dyn}$ .

\bigskip

\textit{Let's remember that all the continuous dynamics on} $\mathbb{P}_{%
\mathbb{C}}(\mathcal{H}\mathbb{)}$ \textit{are expressable as:} $t\mapsto
e^{-itH}$ \textit{where} $H$ \textit{is a self-adjoint operator on} $%
\mathcal{H}$ \textit{and that two operators} $H_{1}$ \textit{and} $H_{2}$
\textit{define the same dynamic in }$\mathbb{P}_{\mathbb{C}}(\mathcal{H}%
\mathbb{)}$ \textit{if and only if} $H_{2}-H_{1}$ \textit{is in} $\mathbb{%
R\cdot }I$. \textit{Therefore the set of all the dynamics of} $\mathbb{P}_{%
\mathbb{C}}(\mathcal{H}\mathbb{)}$ \textit{can be} \textit{represented by
the set:} $SA(\mathcal{H})/\mathbb{R\cdot }I$.

\textit{The following theorem proves that the apparent dynamics induced on} $%
\widehat{\mathbb{S}}$ \textit{by the dynamics} \textit{of} $\mathbb{S}$
\textit{are precisely the natural dynamics for the imprecise observer and }
\textit{moreover that}, \textit{through the identification} $\mathbb{\chi }:%
\mathbb{P}_{\mathbb{C}}(\mathcal{H}\mathbb{)\rightarrow }\widehat{\mathbb{S}}
$, \textit{the} \textit{dynamics for the imprecise observer} \textit{become
the dynamics of \ }$\mathbb{P}_{\mathbb{C}}(\mathcal{H}\mathbb{)}$:

\bigskip

\begin{theorem}
\begin{enumerate}
\item  The map $\Delta :SA(\mathcal{H})/\mathbb{R\cdot }I\rightarrow $ $%
\left\{ \widehat{\Phi }_{\cdot };\widehat{\Phi }_{\cdot }\text{ \ is a dyn.
on }\widehat{\mathbb{S}}\right\} $ defined by $\Delta \left[ H\right]
=\left( \widehat{e^{-itH}}\right) _{t\in \mathbb{R}}$ is bijective.

\item  T\textit{he map:} $\delta :\left\{ \Phi _{\cdot };\Phi _{\cdot }\text{
\ is a dyn. \textit{on }}\mathbb{S}\right\} /\mathcal{R}_{Dyn}\rightarrow
\left\{ \widehat{\Phi }_{\cdot };\widehat{\Phi }_{\cdot }\text{ \ is a dyn.
on }\widehat{\mathbb{S}}\right\} $ \textit{defined by} $\delta \left[ \Phi
_{\cdot }\right] =\left( \widehat{\Phi }_{\cdot }\right) $ \textit{is well
defined and} bijective.
\end{enumerate}
\end{theorem}

\begin{proof}
1.\ If $\Theta :\mathbb{R\rightarrow }\widehat{Aut(\mathbb{S)}}$ is a
one-parameter continuous group then for every $t$ there is\textit{\ }an
automorphism $\Phi _{t}$ such that $\Theta _{t}=\left[ \Phi _{t}\right]
_{Aut_{I}(\mathbb{S)}}$. This implies that for each $\varphi $ in $\mathbb{S}
$ the map: $t\rightarrow \left[ \Phi _{t}(\varphi )\right] _{\widehat{%
\mathbb{S}}}$ is continuous and then is continuous the map: $t\rightarrow
\widehat{\Phi _{t}}\left[ \varphi \right] _{\mathbb{P}_{\mathbb{C}}(\mathcal{%
H}\mathbb{)}}=\left[ \Phi _{t}(\varphi )\right] _{\mathbb{P}_{\mathbb{C}}(%
\mathcal{H}\mathbb{)}}$. Therefore the map $\widehat{\Phi _{\cdot }}:\mathbb{%
R\rightarrow }\mathbf{PROJ}(\mathbb{P}_{\mathbb{C}}(\mathcal{H}\mathbb{))}$
is a one-parameter continuous group and by a theorem (cfr. [Ba]) there is a%
\textit{\ }self-adjoint operator\textit{\ }$H$ such that $\widehat{\Phi _{t}}%
\left[ \varphi \right] _{\mathbb{P}_{\mathbb{C}}(\mathcal{H}\mathbb{)}}=%
\left[ e^{-itH}\varphi \right] _{\mathbb{P}_{\mathbb{C}}(\mathcal{H}\mathbb{)%
}}$ for every $\left[ \varphi \right] _{\mathbb{P}_{\mathbb{C}}(\mathcal{H}%
\mathbb{)}}$. This proves the surjectivity of $\Delta $.

2. The map $\delta $ is obviously injective; if $\Theta :\mathbb{%
R\rightarrow }\widehat{Aut(\mathbb{S)}}$ is a one-parameter continuous group
we already know that exists a self-adjoint operator $H$ such that\ $\Theta
_{t}=\widehat{e^{-itH}}$ for every $t$ in $\mathbb{R}$, therefore $t\mapsto
e^{-itH}$ is a one-parameter continuous group in $Aut(\mathbb{S)}$ whose
class is sent by $\delta $ in $\Theta $.
\end{proof}

\bigskip

\bigskip

\section{\protect\LARGE Bibliography}

\bigskip

[A] \ D. Aerts: A possible explanation for the probabilities of quantum
mechanics.

\qquad J. Math. Phys. -1986-vol. 27, p.202-210

\bigskip

[B] \ G. Boniolo(a cura di): Filosofia della fisica.

\qquad Bruno Mondadori, Milano 1997

\bigskip

[Ba] V. Bargmann: On unitary ray representations of continuous groups.

\qquad Ann. Math. - 1954- vol. 59, p.1-46

\bigskip

[C] \ \ A. Cassa: Quantum physical systems as classical systems.

\qquad J. Math. Phys. -2001-vol. 42, N. 11- p. 5143-49

\bigskip

[CGM] R. Cirelli, M. Gatti and A. Mani\`{a}: On the nonlinear extension of

\qquad quantum superposition and uncertainty principles.

\qquad J. Geom. and Phys.- 1999-vol 29, N. 1-2, p.64-86

\bigskip

[CM] \ B. Coecke and D.J. Moore: Decompositions of probability measures on

\qquad complete ortholattices in join preserving maps.

\qquad Preprint.

\bigskip

[CMP] \ \ R. Cirelli, A. Mani\`{a} and L. Pizzocchero: Quantum mechanics as
an

\qquad infinite-dimensional Hamiltonian system with...- Part I and II.

\qquad J. Math. Phys. -1990-vol. 31, N. 12- p. 2891-2903

\bigskip

[E] G.G. Emch: Mathematical and conceptual foundations of 20th century
physics.

\qquad North-Holland Math. studies, Amsterdam 1984

\bigskip

[G] \ \ M. Gatti: Sull'estensione non lineare dei principi fondamentali

\qquad della meccanica quantistica.

\qquad 1996 - Ph. D. Thesis.

\bigskip

[G-D] G.C. Ghirardi, F. De Stefano: Il mondo quantistico, una realt\`{a}
ambigua.

\qquad in Ambiguit\`{a} (a cura di) C. Magris

\qquad Longo, Moretti e Vitali Bergamo 1996

\bigskip

[Gl] \ A.M. Gleason: Measures on the closed subspaces of a Hilbert space.

\qquad J. Math. and Mech. -1957- N.6, p. 123-133

\bigskip

[K-S] \ \ J.L. Kelley and T.P. Srinivasan: Measure and Integral - Vol. 1.

\qquad Springer-Verlag, NY 1988

\bigskip

[KS] S. Kochen and E. Specker: The problem of hidden variables in Quantum

\qquad Mechanics.

\qquad J. Math. and Mech. -1967- N.17, p. 59-87

\bigskip

[J] \ \ J. M. Jauch: Foundations of quantum mechanics.

\qquad Addison-Wesley P.C., Reading (Mass.) 1968

\bigskip

[J-P1] \ \ J.M. Jauch and C. Piron: Foundations of quantum mechanics.

\qquad John Wiley \& sons, NY 1974

\bigskip

[J-P2] J. M. Jauch and C. Piron: Can hidden variables be excluded in quantum

\qquad mechanics?

\qquad Helv. Phys. Acta 36, 827-837 (1963)

\bigskip

[La] S. Lang: Differential and riemannian manifolds.

\qquad Springer-Verlag, NY 1995

\bigskip

[M] G. Mackey: Mathematical foundations of quantum mechanics

\qquad Benjamin, NY 1963

\bigskip

[N] J. von Neumann: Mathematical foundations of quantum mechanics.

\qquad PUP, Princeton NJ 1955

\bigskip

[Pi] C. Piron: Foundations of quantum physics.

\qquad Mathematical Physics monograph series (ed. A Wightman).

\qquad Benjamin, Reading 1976

\bigskip

[Pr] \ \ E. Prugovecki: Quantum mechanics in Hilbert space.

\qquad Ac. Press Inc., NY 1981

\bigskip

[R] \ \ H.L. Royden: Real analysis.

\qquad The MacMillan Co., Toronto 1968

\bigskip

[U] U. Uhlorn: Representation of symmetry transformations in quantum

\qquad mechanics.

\qquad Arkiv for Fysik 1962, 23 p. 307-340

\bigskip

[V] \ \ V.S. Varadarajan: Geometry of quantum theory. Vol. I

\qquad Van Nostrand, Princeton 1984

\bigskip

[W] \ \ J. Weidmann: Linear operators in Hilbert spaces.

\qquad Springer-Verlag, NY 1980

\end{document}